\PassOptionsToPackage{usenames,dvipsnames}{xcolor}
\documentclass[twocolumn,twocolappendix]{aastex63}
\pdfoutput=1 
\usepackage[T1]{fontenc}
\usepackage{ae,aecompl}
\usepackage[utf8]{inputenc}
\usepackage[english]{babel}
\usepackage{amsmath,amstext}
\usepackage{apjfonts}
\usepackage{chngcntr}
\usepackage[figure,figure*]{hypcap}
\usepackage[hang]{footmisc}
\input{hyperlink-year-only-natbib-patch}
\allowdisplaybreaks
\graphicspath{{figures/}}

\newcommand*{\https}[1]{\href{https://#1}{\nolinkurl{#1}}}
\newcommand*{\http}[1]{\href{http://#1}{\nolinkurl{#1}}}

\defcitealias{Geha2017}{{Paper\,I}}
\newcommand*{\paperone}{\citetalias{Geha2017}}

\newcommand*{\nhosts}{36}
\newcommand*{\nhostsnew}{28}
\newcommand*{\nhoststotal}{115}
\newcommand*{\nhostsinclnoimage}{205}
\newcommand*{\nhostswsats}{34}
\newcommand*{\nhostswsatslimit}{33}
\newcommand*{\nhostsnosats}{two}

\newcommand*{\nsats}{127}
\newcommand*{\nsatslimit}{123}
\newcommand*{\nsatsbelow}{four}
\newcommand*{\nsatssaga}{69}
\newcommand*{\nsatslimitsaga}{66}

\newcommand*{\nzsaga}{25,372}

\newcommand*{\nzsaganew}{12,690}

\newcommand*{\nzother}{28,425}
\newcommand*{\nzothernew}{21,075}

\newcommand*{\nzsaganewrvir}{11,726}

\newcommand*{\kms}{\ensuremath{\mathrm{km}\,\mathrm{s}^{-1}}}
\newcommand*{\msun}{\ensuremath{M_\odot}}

\newcommand*{\LCDM}{$\Lambda$CDM}
\newcommand*{\mueff}{\ensuremath{\mu_{r_o,\mathrm{eff}}}}
\newcommand*{\code}[1]{\ensuremath{\mathtt{#1}}}
\newcommand*{\ditto}{\raisebox{-0.5ex}{''}}

\shorttitle{The SAGA Survey.\ II.}
\shortauthors{Mao et al.}

\begin{document}
\title{The SAGA Survey.\ II.\ Building a Statistical Sample of Satellite Systems around Milky Way--like Galaxies}

\author[0000-0002-1200-0820]{Yao-Yuan~Mao}
\altaffiliation{NASA Einstein Fellow; \href{mailto:Yao-Yuan Mao <yymao.astro@gmail.com>}{yymao.astro@gmail.com}%
}
\affiliation{Department of Physics and Astronomy, Rutgers, The State University of New Jersey, Piscataway, NJ 08854, USA}
\author[0000-0002-7007-9725]{Marla~Geha}
\affiliation{Department of Astronomy, Yale University, New Haven, CT 06520, USA}
\author[0000-0003-2229-011X]{Risa~H.~Wechsler}
\affiliation{Kavli Institute for Particle Astrophysics and Cosmology and Department of Physics, Stanford University, Stanford, CA 94305, USA}
\affiliation{SLAC National Accelerator Laboratory, Menlo Park, CA 94025, USA}
\author[0000-0001-6065-7483]{Benjamin~Weiner}
\affiliation{Department of Astronomy and Steward Observatory, University of Arizona, Tucson, AZ 85721, USA}
\author[0000-0002-9599-310X]{Erik~J.~Tollerud}
\affiliation{Space Telescope Science Institute, 3700 San Martin Drive, Baltimore, MD 21218, USA}
\author[0000-0002-1182-3825]{Ethan~O.~Nadler}
\affiliation{Kavli Institute for Particle Astrophysics and Cosmology and Department of Physics, Stanford University, Stanford, CA 94305, USA}
\author[0000-0002-3204-1742]{Nitya~Kallivayalil}
\affiliation{Department of Astronomy, University of Virginia, 530 McCormick Road, Charlottesville, VA 22904, USA}

\begin{abstract}
We present the Stage~II results from the ongoing Satellites Around Galactic Analogs (SAGA) Survey. Upon completion, the SAGA Survey will spectroscopically identify satellite galaxies brighter than $ M_{r,o} = -12.3 $ around 100 Milky Way (MW) analogs at $ z \sim 0.01 $. 
In Stage~II, we have more than quadrupled the sample size of Stage~I, delivering results from \nsats{} satellites around \nhosts{} MW analogs with an improved target selection strategy and deep photometric imaging catalogs from the Dark Energy Survey and the Legacy Surveys.
We have obtained \nzsaga{} galaxy redshifts, peaking around $ z = 0.2$. These data significantly increase spectroscopic coverage for very low redshift objects in $ 17 < r_o < 20.75 $ around SAGA hosts, creating a unique data set that places the Local Group in a wider context.
The number of confirmed satellites per system ranges from zero to nine, and correlates with host galaxy and brightest satellite luminosities.
We find that the number and the luminosities of MW satellites are consistent with being drawn from the same underlying distribution as SAGA systems. 
The majority of confirmed SAGA satellites are star forming, and the quenched fraction increases as satellite stellar mass and projected radius from the host galaxy decrease. Overall, the satellite quenched fraction among SAGA systems is lower than that in the Local Group.
We compare the luminosity functions and radial distributions of SAGA satellites with theoretical predictions based on cold dark matter simulations and an empirical galaxy--halo connection model and find that the results are broadly in agreement.
\end{abstract}

\keywords{\href{http://astrothesaurus.org/uat/1378}{Redshift surveys (1378)}, \href{http://astrothesaurus.org/uat/416}{Dwarf galaxies (416)}, \href{http://astrothesaurus.org/uat/942}{Luminosity function (942)}, \href{http://astrothesaurus.org/uat/612}{Galaxy physics (612)}, \href{http://astrothesaurus.org/uat/1880}{Galaxy dark matter halos (1880)}}

\section{Introduction}
\label{sec:intro}

\setcounter{footnote}{8}

The Satellites Around Galactic Analogs (SAGA) Survey, first presented in \citet[][hereafter \paperone{}]{Geha2017}, is an ongoing spectroscopic galaxy survey. Its primary science goal is to characterize the satellite populations down to the luminosity of the Leo I dwarf galaxy ($M_{r,o}<-12.3$), around 100 Milky Way (MW) analogs outside of the Local Volume (at 25--40.75\,Mpc).
SAGA \paperone{} included 27 satellites identified around eight MW-like systems.
In this work, we present the Stage II results of the SAGA Survey, which includes \nsats{} satellites identified around \nhosts{} MW-like systems and marks the completion of more than one-third of the full survey. 

The SAGA Survey complements the rich observational data sets of MW satellite galaxies \citep[][and references therein]{Drlica-Wagner191203302}, and provides critical tests of the Lambda Cold Dark Matter (\LCDM{}) paradigm. 
Apparent discrepancies between the observed MW satellite population and \LCDM{} predictions have been considered as ``small-scale challenges'' to \LCDM{} \citep[e.g.,][]{Weinberg2015,Bullock2017}.
These challenges include \LCDM{} overpredicting both the number and the central mass densities of the MW satellites; however, it has long been unclear whether these discrepancies can be explained via a more realistic treatment of baryons \citep[e.g.,][]{Brooks2017, Read2019}, the \LCDM{} paradigm itself is incorrect \citep[e.g.,][]{Polisensky2014}, or other salient simplifications have impacted the comparison. 
These small-scale challenges have been largely based on the MW and M31 \citep[e.g.,][]{Tollerud2014}, just two systems with possibly correlated satellite populations \citep{GK2014}. 
The SAGA Survey addresses these limitations by constraining the distribution of satellites around MW-mass galaxies, putting the MW and its bright satellite population into a cosmological context and allowing us to compare observations and theory statistically.

The SAGA Survey also provides a benchmark for understanding low-mass galaxy evolution. The galaxy--halo connection is reasonably well constrained for galaxies whose stellar mass $M_* > 10^9$--$10^{10}\,\msun$, but much uncertainty still exists below this scale \citep[e.g.,][]{2015ARA&A..53...51S,WechslerTinker}. Unresolved questions include how the interaction between satellite galaxies and their hosts affects satellite star formation histories \citep[e.g.,][]{Simpson17:1705.03018,Hausammann:2019A&A...624A..11H}.
The SAGA Survey focuses on bright and classical dwarf satellite galaxies ($M_* \sim 10^6$--$10^9\,\msun$), providing crucial information for our understanding of the galaxy--halo connection and host--satellite interactions. 
For example, a surprising result in \paperone{} was that the vast majority (26/27) of satellites in eight SAGA systems were star-forming, as compared to two of the five MW satellites in the same luminosity range. 

In recent years, efforts have been undertaken to characterize satellite systems beyond the Local Group, but around nearby systems in the Local Volume ($<20$\,Mpc). Within this distance, one can use resolved stars or surface brightness, in addition to spectroscopic redshifts, to search for satellites. The MW-like satellite systems that have been studied in the Local Volume include NGC\,4258 \citep{Spencer2014}, M94 \citep{Smercina2018}, NGC\,3175 \citep{Kondapally2018}, NGC\,2950 and NGC\,3245 \citep{Tanaka2018}. A handful of satellite systems around galaxies of different masses have also been studied, such as the systems around M81 \citep{Chiboucas2013}, M101 \citep{Danieli:2017ApJ...837..136D,Bennet2019a,Carlsten:2019ApJ...878L..16C}, and Centaurus~A \citep{Crnojevic2019,Muller:1907.02012}. 
Regardless of the host galaxy mass, the number of complete satellite systems in the Local Volume remains limited. It was not until very recently that a statistical study of the Local Volume satellite systems became possible \citep{Carlsten200602443,Carlsten200602444,LBT-SONG:2003.08352}.

Beyond 20\,Mpc, dwarf galaxies are difficult to distinguish from the far more numerous background galaxy population using photometry alone (although attempts have been made; e.g., \citealt{Xi:1805.07407}). Confirming satellite galaxies hence requires extensive spectroscopic follow-up \citep[e.g.,][]{Zaritsky:1997ApJ...478...39Z}, which is the main challenge of the SAGA Survey.
Existing deep spectroscopic surveys either explicitly remove low-redshift galaxies \citep[e.g., DEEP2;][]{Newman2013} or have limiting magnitudes which allow study of only the brightest satellites 
(e.g., Sloan Digital Sky Survey (SDSS), $r<17.77$; \citealt{Strauss2002,Sales2013}; and GAMA, $r < 19.8$; \citealt{Baldry2018:GAMA:DR3}).
The SAGA Survey was hence born to tackle this challenge before ongoing, planned, and proposed deep spectroscopic surveys become available (e.g., DESI; \citealt{1611.00036}; 4MOST; \citealt{2012SPIE.8446E..0TD}; MSE; \citealt{2016arXiv160600060M}).

On the theoretical side, modern hydrodynamical simulations span a wide range of scales, including cosmological volumes that include large samples of MW-mass galaxies but only resolve their most massive satellites \citep[e.g.,][]{Schaye2015,Pillepich2018}, zoom-in simulations of individual MW analogs that resolve many of their satellites  \citep[e.g.,][]{Wetzel2016,Garrison-Kimmel2019:1806.04143,2008.11207}, and high-resolution simulations that resolve individual star forming regions around individual dwarf galaxies \citep[e.g.,][]{Munshi2019,Wheeler2018}. 
SAGA results will serve as observational tests against the physical models implemented in such simulations~\citep[e.g.,][]{Akins:2008.02805,Samuel2019a}. 

Theoretical frameworks that characterize how host halo properties affect satellite populations \citep{Mao150302637,1807.05180}, along with empirical and semianalytical models to produce realistic satellite systems \citep{Nadler180905542,Jiang:2005.05974}, are also rapidly being developed.
These theories and galaxy--halo connection models enable statistical inferences that can marginalize over the uncertainties in baryonic physics, providing a new way to test the \LCDM{} paradigm.
\citet{Nadler190410000,Nadler191203303,Nadler:2008.00022} have tested this approach with MW satellites to constrain the faint end of the galaxy--halo connection and alternative dark matter models. Incorporating SAGA results in this type of analysis will undoubtedly provide more powerful constraints by adding data in a regime that has been sparse to date.

In this paper, we present the Stage II results of the SAGA Survey, including \nsats{} satellites in \nhosts{} MW-like systems. 
We first review our survey design, including host selection and survey progress (\S\ref{sec:survey}), photometric catalog production (\S\ref{sec:photometry}), target selection and spectroscopic instruments (\S\ref{sec:spectroscopy}), and the definition of satellite and survey completeness (\S\ref{sec:systematic}). 
We then present our findings, including the properties and quenched fraction of individual satellites (\S\ref{sec:sat-properties}); the luminosity functions, radial distributions, and corotating signals of the satellite systems (\S\ref{sec:sat-system}); and the comparison between our observation and theoretical predictions (\S\ref{sec:theory}). 
Readers who are familiar with \paperone{} or want to quickly navigate the main results of this paper can read the summary (\S\ref{sec:summary}) first.

As in \paperone{}, all distance-dependent parameters are calculated assuming $H_0 = 70$\,\kms\,Mpc$^{-1}$ and $\Omega_M = 0.27$.
Magnitudes and colors are extinction-corrected (denoted with a subscript `o,' e.g., $r_o$; using a combination of \citealt{schlegel98} and \citealt{Schlafly2011}; see \S\ref{sec:photometry}).  Absolute magnitudes are $k$-corrected to $z=0$ using \citet{Chilingarian2010}.

\begin{figure*}[t]
    \centering
    \includegraphics[width=\textwidth,clip,trim=0 0 0 0]{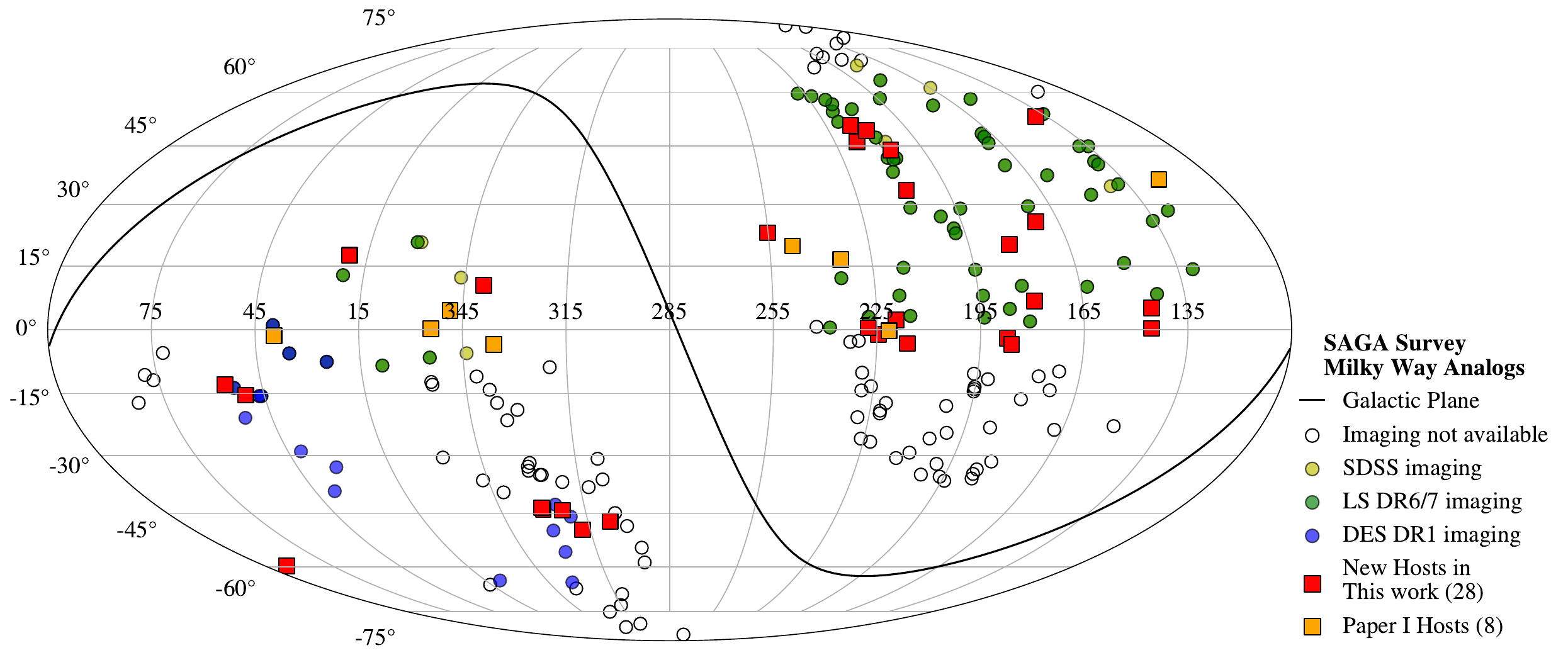}
    \caption{Sky distribution of \nhostsinclnoimage{} MW analog systems identified in the SAGA Survey volume (circles and squares).  There are \nhoststotal{} SAGA systems considered for targeting that have complete imaging over at least 99\% of the projected virial radius in SDSS (yellow), LS DR6+7 (green), and/or DES DR1 (blue).  The solid black line indicates the Galactic plane.  There are \nhosts{} hosts with complete spectroscopic coverage: eight from Paper I (orange squares) and \nhostsnew{} newly presented in this work (red squares).}
    \label{fig:hosts_coverage}
\end{figure*}

\begin{table*}[t]
\begin{tabular}{l|rr|rr|rr}
\hline\hline
SAGA Survey Stage &  \multicolumn{2}{c|}{I} &  \multicolumn{2}{c|}{II} & \multicolumn{2}{c}{III} \\ \hline
Reference &  \multicolumn{2}{c|}{\citet{Geha2017}} &  \multicolumn{2}{c|}{This work (2020)} & \multicolumn{2}{c}{Expected in 2022} \\ \hline
No.\ of complete hosts & \multicolumn{2}{c|}{8} & \multicolumn{2}{c|}{36} & \multicolumn{2}{c}{100} \\ 
Field definition & $\mathbf{<300}$\,\textbf{kpc} & $<1^{\circ}$ & $\mathbf{<300}$\,\textbf{kpc} & $<1^{\circ}$ & $\mathbf{<300}$\,\textbf{kpc} & $<1^{\circ}$ \\ \hline
No.\ of galaxy redshifts SAGA obtained & \textbf{10,587} & 12,682 & \textbf{22,313} & 25,372 & $\sim$\,\textbf{42,000} & $\sim$\,52,000 \\ 
Total no.\ of galaxy redshifts & \textbf{12,670} & 20,032 & \textbf{30,880} & 53,797 & $\sim$\,\textbf{63,000} & $\sim$\,127,000 \\ 
No.\ of satellites discovered & \textbf{14 (2)} & \ditto & \textbf{66 (3)} & \ditto & $\sim$\,\textbf{150} & \ditto \\
Total no.\ of satellites & \textbf{27 (2)} & \ditto & \textbf{123 (4)} & \ditto & $\sim$\,\textbf{300} & \ditto \\ 
No.\ of low-$z$ galaxies discovered & \textbf{75} & 99 & \textbf{246} & 358 & $\sim$\,\textbf{700} & $\sim$\,1100 \\ 
Total no.\ of low-$z$ galaxies & \textbf{112} & 197 & \textbf{434} & 921 & $\sim$\,\textbf{1400} & $\sim$\,2900 \\ \hline
\end{tabular}
    \caption{SAGA Survey Key Numbers. The three column groups represent the three stages of the SAGA Survey. The left two column groups show Stage I and II results, which are presented in \paperone{} and this work, respectively, and the rightmost column group is the projection for Stage III results. The sub-columns in different font weights denote two field definitions: within the virial radius of 300\,kpc (corresponding to 25$'$--42$'$ depending on host distance; bold font) or within 1$^\circ$ (field of view; regular font) around the hosts. Note that satellites are by definition within 300\,kpc of their hosts. The satellite numbers in parentheses correspond to satellites that are dimmer than our formal survey limit ($M_{r,o} = -12.3$). We define ``low-$z$'' as $0.003 < z< 0.03$;  ``discovered'' indicates that the redshift was first obtained by SAGA.}
    \label{tab:roadmap}
\end{table*}


\section{The SAGA Survey Strategy}
\label{sec:survey}

The primary science goal of the SAGA Survey\footnote{\https{sagasurvey.org}\label{fn:saga}} is to characterize the satellite
galaxy populations of 100 Milky Way analogs down to an absolute magnitude of $M_{r,o} = -12.3$.
In this section we describe the survey at a high level: first we describe our criteria to select MW analogs (\autoref{sec:host-selection}), then outline the SAGA Survey strategy and timeline to achieving the project goals (\autoref{sec:roadmap}).

\subsection{MW Analog Selection}
\label{sec:host-selection}

The selection of our MW analogs (``hosts'') is primarily based on $K$-band luminosity and local environment criteria. Our science questions are best answered by comparing the satellite populations of hosts with similar dark matter halo masses, a property that is impossible to directly measure at this mass scale. Among direct observables, $K$-band luminosity and local environment are the best available proxies for halo masses given the available data.

\subsubsection{Updating the Grand List}
\label{sec:grand-list}

As described in \paperone{}, we select MW hosts from a complete nearby galaxy catalog (hereafter the ``Grand List''). This Grand List was constructed when we began the survey in 2014, mainly based on the HyperLEDA database \citep{Makarov2014}\footnote{\http{leda.univ-lyon1.fr}} and supplemented by various literature sources. Since then, the HyperLEDA database has incorporated additional sources and properties from the astronomical literature.   We therefore rebuild our Grand List to take advantage of these new additions while maintaining consistency with our prior results. 

We aim to construct our Grand List so that it is complete out to 60 Mpc in distance and down to $K = 11.9$ (corresponding to $M_K = -22$ at 60 Mpc).  We first obtain all objects in the HyperLEDA database that are galaxies ($\code{objtype}=\code{G}$), have both heliocentric velocity (\code{v}) and distance (\code{modbest}), and have a $K$-band magnitude less than 12 mag ($\code{kt} < 12$). Each object in the HyperLEDA has a unique PGC ID that serves as the main identifier. 

To maintain consistency with our prior results, we opt to use the distance measurement from the NASA--Sloan Atlas (NSA; \citealt{Blanton2011}) and the total $K$-band magnitude from the 2MASS Redshift Survey (2MRS; \citealt{2012ApJS..199...26H}) catalog. More specifically, we spatially match the objects with the NSA v1.0.1 catalog\footnote{\https{www.sdss.org/dr14/manga/manga-target-selection/nsa/}\label{fn:nsa}} and choose the NSA distance (\code{ZDIST}) when available, unless HyperLEDA has a redshift-independent distance measurement and the final error (\code{e\_modbest}) is less than 0.04 dex, in which case we choose the HyperLEDA distance measurement. 
The distance reported in the NSA is corrected for local peculiar velocity with the \citet{1997ApJS..109..333W} model.   While this is not necessarily the best distance measurement for each galaxy, the availability of quality distance measurements in our target distance range (between 25 and 40 Mpc) is scarce, and the available distance estimates usually have an uncertainty of 0.1~dex \citep{2019ApJS..244...24L}.

For the $K$-band magnitude, we use the Extragalactic Distance Database (EDD; \citealt{2009AJ....138..323T})\footnote{\http{edd.ifa.hawaii.edu}} to obtain extinction-corrected total $K$- and $H$-band magnitudes from the 2MRS catalog. We use the prescription in \cite{Chilingarian2010} and $K-H$ color to $k$-correct the $K$-band absolute magnitude to $z=0$. For objects that do not appear in the 2MRS catalog, we use the $K$-band magnitude in HyperLEDA but shift it by $-0.03$ mag. The value of this small shift is derived from the median difference for objects in both catalogs and does not have a significant impact on our host list construction.

\subsubsection{MW Analog Host Criteria}
\label{sec:host-list}

With the compiled information in our Grand List, we define our MW hosts based on four criteria: (1) an absolute $K$-band criterion, (2) a set of cuts to reduce stellar foregrounds, (3) cuts based on the local galactic environment, and (4) distance cuts for our survey.  We detail these criteria below and summarize in \autoref{eq:hostlist_cuts}.

First, we require our MW hosts to have an absolute $K$-band magnitude in the range $-23 > M_K > -24.6$.   We use this magnitude as a proxy for total stellar mass.   As shown in \paperone{}, this range corresponds to the possible magnitude of the central galaxy hosted by a halo of virial mass of~$1.6 \times 10^{12}\,\msun$, when a simple galaxy--halo connection model is assumed (see Figure~2 of \paperone{}). The corresponding stellar mass range for the host galaxy is $10^{10}$--$10^{11}$\,\msun. The virial radius corresponding to this halo mass is $\sim$300\,kpc.

We next require hosts to be sufficiently far from the Galactic plane to reduce stellar foregrounds ($|b| \geq 25^\circ$).   To avoid saturated regions around very bright stars, we search the Hipparcos-2 catalog \citep{Hipparcos} within the virial radius.  We remove hosts with a very bright foreground star, $H_P < 5$ ($H_P$ is the broad optical Hipparcos magnitude), which can significantly reduce the region of usable photometry.

We then examine the candidate host's local galactic environment.  Using our Grand List, we first require there to be no bright galaxies inside the projected virial radius with magnitude of $K < K_\mathrm{host} - 1.6$ or brighter.\footnote{In \paperone{}, this criterion was $K < K_\mathrm{host} - 1$; we have since tightened the criterion but also collected substantial data on a handful of hosts and choose to keep them in our host sample, including NGC\,7166 presented in this work.}  The magnitude difference of 1.6\,mag was chosen to ensure that no two MW host galaxies in our $K$-band magnitude range could be found inside the same virial volume.  This criterion does not exclude MW--M31-like systems, as the MW is slightly beyond two virial radii from M31 (such a configuration is only excluded for $\sim$35\% of the line-of-sight solid angle).
We also use the 2MRS group catalog constructed by \citet[][obtained from EDD]{2017MNRAS.470.2982L} and remove hosts associated with a group whose mass is greater than $M_\text{halo} < 10^{13}\,\msun$.   Finally, we manually flag a small handful of potential host galaxies that are visibly disturbed and/or in merging systems.

We select galaxies in the distance range of 25--40.75\,Mpc with a heliocentric velocity greater than 1400\,\kms. The outer distance limit corresponds to the absolute magnitude limit of $M_{r,o} < -12.3$ given our apparent magnitude limit of $r_o < 20.75$.   The inner distance and velocity limits exclude 80 hosts in fields that are too big for our instruments and that are more efficiently studied using other techniques \citep[e.g.,][]{Danieli2018}. These nearby ($<25$\,Mpc) hosts that also satisfy our stellar mass and environment cuts include the well-studied MW analogs M94 \citep{Smercina2018}, as well as three of the 10 hosts (NGC\,1023, NGC\,2903, and NGC\,4258) presented in \citet{Carlsten2019}. 

To summarize our MW host criteria:
\begin{subequations}%
\begin{align}
     & \text{Stellar Mass: }  -23 > M_K > -24.6; \label{eq:host-MK-cut}\\
     & \text{Stellar Foreground: }  |b| \geq 25^\circ;\\
     & \text{Stellar Foreground: } \text{\it Hp}^{(< 300\,\text{kpc})}_\text{brightest star} > 5;\\
     & \text{Environment: }  K < K^{(< 300\,\text{kpc})}_\text{brightest gal.} - 1.6; \label{eq:env-cut-K}\\
     & \text{Environment: }  M_\text{halo} < 10^{13} \msun;\\
     & \text{Distance: } 25-40.75\,\text{Mpc; and}\\
     & \text{Distance: } v_\text{helio} > 1400\,\kms.
\end{align}\label{eq:hostlist_cuts}%
\end{subequations}

The distance and environmental cuts above are slightly more stringent than those used in \paperone{} but include all previously published hosts. 
From our Grand List, we identify \nhostsinclnoimage{} MW analogs that pass these selection criteria.  Of these hosts, \nhoststotal{} pass the available imaging requirements described in 
\autoref{sec:host-imaging-coverage} to be considered for targeting.
We plot the distribution of these hosts on the sky in \autoref{fig:hosts_coverage}.  
For each host, the SAGA Survey searches for satellites within a projected radius of 300\,kpc, approximately corresponding to the virial radius of a MW-mass halo. 
Given the distance range of SAGA hosts (25--40.75\,Mpc, or $0.005 < z < 0.01$ in redshift), the search radius of 300\,kpc corresponds to 25$'$--42$'$ on the sky.

Among the \nhoststotal{} hosts that meet all of our selection criteria, we identify four Local Group--like pairs (two MW-mass hosts within 1\,Mpc of each other).  One of these pairs is presented in this work: NGC\,7166 and ESO\,288-025 (separated by 668\,kpc). 
We additionally identify 11 hosts that are in a Local Group--like environment but whose companions do not meet our selection criteria or do not have imaging available. Three of these hosts are presented in this work: NGC\,5602, NGC\,5750, and NGC\,5962.
We plan to observe more fields around these Local Group--like hosts in SAGA Stage III to enable studies of the correlations between satellite systems and the host environment.

\subsection{Road Map of the SAGA Survey}
\label{sec:roadmap}

The SAGA Survey is a long-term project that we consider in three main stages, summarized in \autoref{tab:roadmap}:  (I) determine the complete luminosity function for a handful of MW analogs using  basic criteria, (II) design targeting strategies that more efficiently identify high-probability satellites without compromising completeness, and (III) apply these criteria to 100 MW-analog systems.   In \paperone{} we presented the results of SAGA Survey Stage I: eight complete systems with relaxed target selection criteria, over 12,682 newly obtained galaxy spectra around the eight systems, 16 newly discovered satellites, and 99 newly discovered very low redshift ($z < 0.03$) galaxies (see also \autoref{tab:roadmap}).

In this work, we present results from Stage II of the SAGA Survey.  This includes two major improvements over Stage I.    First, we significantly improved our target selection strategies.  Based on data from SAGA Stage I, we reduced the average number of targets per host by more than a factor of 4 (from $\sim$1500 to $\sim$400). We detail these target selection strategies in \autoref{sec:target-selection}. Second, we expanded our photometric data to include both the Dark Energy Survey (DES) and the Legacy Surveys (LS) in addition to SDSS (Stage I used SDSS alone). This expansion provides us access to many more hosts, especially in the southern hemisphere (\autoref{fig:hosts_coverage}).  In expanding our photometric coverage, some hosts have photometric coverage across multiple surveys.  These data provide excellent test cases to understand differences among the photometric catalogs (see also \autoref{sec:phot-validation}). 

\autoref{tab:roadmap} provides key numbers regarding the progress of the SAGA Survey. 
The three column groups represent the three SAGA Survey stages. 
We show the numbers for two different field definitions in subcolumns of different fonts: within 300\,kpc in bold (corresponding to the virial radius), and within 1$^\circ$ in regular (corresponding to the field of view) around the SAGA hosts. 
The Stage II results presented here are based on \nhosts{} complete systems in total, reaching just over one-third of the goal of 100 complete systems (see \autoref{sec:completeness-def} for the definition of a ``complete system"). 
During Stage II, we completed \nhostsnew{} hosts with only \nzsaganewrvir{} new redshifts (on average, $\sim$430 redshifts within 300\,kpc around each host). 
In SAGA Stage III, we will continue to improve our targeting efficiency, especially in the low surface brightness regime, and aim to complete the remaining hosts with an average of $\sim$300 redshifts within 300\,kpc around each host. 

In the \nhosts{} complete systems of SAGA Stage II, we have identified \nsatslimit{} satellites down to $M_{r,o} = -12.3$ (and \nsatsbelow{} more below the threshold). The SAGA Survey provides the first redshift measurements to \nsatslimitsaga{} of these satellites (54\%). 
Among the fainter half of the satellites (62 satellites with $-15.77 < M_{r,o} < -12.3$), $53/62$ (87\%) have the first redshift measurement from SAGA. 
The SAGA Survey has also discovered more than 200 additional very low redshift ($z<0.03$) galaxies around the \nhosts{} SAGA hosts. 

As of 2020 July, we had obtained close to 40,000 galaxy redshifts in total across all SAGA hosts. We are on track to complete spectroscopic follow-up for our Stage III goal by the end of 2021.

\section{Photometric Object Catalogs}
\label{sec:photometry}

We next detail our pipeline to build the photometric object catalog for each SAGA host.   Our photometric object catalogs provide the base catalogs for our target selection.  In SAGA Stage I, we used only the SDSS DR12 catalogs as our photometric object catalogs. In SAGA Stage II, we expand to include deeper data sets from the DES and LS.\footnote{\http{darkenergysurvey.org} and \http{legacysurvey.org}} Our target selection concentrates on galaxies brighter than an extinction-corrected $r_o < 20.75$.  This is 2 or more mag brighter than the DES or LS photometric magnitude limits.

We first describe the source catalogs from the three surveys (SDSS DR14, DES DR1, and LS DR6/7; \autoref{sec:sdss}--\ref{sec:decals}). We then describe the process used to merge these into a single photometric object catalog for each host (\autoref{sec:phot-merge}), the cleaning process (\autoref{sec:phot-clean}), and the validation of our final object catalog (\autoref{sec:phot-validation}). Our photometric object catalog-building pipeline is developed in an open-source repository.\footnote{\https{github.com/sagasurvey/saga}\label{fn:saga-code}}

\subsection{SDSS DR14 Photometry}
\label{sec:sdss}

We use the SDSS DR14 photometry catalog \citep{SDSS_DR14} and ingest all photometric objects around each host within a 1$^\circ$ radius. We use \code{modelMag} for galaxy magnitudes and \code{PETROR50\_R} as the effective photometric radius $R_{r,\text{eff}}$ in our source catalog.
To correct for extinction, we use the value reported in \code{EXTINCTION}, which is derived from  \citet{Schlafly2011} for DR14.
We use the \code{PHOTPTYPE} flag for star/galaxy separation and select only galaxy targets.

We then apply a set of selection criteria to identify ``good'' galaxy targets:
\begin{align*}
     & \code{BINNED1} \neq 0, \\
     & \code{SATURATED} = 0, \\
     & \code{BAD\_COUNTS\_ERROR} = 0, \\
     & \code{FIBERMAG\_R} \leq 23, \\
     & |\sigma_g| < 0.5, |\sigma_r| < 0.5, |\sigma_i| < 0.5 \text{ (at least 2)}, \\
     & |g_o-r_o| \leq 10, |r_o-i_o| \leq 10, |u_o-r_o| \leq 10 \text{ (all)},
\end{align*}
where $\sigma_{g,r,i}$ are the reported 1$\sigma$ photometric errors in each band. The first three criteria are standard SDSS quality cuts. The fourth criterion flags galaxies that do not have accurate SDSS photometry to favor photometry from deep images. The last two sets of criteria aim to remove objects that are anomalously bright due to sky oversubtraction near bright stars.  

\subsection{DES DR1 Photometry}
\label{sec:des}

We use the DES DR1 photometry catalog \citep{DES_DR1} and ingest all photometric objects around each host within a 1$^\circ$ radius. We use \code{MAG\_AUTO\_*\_DERED} for dereddened magnitudes (derived from \citealt{schlegel98}),\footnote{\https{des.ncsa.illinois.edu/releases/dr1/dr1-faq}\label{fn:des-faq}} and use \code{FLUX\_RADIUS} as our effective photometric radius $R_{r,\text{eff}}$. We convert DES $griz$ magnitudes to approximate SDSS magnitudes using the following formulae (see footnote \ref{fn:des-faq}):
\begin{align*}
g_\text{SDSS} & = g_\text{DES} -0.0009 + 0.055(g-i)_\text{DES}, \\
r_\text{SDSS} & = r_\text{DES} -0.0048 + 0.0703(g-i)_\text{DES}, \\
i_\text{SDSS} & = i_\text{DES} -0.0065 - 0.0036(g-i)_\text{DES} + 0.02672(g-i)^2_\text{DES}, \\
z_\text{SDSS} & = z_\text{DES} -0.0438 + 0.02854(g-i)_\text{DES}. 
\end{align*}

For star--galaxy separation, we use the \code{EXTENDED\_COADD} parameter in the $r$ band, which is defined as (see footnote \ref{fn:des-faq})
\begin{align*}
& \code{EXTENDED\_COADD} = \\
& \; (\code{SPREAD\_MODEL\_R}+ 3 \times \code{SPREADERR\_MODEL\_R} > 0.005) + \\
& \; (\code{SPREAD\_MODEL\_R}+\code{SPREADERR\_MODEL\_R} > 0.003) + \\
& \; (\code{SPREAD\_MODEL\_R}-\code{SPREADERR\_MODEL\_R} > 0.003),
\end{align*}
and we require galaxies to have $\code{EXTENDED\_COADD} = 3$. 
However, for very bright objects ($r_o < 17$), the DES pipeline sometimes misclassifies stars as galaxies. We hence applied additional cuts to bright objects to remove stars. In particular, we relabel objects that satisfy the following criteria as stars:
\begin{align*}
& 0.7 \cdot (r_o + 10.2) > r_o + 2.5 \log \left(2\pi R_{r_o,\text{eff}}^2\right), \\
& g_o - r_o < 0.6, \text{ and}\\
& r_o < 17.
\end{align*}

Finally, we apply the following standard DES quality cuts  to identify ``good'' galaxy targets:
\begin{align*}
 & \code{IMAFLAGS\_ISO\_R} = 0\text{, and}\\
 & \code{FLAGS\_R} < 4. 
\end{align*}

\subsection{LS DR6/7 Photometry}
\label{sec:decals}

We use the combined LS DR6 (BASS/MzLS) and DR7 (DECaLS) photometry catalogs \citep{Dey2019} and ingest all photometric objects around each host within a 1$\circ$ radius from the ``sweep'' files. 

We use the \code{TYPE} flag to identify galaxies as all objects whose $\code{TYPE} \neq \code{PSF}$.
We use \code{FLUX} and \code{MW\_TRANSMISSION} to calculate dereddened magnitudes, where \code{MW\_TRANSMISSION} is calculated using the coefficients in \citet{Schlafly2011} and the map from \citet{schlegel98}.\footnote{\http{legacysurvey.org/dr7/catalogs/\#galactic-extinction-coefficients}}

We use \code{SHAPEEXP\_R} (or \code{SHAPEDEV\_R}) as our effective photometric radius, $R_{r,\text{eff}}$, for objects that are better fitted with an Exponential (or De Vaucouleurs) profile. For objects that are better fitted with a composite profile, we linearly interpolate between these two radii according to the value of \code{FRACDEV} to set $R_{r,\text{eff}}$. 

We approximate the SDSS $grz$ magnitudes as
\begin{align*}
g_\text{SDSS} & = g_\text{LS} + 0.09, \\
r_\text{SDSS} & = r_\text{LS} + 0.1\text{, and} \\
z_\text{SDSS} & = z_\text{LS} + 0.02.
\end{align*}
These photometric shifts were determined in the several-degree region of overlapping coverage between the two surveys.  

We define ``good'' galaxy targets in the LS catalogs as objects that satisfy all of the following criteria. Note that each of the following criteria is applied to all three bands ($grz$) unless otherwise noted:
\begin{align*}
&\code{NOBS} \geq 1 \text{ ($g$ and $r$ bands)}, \\
&\code{ALLMASK} = 0, \\
&\code{FRACMASKED} < 0.35, \\
&\code{FRACFLUX} < 4, \\
&\code{RCHISQ} < 10, \\
&\code{RCHISQ} < 4 \text{ (any one band)},\\
&\code{FRACIN} > 0.7 \text{ (any one band), and}\\
&\sigma(\text{magnitude}) < 0.2.
\end{align*}
Here the first criterion requires the presence of $g$- and $r$-band measurements, the second criterion is a standard quality cut, and the remaining criteria remove sources that are not well described by the model due to either bad fits or significant blending.\footnote{Definitions of these flags can be found at \https{www.legacysurvey.org/dr7/files/}.} 

Even with the above selection criteria, a noticeable number of objects have unrealistic model radii. In most cases, the model is attempting to compensate for nearby sources that were not identified as objects and creates an elongated light profile. 
Hence, we further require our galaxy targets to satisfy
\begin{align*}
&R_{r_o,\text{eff}} < \min \left\{\,15 \,,\, 10 ^ {-0.2 (r_o - 23.5)}\,\right\} \text{ arcsec}, \\
&R_{r_o,\text{eff}} > 10 ^ {-0.2 (r_o - 17)}  \text{ arcsec, and} \\
&(g-r)_o > -0.5.
\end{align*}

The criteria for selecting ``good'' galaxy targets from LS catalogs that we described here are rather aggressive; they are needed to avoid a large number of spurious objects in our source catalogs. To ensure our source catalogs are still complete down to our magnitude limit, we only use LS catalogs to supplement SDSS/DES catalogs in this stage of the SAGA Survey (see the next section on the merging process); that is, we do not observe hosts that have only LS coverage.   These issues have been mitigated in the recently released LS DR8, and we plan to update our LS photometric catalogs in Stage III.

\begin{figure*}[t]
\centering
\includegraphics[width=0.85\textwidth]{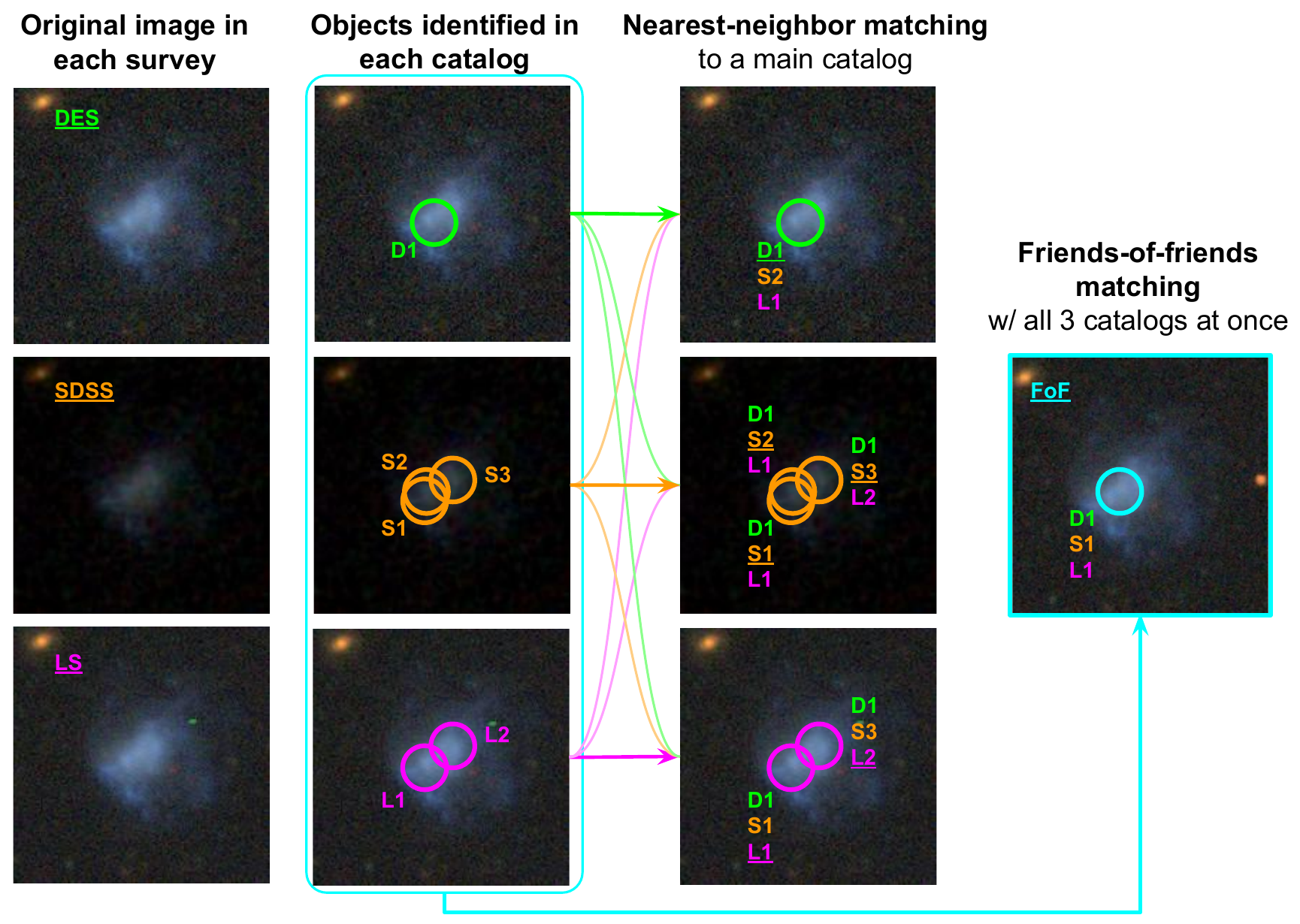}
\caption{Illustration of the FoF algorithm with variable linking lengths, used to merge multiple photometric catalogs (\autoref{sec:phot-merge}).
The galaxy shown in the first column is identified as a single object in some photometric catalogs (in this example, DES; top panel in the second column), and as multiple shredded objects in other photometric catalogs (SDSS and LS; middle and bottom panels in the second column). 
The numerical indices are ordered by the object magnitudes. 
If we choose only one particular photometric catalog as the ``main catalog'' and do a nearest-neighbor match, we will inherit the shredded objects (middle and bottom panels in the third column, or mismatch to a shredded object instead of the main object due to slight difference in position (top panel in the third column).
Note that DES is not always the best-performing catalog; sometimes SDSS or LS performs better, and sometimes all of them produce shredded objects.
By using the FoF matching method, all of these objects are joined together, and the brightest objects in each photometric catalog are chosen to be the main object for that catalog (rightmost panel).}
\label{fig:fof_illustration}
\end{figure*}

\subsection{Merging Source Catalogs with FoF}
\label{sec:phot-merge}

Sky catalogs are traditionally matched using the nearest-neighbor method. While this method is straightforward, it has difficulty dealing with some edge cases, such as when a single object is shredded into different numbers of objects in different catalogs.  Low-redshift galaxies tend to be several arcseconds or larger in size and are often shredded in the above catalogs.   Thus, we adapt a different approach to combine available photometric catalogs.  We  concatenate all objects from all catalogs and then run a friends-of-friends (FoF) group finder.  When the linking length is chosen properly, each FoF group represents a single object in the combined catalog. Within each group, we identify one group member as the primary photometry source for that object.
The FoF algorithm and its main advantage over the nearest-neighbor method are demonstrated in \autoref{fig:fof_illustration}. 
Our code for FoF sky catalog matching has been extracted as a stand-alone, publicly available package.\footnote{\https{github.com/yymao/FoFCatalogMatching}}

To ensure that we both (1) correctly merge objects that have slightly different positions in different catalogs and (2) do not accidentally combine distinct objects into one, we use a dynamic linking length for the FoF merging algorithm. We start with a linking length set to $3''$; if there is any group in which there is more than one object from all available surveys, we reduce the linking length by $0.5''$ and repeat an FoF search on that group, repeating until the linking length reaches $0.5''$ or until at least one survey has only one object in that group. 

As demonstrated in \autoref{fig:fof_illustration}, each FoF group represents a single object in our final catalog, but it may include multiple photometric objects from multiple surveys. We identify one primary photometry source for that object.
Among the choices, we follow the order of DES, LS, and SDSS to select the primary source.  If there are multiple good objects from the same survey within the group, we select the object with the brightest magnitude.  We use the galaxy properties (magnitude, radius) reported in the primary photometric catalog.  

After the FoF matching process, we supplement our source catalog using the NSA catalog v1.0.1 (\citealt{Blanton2011}; see footnote \ref{fn:nsa}).
The NSA catalog is a reprocessing of the SDSS photometry for galaxies with $z < 0.15$ using an improved background subtraction technique important for large, extended galaxies.  If a merged source catalog object is identified in the NSA v1.0.1 catalog, we use the NSA photometry as the primary photometric information. These objects include many of our host galaxies and their brighter satellites.

\subsection{Cleaning Photometric Catalogs}
\label{sec:phot-clean}

In the photometric catalogs, spatially extended galaxies are often shredded into multiple objects. Galactic cirrus or diffraction spikes from bright stars can be misidentified as galaxies.   Although our FoF merging method described above mitigates many of these problems, spurious objects still contaminate our source catalogs.   This issue exists in all three photometric surveys.

We clean these remaining spurious objects both using available spectroscopic data (automated removal) and by eye (manual removal).  For automated removal, we first identify all galaxies that have a confirmed spectroscopic redshift and have an error on effective photometric radius less than 50\%; we then remove objects that are within twice the effective photometric radius of such galaxies. If the confirmed galaxy is in the NSA catalog, we use the elliptical Petrosian light profile instead of the simple circular radius. Since these ellipses sometimes overlap with other distinct background galaxies, we only remove objects that do not have redshift information or have a redshift within 200\,\kms{} of the said galaxy's heliocentric velocity.  As a final check, we visually inspect all objects in our primary targeting region (defined in \autoref{sec:target-selection}) and manually flag the handful of spurious objects that remain after automated cleaning. We developed an image list tool\footnote{\https{yymao.github.io/decals-image-list-tool}} based on the cutout service of the LS Viewer\footnote{\http{legacysurvey.org/viewer}\label{fn:ls-viewer}} to facilitate the visual check.

\subsection{Computed Catalog Quantities}
\label{sec:phot-quantities}

For each object in our cleaned photometric catalogs, we calculate the effective surface brightness \mueff{} defined as
\begin{align}
    \mueff = r_o + 2.5 \log \left(2\pi R_{r_o,\text{eff}}^2\right),
\label{eq:sb}
\end{align}
where $r_o$ is the total magnitude and $R_{r,\text{eff}}$ is the effective half-light radius in the $r$ band from the primary photometric survey in units of arcseconds. The effective half-light radii we adopted from the three surveys are consistent in definition.

Across the photometric surveys described in \autoref{sec:photometry}, all of our objects have measured $g$- and $r$-band photometry. We therefore use \cite{Chilingarian2010} and $g-r$ color to $k$-correct the $r$-band absolute magnitude to $z=0$ and estimate stellar mass based only on these two bands.
We begin with the \citet{bell2003} relation to determine the mass-to-light ratio based on the $g-r$ color. We adopt a Kroupa initial mass function (IMF) and subtract 0.15\,dex from the Bell et al.~value which assumes a modified Salpeter IMF.   We compare these stellar masses to recent estimates from \citet{Zibetti2009} and \citet{Taylor2011}. We find that an additional $-0.15$\,dex is consistently needed to match to these two data sets. We calculate stellar mass assuming an absolute solar $r$-band magnitude of 4.65 \citep{Willmer2018}. The resulting stellar mass conversion is $\log[M_*/\msun] = 1.254 + 1.098 \,(g-r)_o - 0.4 \, M_{r,o}$.  We assume a systematic error of 0.2\,dex which is larger than the random errors due to propagating errors in photometry and distance. 

\subsection{Photometric Catalog Completeness and Validation}
\label{sec:phot-validation}

We compare our merged and cleaned photometric catalogs with various sources to ensure their completeness and validity. 
For SAGA fields that are in the DES footprint, we compare to \cite{Tanoglidis200604294}, who reanalyzed low surface brightness galaxies (LSBGs) using \textsc{galfitm}.  We confirm that all LSBGs analyzed in \cite{Tanoglidis200604294} and in the SAGA footprint are present in the SAGA photometric catalogs with consistent properties.

We also compare our object catalogs that are in the Hyper Suprime-Cam Subaru Strategic Program (HSC-SSP) footprint with the HSC-SSP Public Data Release 2 \citep[overlapping with nine SAGA complete fields, two of which we compared here]{hsc-pdr2:1905.12221} and the HSC-SSP LSBG catalog compiled by \citet[overlapping with three SAGA complete fields, all of which we compared here]{Greco18:1709.04474,Greco:2004.07273}. 
We find that our object catalogs indeed include all real galaxy objects in both HSC catalogs down to our magnitude limit of $r_o < 20.75$.
In other words, partly thanks to the inclusion of deep DECam data (DES and LS), our object catalogs do not miss low surface brightness objects at the catalog level. 

\subsection{Imaging Criteria for Targeting a SAGA Host}
\label{sec:host-imaging-coverage}

For a host to be considered for SAGA targeting, it must have sufficient imaging data within a projected radius of 300\,kpc.  Our photometric coverage criteria for targeting are (1) greater than 99\% imaging coverage in at least one of the photometric surveys (SDSS, DES, LS), and (2) if the best coverage comes from LS, the area must also have at least 85\% imaging coverage from SDSS or DES.  We apply the second criterion because we have used a set of more aggressive cuts to remove problematic photometric objects in LS (\autoref{sec:decals}), and the supplemental SDSS or DES coverage helps ensure that we do not systematically remove good objects.   Of the \nhostsinclnoimage{} MW analogs identified in the SAGA volume, \nhoststotal{} hosts have sufficient photometric coverage for spectroscopic targeting (\autoref{fig:hosts_coverage}).

\begin{figure*}[t]
    \centering
    \includegraphics[width=\textwidth,clip,trim=0 0.3cm 0 0]{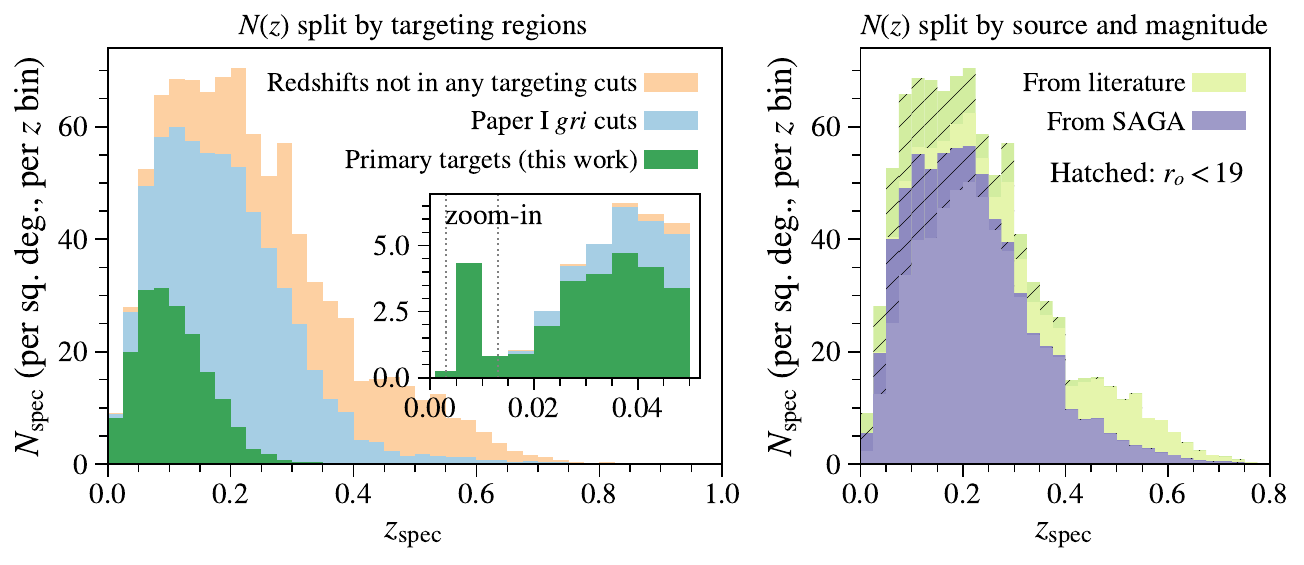}
    \caption{Redshift distribution, $N(z)$, of galaxies that are brighter than $r_o = 20.75$ and within $<300$\,kpc around all \nhosts{} hosts ($\sim$30 deg$^2$ in total). These redshifts account for about 60\% of the total photometric galaxy sources to that magnitude limit. The overall (stacked) histograms in the two panels are identical (representing all redshifts) but split differently in the two panels. \textit{Left}: $N(z)$ split by target samples. The dark green (inner) histogram shows all galaxies that are in our primary targeting region, as specified in \autoref{eq:targeting-cuts}.
    The blue (middle) histogram shows all galaxies that satisfy the \paperone{} $gri$ cuts but not the primary cuts used in this work. The orange (outer) histogram shows the remaining redshifts (i.e., not satisfying any targeting cuts). The inset shows a zoom-in version ($0 < z < 0.05$) of the same set of histograms. The SAGA satellites are in the redshift range ($0.003 < z < 0.013$) indicated by the dotted vertical lines.
    \textit{Right}: $N(z)$ split by redshift source and apparent magnitude. The yellow (outer) histogram shows all galaxies brighter than $r_o = 20.75$ that have redshifts from the literature and preexisting surveys.
    The purple (inner) histogram shows those from the SAGA Survey.  The hatched (darker) portions denote galaxies brighter than $r_o = 19$. In both panels and the inset, the histograms are normalized so that the height shows the number of galaxies per square degree per redshift bin.}
    \label{fig:z_distribution}
\end{figure*}


\section{Spectroscopic Data}
\label{sec:spectroscopy}

In SAGA Stage II, we have improved our target selection strategy (\autoref{sec:target-selection}) and obtained \nzsaganew{} galaxy redshifts (\autoref{sec:mmt}--\ref{sec:dbsp}), bringing the total number of SAGA-obtained redshifts within $1^\circ$ around \nhosts{} SAGA hosts to \nzsaga{}.
We also compile \nzothernew{} additional redshifts from the literature (\autoref{sec:spec-literature}).
We describe the association of spectroscopic and photometric objects in \autoref{sec:spec-merge}, 
and the resulting redshift distribution in \autoref{sec:spec-distribution}.

\subsection{Target Selection}
\label{sec:target-selection}

Improving our target selection efficiency is one of the major goals of Stage II of the SAGA Survey.
Our aim is to complete a host with one or two pointings (200--400 targets) with MMT/Hectospec or the Anglo-Australian Telescope (AAT) 2dF while maintaining  high satellite completeness.
Achieving this goal requires carefully characterizing the photometric properties of very low redshift galaxies (the redshift range of SAGA hosts is $0.005 < z < 0.01$, and that of SAGA satellites is $0.003 < z < 0.013$ when peculiar velocities are included), a regime in which traditional photometric redshifts are not optimized.  There exist very few galaxies at these low redshifts with spectroscopic confirmation or distance measurements that are dimmer than $r_o=18$, and the SAGA Survey aims to be complete down to $r_o = 20.75$. Hence, it is necessary that we develop our own target selection strategy. 

In SAGA Survey Stage I, we employed a set of color cuts (the $gri$ cuts; Equations (1) and (2) of \paperone{}) to exclude objects that are extremely unlikely to be very low redshift galaxies. These cuts still leave over 1000 targets per host, on average.  In that work, we attempted to obtain spectroscopic confirmation for all galaxies satisfying these cuts (within 300\,kpc of the host and down to $r_o = 20.75$). As described in \paperone{}, we achieved more than 90\% spectroscopic coverage around six hosts, and more than 80\% around the remaining two hosts.
This effort resulted in an invaluable training set that allows us to develop new strategies to efficiently identify very low-redshift galaxy candidates; we have applied our strategies to this work and to the low redshift galaxy program of the Southern Stellar Stream Spectroscopic Survey \citep{10.1093/mnras/stz2731}. 
This training set also enables development of machine learning-based methods to select low-redshift galaxies based on photometry (e.g., \citealt{Wu:2001.00018}).

In SAGA Stage II, we improve our target selection strategy by adapting a twofold approach and devote about half of our spectroscopic resources to each part. 
We first identify a ``primary targeting region'' using simple cuts in the surface brightness--magnitude and color--magnitude planes,
\begin{subequations}%
\begin{align}%
\mueff + \sigma_{\mu} - 0.7 \, (r_o - 14) &> 18.5, \label{eq:targeting-cuts-sb-r} \\
(g-r)_o - \sigma_{gr} + 0.06\,(r_o - 14) &< 0.9, \label{eq:targeting-cuts-gr-r}
\end{align}\label{eq:targeting-cuts}%
\end{subequations}%
where \mueff{} is the effective surface brightness defined in \autoref{eq:sb}, $\sigma_{\mu}$ is the error on \mueff{}, and $\sigma_{gr} \equiv \sqrt{\sigma_g^2 + \sigma_r^2}$ is the error on the $(g-r)_o$ color. 
We only use $(g-r)_o$ color, as the LS photometry does not include the $i$ band.

This primary targeting region is designed to encompass all $z<0.015$ galaxies, based on all available redshifts, especially those taken during SAGA Stage I where we obtained a large number of redshifts without any photometric selection. 
Hence, the probability of a photometric object in this region being a satellite is very high,
and we devoted about half of our spectroscopic resources to achieve high spectroscopic coverage for all ``good'' galaxy targets (as defined in \autoref{sec:sdss}--\ref{sec:decals}) within this region. 
Down to our magnitude limit of $r_o = 20.75$, the number of targets that are within the primary targeting region and between 10 and 300 kpc of each host in projection is 160 per host, on average ($\sim$200 targets deg$^{-2}$). 
We do not attempt to target galaxies within 10\,kpc of the host galaxy because that region is dominated by the host galaxy light. 

The other half of our spectroscopic resources are devoted to addressing the potential issue of selection bias and ensuring satellite completeness. 
This is done by obtaining redshifts for objects outside of the primary targeting region with various selection methods, including random selection, human selection, and selection based on photometric properties (i.e., selecting objects with similar properties as other low-redshift galaxies). 
For example, for each pointing that we observe, we allocate 50 or more fibers in ``discovery mode,'' where they are randomly assigned to targets within \paperone{} $gri$ cuts but outside of the primary targeting region defined in \autoref{eq:targeting-cuts}. 
To date, among more than 20,000 redshifts outside of the primary targeting region, we have not found any satellites.   All galaxies that have confirmed redshifts in $0.003 < z < 0.015$ are within the primary targeting region. 
This result demonstrates the robustness of our targeting strategy.

The left panel of \autoref{fig:z_distribution} shows the redshift distribution within (green histogram) and outside of (blue and orange histograms) the primary targeting region among all of the Stage~I and II redshifts and compares it with the $gri$ cuts in \paperone{} (blue and green histograms).
The $gri$ cuts effectively remove galaxy targets beyond $z \sim 0.5$ but include most galaxies up to $z \sim 0.05$. 
In the inset of \autoref{fig:z_distribution}, the visible bump around redshift $0.005 < z < 0.01$ is due to the presence of satellites in SAGA fields. 
The definition of our primary targeting region further removes galaxy targets beyond $z \sim 0.3$, while including all galaxies within $z < 0.015$ (and hence all SAGA satellites).
In the slightly higher redshift range of $0.015 < z < 0.03$, 83\% of galaxies are still in the primary targeting region.  The small fraction of $0.015 < z < 0.03$ galaxies that sit outside of the primary targeting region motivates our twofold effort to obtain redshifts both within and outside of the primary targeting region, ensuring that we are not missing satellite candidates during the target selection phase.

\subsection{Multifiber Spectroscopy: MMT/Hectospec}
\label{sec:mmt}

Stage II of our survey includes redshifts taken with the fiber-based MMT/Hectospec between 2017 May and 2019 October. The MMT/Hectospec deploys 300 fibers over a 1$^\circ$ diameter field \citep{2005PASP..117.1411F}.  We used Hectospec with the 270 lines\,mm$^{-1}$ grating, resulting in wavelength coverage of 3650--9200\,\AA{} and spectral resolution of 1.2\,\AA{}\,pixel$^{-1}$ ($R\sim1000$). 
As in \paperone{}, the data were reduced using the HSRED pipeline \citep{10.1086/589642}, which finds candidate redshifts by fitting spectral templates as a function of lag using the `zfind' algorithm from the SDSS spectroscopic pipeline \citep[see][]{2012AJ....144..144B}. The spectra were then visually inspected to verify the correct template fit and assign a quality code using an interactive tool, qplot.\footnote{\https{github.com/bjweiner/qplot}}
Typical redshift errors are 25\,\kms.

\subsection{Multifiber Spectroscopy: AAT/2dF}
\label{sec:aat}

Stage II of our survey includes redshifts taken with the fiber-based AAT/2dF observations between 2018 June and 2020 May. The AAT/2dF deploys 400 fibers over a 2$^\circ$
diameter field.  We used the 580V and 385R gratings in the blue and
red arms, respectively, both providing a resolution of $R = 1300$ (between 1 and 1.6\,\AA{}\,pixel$^{-1}$) over a maximum wavelength range of 3700--8700\,\AA{}. As in \paperone{}, the data were reduced using the facility software 2dfdr and Marz \citep{10.1016/j.ascom.2016.03.001}. Spectra were visually inspected to verify the correct template fit. Typical redshift errors are 25\,\kms.

\subsection{Single-slit Spectra: Palomar DBSP}
\label{sec:dbsp}

The primary goal of our single-slit work is to observe objects for which we failed to measure a redshift using the facilities above due to low signal-to-noise, bad sky subtraction, or other issues.   We obtained data on the 200 inch Hale Telescope at Palomar Observatory with the Double Spectrograph (DBSP) between 2019 June to 2020 June.  The DBSP is a single-slit spectrograph that uses a dichroic to split light into separate, simultaneously observed red and blue channels \citep{oke1982}.   For the blue side, we used the 600/4000 grating at an angle of $27.3^\circ$.  For the red side, we used the 316/7500 grating at an angle of $24.6^\circ$.  This results in wavelength coverage of 3800--5500 and 5500--9500\,\AA{} and spectral resolution of 1.1 and 1.5\,\AA{}\,pixel$^{-1}$, respectively, for the blue and red sides.
We determine redshifts using a modified version of the MMT redshift algorithm above; typical errors are 30\,\kms.

\subsection{External Spectroscopic Data Sets}
\label{sec:spec-literature}

We include spectroscopic data in the regions of our hosts from a variety of publicly available surveys. We include spectroscopic data from SDSS DR14 \citep{SDSS_DR14}, NSA v1.0.1 \citep{Blanton2011}, and GAMA DR3 \citep{Baldry2018:GAMA:DR3}; these sources were included in \paperone{}, but now updated versions/data releases are used.    In this work, we also include data from the following surveys, listed in decreasing order of number of sources in complete SAGA fields:  the WiggleZ Dark Energy Survey \citep{WiggleZ}, the 2dF Galaxy Redshift Survey \citep[2dFGRS;][]{2dFGRS}, the Australian Dark Energy Survey \citep[OzDES, Data Release 2;][]{OzDES}, the 6dF Galaxy Survey \citep[6dFGS;][]{6dF}, the 2-degree Field Lensing Survey \citep[2dFLenS;][]{2dFlens}, and the Las Campanas Redshift Survey \citep[LCRS;][]{1996ApJ...470..172S}.   Sources from these surveys are included for quality flag $q_z \ge 3$ and include a few thousand galaxies in the fields around the SAGA hosts presented here.   A handful of galaxies that do not have optical redshifts have velocities measured in \textsc{Hi} by the ALFALFA survey \citep{ALFALFA100}; we followed up these objects and found no inconsistency.  We also retarget galaxies in cases where a literature optical redshift indicates a satellite galaxy to verify the literature velocity measurement and to ensure homogeneity, but we do not count this as a SAGA discovery.
When an object has a consistent redshift from multiple sources, we prefer optical sources and telescopes with a larger aperture.

\begin{figure*}[t]
\centering
\includegraphics[width=\textwidth,clip,trim=0.2cm 0.2cm 0cm 0cm]{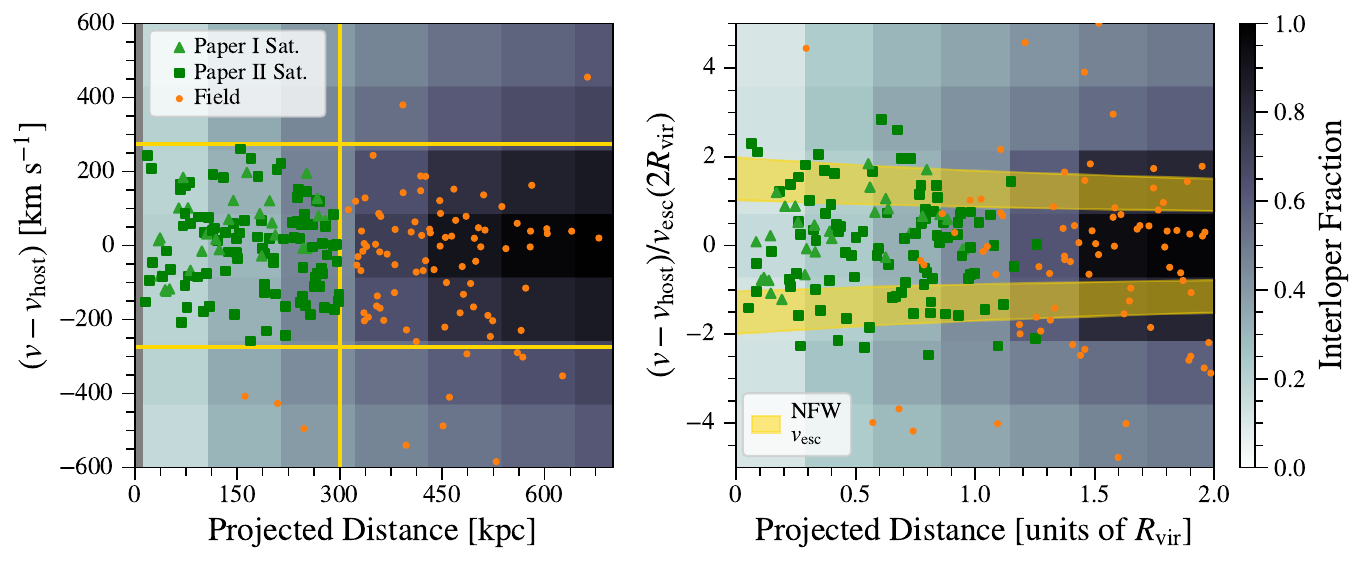}
\caption{Illustration of SAGA satellite definition. \emph{Left panel}: galaxies around SAGA hosts, including SAGA \paperone{} satellites (light green triangles), satellites newly identified in this work (green squares), and nonsatellite galaxies in SAGA fields (orange circles), shown in their heliocentric velocity difference and projected radial distance with respect to their hosts. Satellites are defined to lie within 300\,kpc and $\pm 275\,\kms$ of their host galaxies in projection; this region is indicated by yellow lines. The dark gray band within 10\,kpc indicates the region in which we do not attempt to search for satellites. \emph{Right panel}: same as the left panel but plotted with projected distances in units of their mean estimated host halo virial radius and projected velocities in units of their mean estimated host halo escape velocity. These halo quantities are estimated using the abundance-matching procedure in \paperone. In both panels, the color maps show the fraction of interloping galaxies, estimated by applying our galaxy--halo connection model to a cosmological simulation (see \autoref{sec:satellite-def}). The yellow bands show $\pm 1\sigma$ escape velocity curves calculated based on the NFW profiles of the potential SAGA host halos in our analysis.}
\label{fig:velocity}
\end{figure*}


\subsection{Associating Spectroscopic and Photometric Objects}
\label{sec:spec-merge}

We need to robustly join all of the available spectroscopic data and associate these data with the appropriate objects in our photometric object catalogs. While a simple sky coordinate match would serve our purposes in most cases, it does not work as well for extended galaxies, as the separation between the spectroscopic object and the corresponding photometric object can be several arcseconds.\footnote{This situation happens when the spectroscopic object (especially if it is from literature) has an offset center from the photometric object, usually when the spectroscopy was done with a different photometric catalog or when an off-center star-forming region was targeted.} Hence, for each spectroscopic object, we search within $20''$ to find the best-matching photometric object based on the separation and radii of the photometric objects. Each spectroscopic object is preferably matched to the nearest point source that is within $1''$, then to the brightest extended source within $20''$ and twice the effective radius of the extended source. If there is no photometric object that satisfies the above criteria, we match to the nearest photometric object within $20''$.

A single photometric object may be associated with multiple spectra from different sources. We adopt the redshift value that has the best reported quality, or, when multiple sources have the same quality, the source with the largest aperture. When multiple good redshift values are inconsistent with each other (differ by 150\,\kms{} or more), they will not be associated with the same photometric object. The best, nearest redshift value will first be associated with a photometric object, and other redshift values will be matched to the remaining photometric objects that have not yet been matched to a redshift value.

Our code to associate spectroscopic and photometric objects is also publicly available as a component of our survey software (see footnote \ref{fn:saga-code}).

\subsection{Distribution of SAGA Redshifts}
\label{sec:spec-distribution}

\autoref{fig:z_distribution} shows the redshift distribution within 300\,kpc around all \nhosts{} SAGA hosts, split by different targeting regions (left panel; described in \autoref{sec:target-selection}) and redshift sources and magnitude (right panel). 
Overall, we see that the \nzsaga{} SAGA redshifts (purple histogram) peak around $z=0.2$. The majority of SAGA-obtained redshifts are for faint galaxies $19 < r_o < 20.75$. Conversely, almost all redshifts for dim, low-redshift galaxies ($19 < r_o < 20.75$ and $z<0.2$) were obtained by SAGA.
Specifically, the relative fraction of SAGA to literature redshifts is 46\% for bright galaxies ($r_o < 19$) and 82\% for faint galaxies ($r_o > 19$). 
For an in-depth discussion of the SAGA redshift distribution, see \autoref{app:redshift-stats}.


\begin{figure*}[t]
    \centering
    \includegraphics[width=\textwidth,clip,trim=0.2cm 0.2cm 0 0]{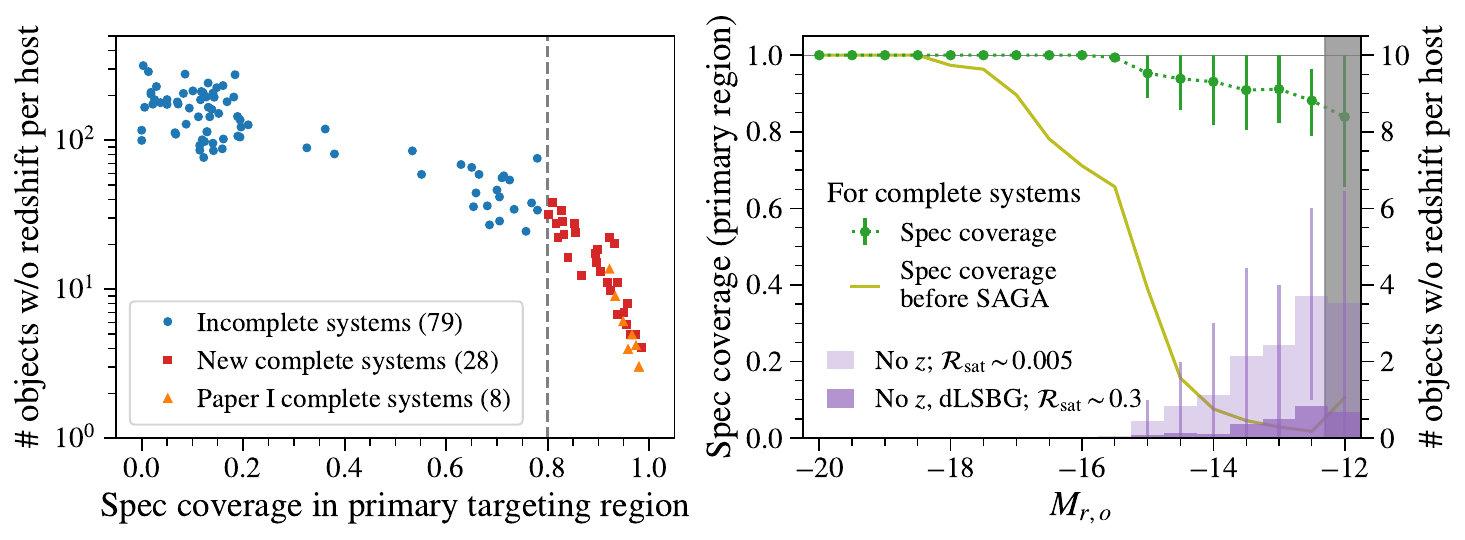}
    \caption{\textit{Left panel}: spectroscopic coverage within the primary targeting region ($x$-axis) and number of remaining targets that do not yet have a redshift ($y$ axis), for all \nhoststotal{} SAGA systems. The vertical dashed line denotes our targeting completeness criteria; systems to the right are designated ``complete systems.'' We plot the \nhostsnew{} newly completed systems as red squares; the eight complete systems reported in \paperone{} are shown as orange triangles. The remaining 79 incomplete systems are shown as blue dots.
    \textit{Right panel}: spectroscopic coverage among the \nhosts{} complete systems as a function of absolute magnitude (assuming all objects are at the host distance).
    The top green points show the median spectroscopic coverage within the primary targeting region among the complete systems (left $y$-axis), while the lower purple histograms show the mean number of targets that have no redshift within the primary targeting region (right $y$-axis). In both cases, the error bars shows the 1$\sigma$ host-to-host scatter. The darker part of the purple histogram shows the objects that are dLSBGs, which have higher chances of being satellites (see \autoref{sec:incompleteness-correction}).
    }
    \label{fig:host_completeness_def}
\end{figure*}

\begin{figure*}[t]
\centering
\includegraphics[width=\textwidth,clip,trim=0.2cm 0.3cm 0 0]{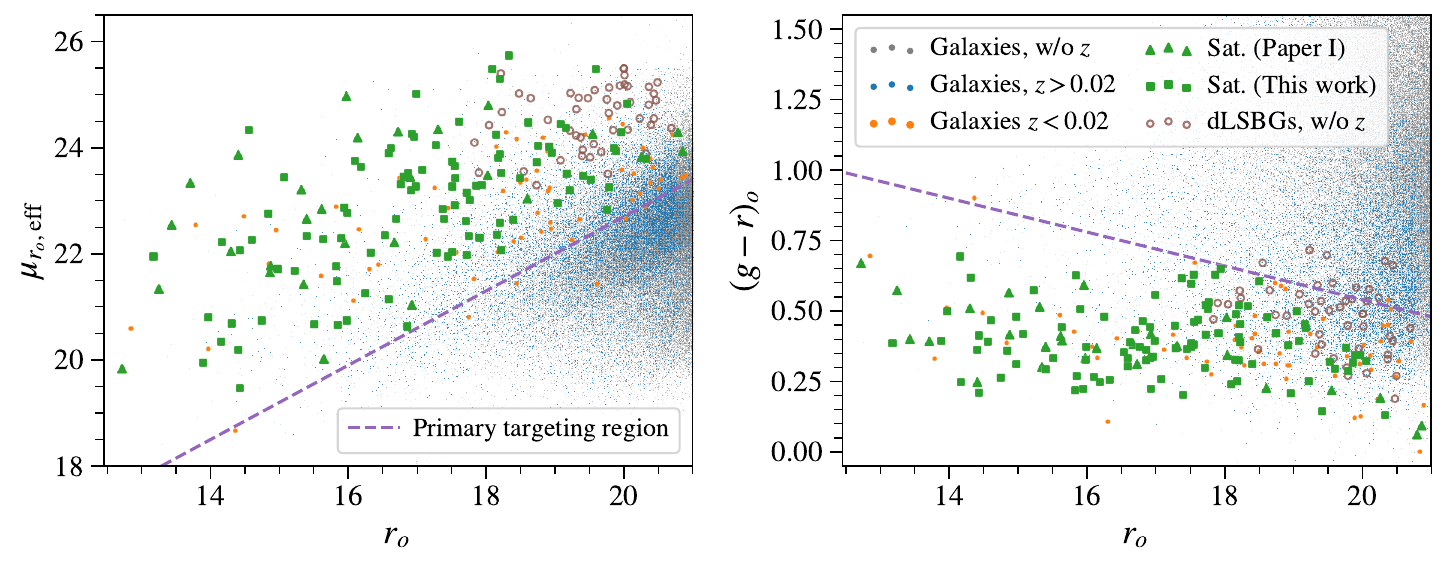}
\caption{SAGA targets and satellites in the surface brightness \mueff{} vs.\ magnitude $r_o$ plane (\textit{left}) and the color $(g-r)_o$ vs.\ magnitude $r_o$ plane (\textit{right}). 
Confirmed SAGA satellites from the \nhosts{} hosts are plotted as green squares or, if they were known from \paperone{}, as green triangles. 
Nonsatellite galaxies that have confirmed redshifts $z<0.02$ are plotted as orange dots. All other galaxies are shown as blue dots if they have confirmed redshifts or as gray dots if they do not.
Brown open circles represent dLSBGs that we targeted but did not obtain a secure redshift for. 
The number of brown open circles is representative of the expected number of unconfirmed satellites.
The boundaries of the ``primary targeting region" (as defined in \autoref{sec:target-selection}, \autoref{eq:targeting-cuts}) are shown as purple dashed lines.}
\label{fig:target_sel}
\end{figure*}

\begin{figure*}[t]
\centering
\includegraphics[width=0.75\textwidth,clip,trim=0 0.2cm 0.1cm 0.1cm]{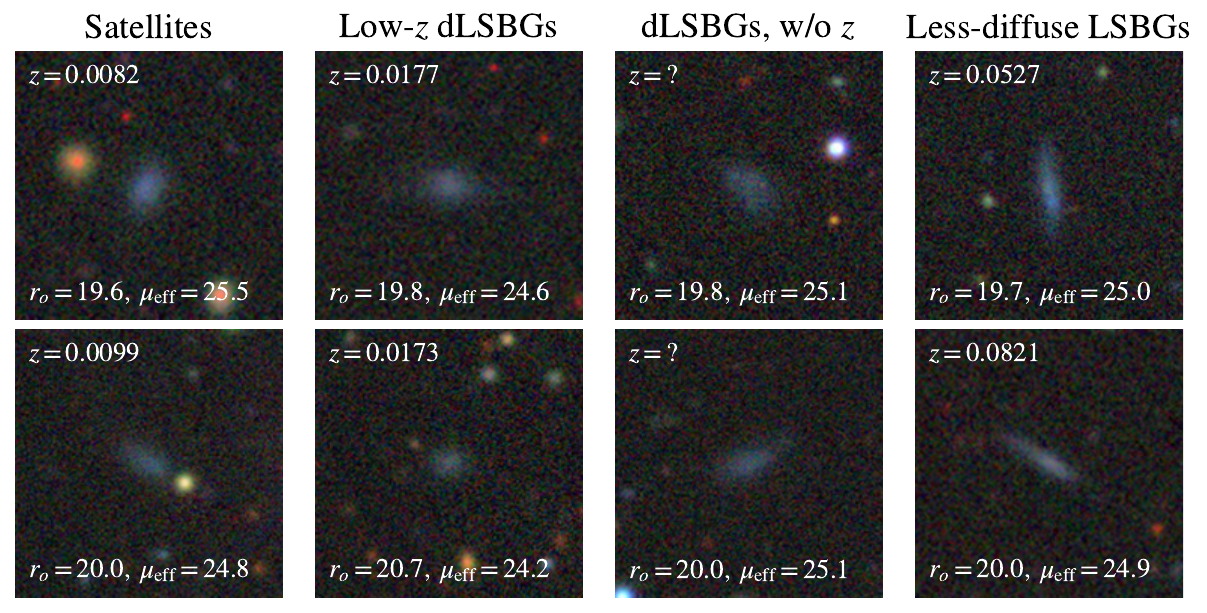}
\caption{Images of LSBGs with similar photometric properties. The four columns, from left to right, show (i) dLSBGs that are confirmed satellites, (ii) low-redshift dLSBGs that are not satellites, (iii) dLSBGs that we failed to obtain redshifts for due to low fiber flux, and (iv) less-diffuse LSBGs that have similar magnitudes, colors, and surface brightnesses in the photometric catalog and slightly higher redshifts. Images are taken from Legacy Surveys DR8 (see footnote \ref{fn:ls-viewer}). The width and height of each image correspond to 40$''$.}
\label{fig:lsbg}
\end{figure*}

\section{Satellite Definition and Survey Completeness}
\label{sec:systematic}

In this section, we examine two systematic uncertainties of our survey design. We first review our satellite definition and estimate the interloper fraction (\autoref{sec:satellite-def}). We then define survey completeness (\autoref{sec:completeness-def}) and carefully develop methods to correct for spectroscopic failures or targeting incompleteness (\autoref{sec:incompleteness-correction}).

\subsection{Satellite Definition and Interloper Fraction}
\label{sec:satellite-def}

We define a ``satellite'' as a galaxy that is within a projected distance of 300\,kpc of its host and within $\pm 275\,\kms$ of the host's redshift, as shown in the left panel of \autoref{fig:velocity} (yellow lines). The projected distance cut roughly corresponds to the virial radius for a typical MW-mass host halo.
The velocity cut is slightly larger than the value used in \paperone{} ($\pm 250\,\kms$) and is motivated by the clear boundary of low-velocity objects seen in \autoref{fig:velocity}.
This observationally motivated satellite definition selects galaxies that lie within the three-dimensional virial radius of their dark matter host halo and conservatively excludes unbound galaxies under the assumption that SAGA host galaxies reside in typical MW-mass dark matter halos.
We list these satellites and describe their properties in \autoref{sec:sat-properties}.

We apply the same satellite definition when comparing our results with simulations; thus, there is no need to correct for interloping galaxies in our comparisons with theoretical predictions.
Nevertheless, we still estimate the fraction of our satellites that may be interloping galaxies in three-dimensional space. 
To calculate this interloper fraction, we populate the \code{c125-2048} cosmological gravity-only simulation presented in \cite{Mao150302637} with galaxies using the subhalo abundance-matching (SHAM) technique, which has been thoroughly validated for MW-mass galaxies
\citep[e.g.,][]{Reddick2013,WechslerTinker,Cao2019, Maccio200600818}.
This procedure yields both mock host halos and mock satellite systems (including interlopers) for each SAGA host. We describe this procedure in more detail in \autoref{sec:theory}.

We use these predictions to compute the fraction of mock galaxies that lie within the three-dimensional splashback radius (the radius where particles reach the apocenter of their first orbit) of each mock host halo divided by the total number of mock galaxies in the bin, applying an absolute magnitude cut of $M_{r,o}<-12.3$ to mimic SAGA observations. 
Here we choose the splashback radius to be 1.5 times the virial radius for the MW-mass host halos; this radius is a physically motivated boundary within which satellite properties are potentially influenced relative to field galaxies. \citep[e.g.,][]{Adhikari14094482,Diemer:2014xya,More150405591,buck:2019MNRAS.483.1314B,2008.11207}.
The result of this procedure is shown by the gray-scale map in \autoref{fig:velocity}. We plot the interloper fraction in both ``observational units'' (i.e., projected distance in units of kiloparsecs and velocity in units of kilometers per second; left panel) and ``scaled units'' (i.e., projected distance in units of the true virial radius of the mock host halo , and velocity in units of the host halo escape velocity evaluated at a characteristic distance; right panel).
 
We find that the galaxies observed by SAGA that pass our satellite definition are very likely to lie within the splashback radius of their corresponding dark matter halo.  In addition, nearly all galaxies that pass our satellite criteria are predicted to be bound to their corresponding halos, assuming Navarro--Frenk--White (NFW) host halo density profiles \citep[yellow bands;][]{nfw1996,nfw1997}. 
Within the SAGA satellite definition, the average interloper fraction is 30\%, and the fraction varies with projected distance.
A handful of the galaxies we observe lie at low projected distance but high relative velocity with respect to their predicted host halos and therefore deserve further study as potential ``fly-by'' galaxies \citep{Sinha11031675,An191111782}. Follow-up observations of their gas content and the reconstruction of their star formation history can help distinguish whether these galaxies have experienced stripping and their nature as interlopers or satellites.

\subsection{Survey Completeness}
\label{sec:completeness-def}

In \autoref{sec:target-selection} we designated a ``primary targeting region'' (\autoref{eq:targeting-cuts}) in which we aim to achieve high spectroscopic coverage.
While we have obtained, and continue to obtain, redshifts for targets outside this region, the chance of finding satellites outside of the primary targeting region is extremely low based on our existing data. 
Following the terminology used in \paperone{}, we consider a SAGA satellite system ``complete'' for our survey if its spectroscopic coverage is higher than 80\% for all galaxies that are both within 300\,kpc in projection to the galaxy host and within the primary targeting region and down to $r_o = 20.75$ (corresponding to $M_{r,o}=-12.3$ at 40.75\,Mpc).

\autoref{fig:host_completeness_def} demonstrates our completeness definition. The left panel shows the number of targets without redshifts and the spectroscopic coverage within the primary targeting region for all of the \nhoststotal{} SAGA systems passing our targeting criteria (\autoref{sec:host-imaging-coverage}). Among \nhoststotal{} systems, \nhosts{} have spectroscopic coverage higher than 80\% (and 22 have coverage higher than 90\%). We have obtained some SAGA redshifts for an additional 19 systems; these systems are not presented in this work, except for being shown as the incomplete systems in \autoref{fig:host_completeness_def}.

The list of \nhosts{} complete SAGA systems is given in \autoref{tab:complete-hosts}. Each of our complete systems is assigned a ``SAGA name'' for internal reference. The table also lists, for each SAGA system, the spectroscopic coverage within the SAGA primary targeting region, the number of satellites (defined in \autoref{sec:satellite-def}), and the estimated number of missed satellites (defined in \autoref{sec:incompleteness-correction}).
\autoref{tab:complete-hosts} is also available in machine-readable format on the SAGA website (see footnote \ref{fn:saga}).

The right panel of \autoref{fig:host_completeness_def} shows the average spectroscopic coverage and the number of targets without redshifts as a function of absolute magnitude. For this plot, we compute absolute magnitude as if all targets were at the distance of their host. All \nhosts{} complete systems reach 100\% spectroscopic coverage within the primary targeting region for galaxies brighter than $M_{r,o} = -15.5$. For galaxies fainter than $M_{r,o} = -15.5$, we maintain, on average, a 90\% spectroscopic coverage down to $M_{r,o} = -12.3$, with completeness slightly decreasing toward fainter magnitudes. 
For comparison, we show the average spectroscopic coverage from the literature and preexisting surveys before SAGA; the coverage declines rapidly below $M_{r,o} = -17.5$, and is not higher than 20\% below $M_{r,o} = -14.5$. 

\autoref{fig:host_completeness_def} also shows that, on average, there are about 10 galaxies per host with $ -15.5 < M_{r,o} < -12.3$ for which we do not have redshifts. If all targets within the primary targeting region have exactly the same probability of being identified as a satellite, then we expect that there will be $\sim$0.2 satellites for every 10 objects. 
In the next section, we will utilize a more sophisticated method to estimate and correct for incompleteness. 

\subsection{Incompleteness Correction}
\label{sec:incompleteness-correction}

While quantifying the spectroscopic coverage and the number of incomplete targets provides a robust way to evaluate our survey completeness, it is important to note that not all targets are equally likely to be satellites.
\autoref{fig:target_sel} clearly demonstrates this statement by showing the distributions of satellites and targets in the surface brightness--magnitude (left panel) and $gr$ color--magnitude (right panel) planes. The boundaries of our primary targeting region are shown as purple dashed lines. 

As stated in \autoref{sec:target-selection}, all confirmed satellites are in the primary targeting region, where we find \nsats{} satellites among 5419 redshifts. Outside of the primary targeting region, we find no satellites among 21,308 redshifts. If we assume that the rate of finding a satellite within (or outside of) the primary targeting region is constant, then its value would be $\sim$2\% (or $< 0.005$\% for objects outside).

Of course, the chance of finding a satellite should vary smoothly rather than behave like a step function in photometric space.  
We hence build a model to evaluate ``satellite rates,'' denoted by $\mathcal{R}_\mathrm{sat}$ and defined as the fraction of satellites among the targets with redshifts at a specific point in the three-dimensional photometric space ($r_o$, \mueff, $(g-r)_o$). We describe the details of our model construction in \autoref{app:completeness-model} and calculate the satellite rate $\mathcal{R}_\mathrm{sat}$ for each object for which we have not obtained a redshift.
These satellite rate values can then be used to correct for our incompleteness; the sum of $\mathcal{R}_\mathrm{sat}$ would be the expected number of unconfirmed satellites. 

There is an additional concern about the situation in which we failed to measure redshifts for faint, diffuse, low surface brightness objects using multifiber spectroscopy facilities; this occurs when the flux incident on the fiber is too low (i.e., for objects with low fiber magnitudes). While we do attempt to follow up these targets with single-slit spectroscopy, we do not always recover redshifts due to our limited single-slit resources. 
In \autoref{fig:target_sel}, we show these diffuse LSBGs (dLSBGs) for which we failed to obtain redshifts with AAT or MMT as brown open circles; they cluster in the faint, low surface brightness, and slightly redder regime.  
Note that within the same photometric regime, we did successfully obtain many redshifts (blue dots), including absorption line--only galaxies; some of these are LSBGs that have a slightly higher central surface brightness and are at a slightly higher redshift than the dLSBGs. 
While the distinction between the dLSBGs and less-diffuse LSBGs is often discernible by the human eye (see examples in \autoref{fig:lsbg}), they are not easily separable using photometric catalog properties alone \citep{Tanoglidis200604294}. 

To ensure that we have a conservative estimate of our completeness, we visually inspected targets in each SAGA field to identify all dLSBGs. Among the \nhosts{} SAGA fields, we found 70 of these objects that we have not obtained redshifts for (down to $M_{r,o} < -12.3$); 13 of them are slightly outside the primary targeting region.
The counts of these dLSBGs without redshifts are shown in \autoref{fig:host_completeness_def} as a darker histogram in the right panel.
Based on the redshifts that we have obtained for some dLSBGs and on the radial distribution of these dLSBGs (i.e., accounting for the fact that satellites concentrate closer to host galaxies, while background low-redshift galaxies should be uniform), we estimate that about 25--30\% of these dLSBGs may be satellites.

We incorporate this dLSBG estimate in our satellite rate model and then calculate $\mathcal{R}_\mathrm{sat}$ for each target for which we have not obtained a redshift. Among them, the average $\mathcal{R}_\mathrm{sat}$ for targets that are within the primary targeting region but not dLSBGs is $\sim$0.005; while the average $\mathcal{R}_\mathrm{sat}$ for dLSBGs is $\sim$0.3. 
The sum of these $\mathcal{R}_\mathrm{sat}$ values gives us the expected total number of missed satellites. 
We found that, on average, we may be missing about 0.7 faint satellites ($-15.4 < M_{r,o} < -12.3$) per SAGA host (24 in total among \nhosts{} hosts). All SAGA hosts are close to 100\% complete above $ M_{r,o}=-15.4$. 

In the analysis below, we interpret $\mathcal{R}_\mathrm{sat}$ as the probability of each target without a redshift being a satellite, and we incorporate this correction in binned statistics such as the luminosity function and quenched fraction. 
For scatter plots that show distributions of individual satellites, we include a representative set of unconfirmed potential satellites. This set contains 24 potential satellites in $-15.4 < M_{r,o} < -12.3$ and 11 below $M_{r,o} = -12.3$, selected by a random realization based on $\mathcal{R}_\mathrm{sat}$. Potential satellites are always marked distinctly from confirmed satellites. 

\begin{figure*}[t]
    \centering
    \includegraphics[width=\textwidth,clip,trim=0.4cm 0.8cm 0.4cm 0.2cm]{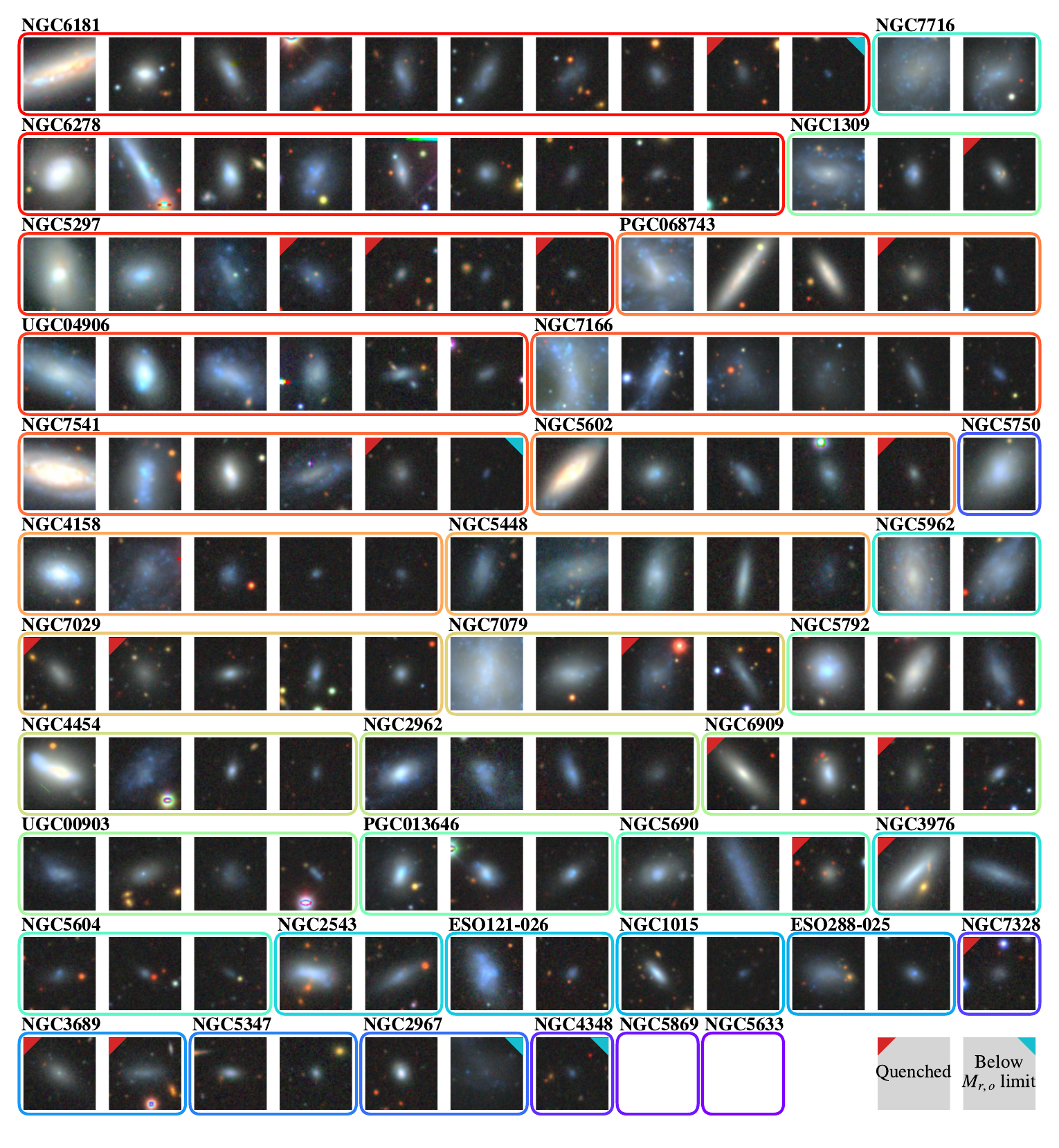}
    \caption{Images of \nsats{} satellites, sorted by $M_{r,o}$ within each of the \nhostswsats{} complete systems that have satellites. One system shown (NGC\,4348) has one confirmed satellite fainter than our absolute magnitude limit, and two complete systems (NGC\,5896 and NGC\,5633) do not have confirmed satellites. 
    A red triangle in the upper left corner indicates a quenched satellite, and
    a cyan triangle in the upper right corner indicates a satellite fainter than our absolute magnitude limit, $M_{r,o} = -12.3$. 
    Images are taken from LS DR8 (see footnote \ref{fn:ls-viewer}). The width and height of each image correspond to 40$''$.}
    \label{fig:all-sats}
\end{figure*}

\begin{figure*}[t]
    \centering
    \includegraphics[width=\textwidth,clip,trim=0.4cm 0.6cm 2.1cm 0.3cm]{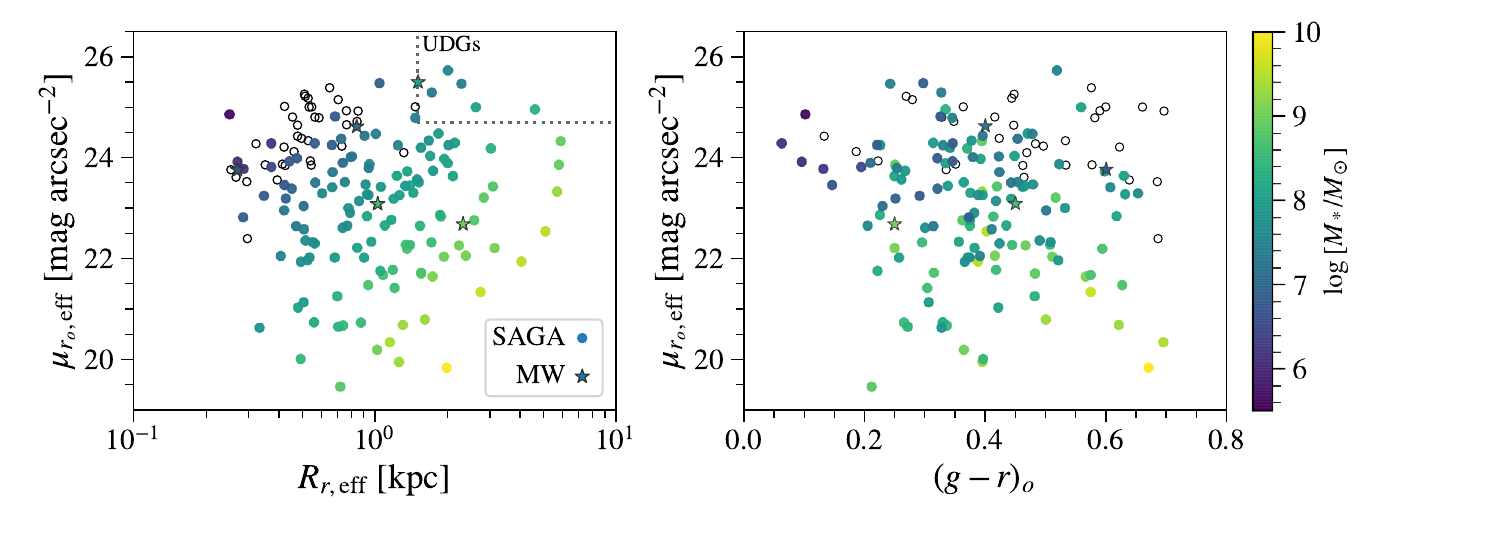}
    \caption{Effective surface brightness vs.\ physical radius (\textit{left}) and color (\textit{right}) for all satellites in our \nhosts{} complete SAGA hosts.  Confirmed satellites are color-coded based on their stellar mass, and a representative set of potential satellites is shown as open black circles to demonstrate the photometric regions where we may be incomplete. For comparison, MW satellites are shown as stars. In the left panel, the galaxies within the dotted lines are considered UDGs.}
    \label{fig:sat_properties}
\end{figure*}

\begin{figure*}[t]
    \centering
    \includegraphics[width=\textwidth,clip,trim=1.7cm 0.4cm 1.1cm 1.8cm]{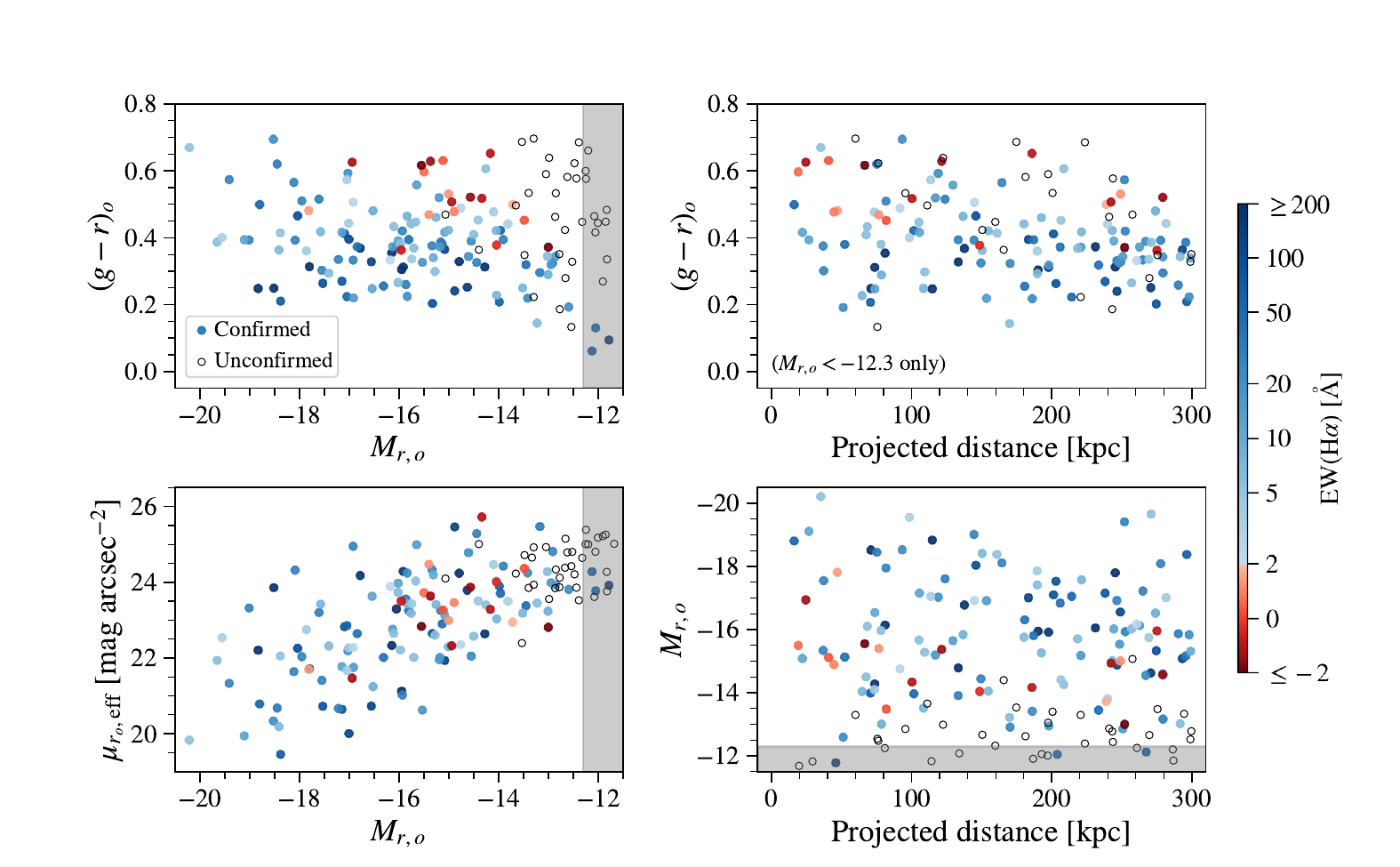}
    \caption{\textit{Left column}: satellite color (\textit{upper}) and  effective surface brightness (\textit{lower}) vs.\ absolute magnitude. \textit{Right column}: Satellite color (\textit{upper}) and  absolute magnitude (\textit{lower}) vs.\ projected distance to host. In all panels, the symbols are color-coded by H$\alpha$ EW as measured in our fiber-based spectra for both emission- and absorption-line galaxies.  The color scale transitions from blue to red at $\mathrm{EW}(\mathrm{H}\alpha) = 2\,\text{\AA{}}$, the threshold we define as a quenched satellite. Open black circles show a representative set of potential satellites that have no confirmed redshifts to demonstrate the photometric regions where we may be incomplete. Gray shaded regions indicate satellites below our magnitude limit, $M_{r,o}=-12.3$.}
    \label{fig:sat_ew_radial}
\end{figure*}

\begin{figure}[t]
    \centering
    \includegraphics[width=\columnwidth,clip,trim=0.1cm 0.3cm 0cm 0cm]{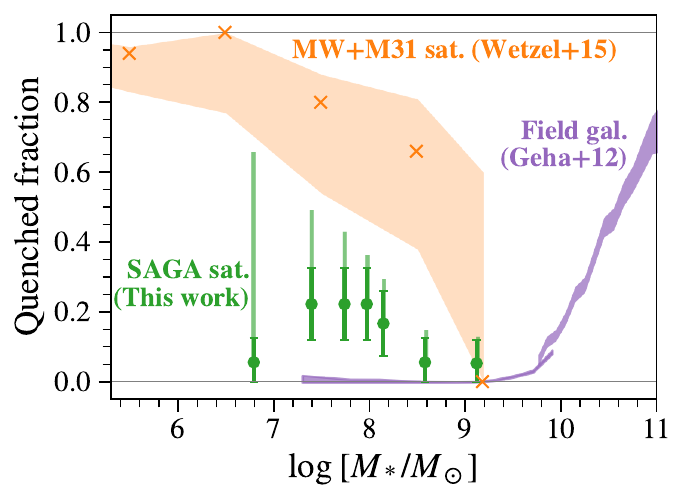}
    \caption{Satellite quenched fraction as a function of stellar mass for SAGA satellites (green points with error bars), MW and M31 satellites (orange crosses and shaded region; \citealt{Wetzel:1503.06799}), and isolated SDSS field galaxies (purple shaded region;  \citealt{geha2012}). See \autoref{sec:quenching} for quenching definition. For SAGA satellites, the darker error bars with caps denote shot noise, while the lighter error bars denote the combined correction for both incompleteness and interlopers (dominated by incompleteness correction). 
    The quenched fraction increases as stellar mass decreases for both SAGA and Local Group satellites, but the increase is steeper for Local Group satellites.
    }
    \label{fig:quenched_frac}
\end{figure}

\section{Properties of Satellite Galaxies}
\label{sec:sat-properties}

We now turn to the main findings from SAGA Stage II. In this section, we investigate the properties of individual satellite galaxies.  In \autoref{sec:sat-system}, we study these galaxies as satellite systems and discuss the implications of our findings for the galaxy--halo connection in \autoref{sec:theory}. 

In the \nhosts{} complete systems, we have identified \nsats{} satellites using the definition in \autoref{sec:satellite-def}. There are \nsatslimit{} within the absolute magnitude limit of the survey, $M_{r,o} < -12.3$. 
The \nsats{} satellites are distributed around  \nhostswsatslimit{} hosts; \nsatssaga{} of these satellites did not have preexisting redshifts.
One host, NGC\,4348, has only one confirmed satellite that is fainter than the magnitude limit, and the other \nhostsnosats{} complete systems (NGC\,5896 and NGC\,5633) have zero confirmed satellites. 
We find two satellites around ESO\,288-025 that are also within 300 kpc in projection of NGC\,7166 and have consistent redshifts (recall that these two hosts are a Local Group--like pair). The two satellites sit on the outskirts of NGC\,7166; hence, we assign them to ESO\,288-025.
 
\autoref{fig:all-sats} shows the images of all \nsats{} satellites. The images are taken from the LS Viewer (See footnote \ref{fn:ls-viewer}) with the DR8 layer (this layer includes DES imaging where available). 
Quenched satellites (defined in \autoref{sec:quenching}) and satellites fainter than our absolute magnitude limit $M_{r,o} = -12.3$ are indicated by a red or cyan triangle in the upper corners, respectively. 
The coordinates and basic properties of these satellites are listed in \autoref{tab:satellites} and are available in machine-readable format on the SAGA website (see footnote \ref{fn:saga}).
We also plot the positions and velocities of all SAGA satellites in \autoref{fig:sat-pos-vel}, \autoref{app:sat-pos-vel}.

For satellites that were reported in \paperone{}, some properties may differ slightly in this work, due to our adoption of new photometric catalogs and the rebuilding of the host list. 
In addition, we have discovered two more satellites around two of the \paperone{} ``complete'' hosts: LS-444338-475 in the system of NGC\,6181 and LS-311554-3218 in the system of PGC\,068743. We excluded one satellite reported in \paperone{} (1237666408439677694 in the system of NGC\,7716), due to its poor redshift quality; however, this object is below our survey magnitude limit. 
We also obtained an accurate redshift for NGC\,5962b, which was reported to be a satellite of NGC\,5962 in \citet{Zaritsky:1997ApJ...478...39Z}. Its redshift is at 0.338; hence, it is not a satellite of NGC\,5962.

\subsection{Distributions of Sizes, Colors, and Stellar Masses}
\label{sec:sat-properties-dist}

In \autoref{fig:target_sel}, we showed how SAGA satellites distribute in the surface brightness--magnitude and color--magnitude planes.   In \autoref{fig:sat_properties}, we plot the distributions of satellite effective surface brightness against physical effective radius and color and color-code each satellite point based on its stellar mass. In \autoref{fig:sat_ew_radial} we plot the distributions of satellite color, effective surface brightness, absolute magnitude, and projected distance to the host in four different views (the color code of \autoref{fig:sat_ew_radial} is explained in \autoref{sec:quenching}).
In both \autoref{fig:sat_properties} and \ref{fig:sat_ew_radial}, we include a representative set of potential satellites that do not yet have measured redshifts (as open circles) based on our satellite rate model prediction. 

In \autoref{fig:sat_properties}, we include the MW satellites using quantities from references listed in \paperone{}.  In the left panel, the SAGA satellites show the same general trend (color gradient) as the MW satellites; intrinsically fainter satellites are, on average, of lower surface brightness and have smaller physical radii.
In the lower left panel of \autoref{fig:sat_ew_radial}, the surface brightness relation appears to flatten near the faint end, even when those unconfirmed satellites are considered.  While this flattening behavior may be due to poorly measured photometry, it is nevertheless consistent with the satellites in the Local Group \citep[Figure~7]{mcconnachie12}. We were able to identify satellites of similar surface brightness as the lowest surface brightness Local Group dwarf galaxies within the SAGA magnitude range.

Five of the SAGA satellites are formally in the ultradiffuse regime, defined by \citet{vandokkum2015} as $R_{\rm eff} > 1.5$\,kpc and $\mu_{g,0} > 24$\,mag\,arcsec$^{-2}$, which translates to $\mueff > 24.7$\,mag\,arcsec$^{-2}$ assuming an $n$=1 S{\'e}rsic profile \citep{Graham2005} and average $(g-r)_o$ = 0.4.  This region is in the upper right corner of  \autoref{fig:sat_properties}  (left panel; shown as dotted lines) and includes the Sagittarius dwarf spheroidal galaxy of the MW.  None of the SAGA satellites approach the size or surface brightness of M31's ultradiffuse satellite And\,XIX although this galaxy is below our absolute magnitude limit \citep{Collins2020}.  We also find a small number of ultradiffuse galaxies (UDGs) in our larger low-redshift galaxy sample.

The right panel of \autoref{fig:sat_properties} and the upper left panel of \autoref{fig:sat_ew_radial} show, respectively, the color--surface brightness and color--magnitude relations of the SAGA satellites. In both cases, slight trends exist when only confirmed satellites are considered, but the trends diminish when the unconfirmed satellites are taken into account. The slight trends are likely driven by a few very faint and blue confirmed satellites ($(g-r)_o < 0.2$; note that some of these are below our formal magnitude limit). 
Overall, the color distribution of the SAGA satellites is similar to that of the MW satellites.
Due to the luminosity--metallicity relationship \citep{Kirby2013},  we might expect a trend of bluer colors with decreasing satellite luminosity.  We do not see this trend when the unconfirmed satellites are included, likely due to the competing trend of increased quenching for the fainter satellites. We will discuss this in detail in the next subsection.

For completeness, here we list $(\rho_s, p)$, the Spearman rank correlation coefficients and corresponding $p$-values, for the SAGA satellite samples without/with unconfirmed satellites (i.e., open circles): Fig.~\ref{fig:sat_properties}, left panel $(0.1, 0.2) / (-0.1, 0.2)$; Fig.~\ref{fig:sat_properties}, right panel $ (-0.2, 0.02) / (-0.03, 0.7)$;
Fig.~\ref{fig:sat_ew_radial}, upper left  panel $ (-0.2, 0.01) / (-0.003, 1)$;
Fig.~\ref{fig:sat_ew_radial}, lower left  panel $ (0.6, 1 \times 10^{-13}) / (0.7, 5 \times 10^{-24})$;
Fig.~\ref{fig:sat_ew_radial}, upper right  panel $ (-0.3, 0.002) / (-0.2, 0.003)$; and 
Fig.~\ref{fig:sat_ew_radial}, lower right panel $ (0.08, 0.4) / (0.08, 0.3)$.

\subsection{Quenched Satellites}
\label{sec:quenching}

We measure the equivalent width (EW) of the H$\alpha$ line in our spectra for all SAGA satellites; the H$\alpha$ line probes star formation within $\sim$5\,Myr \citep{2008.08582}.
We define quenched satellites as having $\mathrm{EW}(\mathrm{H}\alpha) < 2\,\text{\AA{}}$, i.e., having a nonsignificant H$\alpha$ emission line or H$\alpha$ in absorption (negative values), following several authors, including \citealt{geha2012}.
Since the majority of our spectra are fiber-based, and the fiber coverage is usually a small fraction of the overall galaxy, our EWs cannot be directly translated into star formation rates without correcting for aperture effects. 
Based on our definition, a galaxy labeled as star-forming is secure, while a quenched galaxy could be star forming in the rare cases where our fiber missed a star-forming region.
We color-code the satellite points in \autoref{fig:sat_ew_radial} based on the measured value of $\mathrm{EW}(\mathrm{H}\alpha)$. 
The representative set of potential satellites do not have spectra and are shown as open circles.
The color bar in \autoref{fig:sat_ew_radial} transitions from blue to red at $\mathrm{EW}(\mathrm{H}\alpha) = 2\,\text{\AA{}}$ to indicate quenched satellites in red.

Down to our magnitude limit of $M_{r,o} = -12.3$, there are 18 quenched satellites out of \nsatslimit{} confirmed satellites (15\%).   This is slightly higher than the 1/27 quenched satellites (4\%) presented in \paperone{};  these numbers are statistically consistent given the small number of satellites presented in \paperone{}.   Since the AAT/MMT fibers cover only a fraction of each galaxy, it is possible we missed a star-forming region in a few cases.  All quenched satellites are indicated in \autoref{fig:all-sats} by a red triangle in the upper left corner.

As mentioned in \autoref{sec:sat-properties-dist}, we do not see a clear color--magnitude trend, likely due to the competing trend of increased quenching for the fainter satellites  (upper left panel of \autoref{fig:sat_ew_radial}). 
Nevertheless, we do see that quenched satellites tend to be redder when compared with satellites of the same absolute magnitude. 
There is a modest trend between color and projected distance from the host, such that inner satellites have redder colors (upper right panel of \autoref{fig:sat_ew_radial}). Quenched satellites also tend to be closer to their host galaxies; they have a median projected distance of 110\,kpc, more centrally concentrated than the median projected distance of 180\,kpc for star-forming satellites.

\autoref{fig:quenched_frac} shows the satellite quenched fraction as a function of stellar mass for the SAGA satellites, MW and M31 satellites \citep{Wetzel:1503.06799}, and for isolated SDSS field galaxies \citep{geha2012}. 
For the SAGA satellites, we show the correction due to incompleteness and interlopers (as light green error bars) in addition to the shot noise (as dark green capped error bars). 
The incompleteness correction conservatively assumes that all potential satellites are quenched. The interloper correction downweights each confirmed satellite based on its interloper probability.

In \autoref{fig:quenched_frac}, the quenched fraction increases as stellar mass decreases for both SAGA and Local Group satellites, but the increase is steeper for Local Group satellites.
All but two of our quenched satellites are fainter than $M_{r,o} = -16$, with stellar masses less than $10^{8.2}$\,\msun.  Interestingly, the two bright quenched satellites are both very close to their respective hosts.
Around $M_* = 10^{8.5}\,\msun{}$, almost all SAGA satellites are still star-forming, while more than half of the Local Group satellites (dominated by the M31 satellites) are quenched. 
At $M_* = 10^7 \,\msun{}$, the quenched fraction goes to about 0.6 for SAGA satellites if we consider incompleteness and interloper corrections, while the majority of Local Group satellites are quenched at this stellar mass. 

We note that in \autoref{fig:quenched_frac}, quenching is defined as $\mathrm{EW}(\mathrm{H}\alpha) < 2\,\text{\AA{}}$ for SAGA satellites and field galaxies and as $M_\mathrm{gas}/M_* < 0.1$ for Local Group satellites.  This difference in quenching definition would not reconcile the discrepancy, as a few of our H$\alpha$-based quenched galaxies may be star-forming due to fiber coverage, but gas-depleted galaxies would not be star-forming. Related work by \citet{Bennet2019a} measured the quenched fraction of the satellite galaxies of four Local Volume galaxies using the presence of bright blue main-sequence stars as an indicator of star formation and found that, down to the SAGA magnitude limit, the quenched fraction is about 0.6--0.7 for M81 and Cen\,A, but zero for M94 and M101.

These apparently different quenched fractions around different host galaxies may indicate that the host environment or other host properties impact the star formation of Local Group satellites.
The rapid variation (``burstiness'') of star formation in dwarf galaxies may also cause a large scatter in the quenched fraction from host to host \citep[e.g.,][]{Weisz2012}.
Comparing our results with numerical simulations will shed light on the relation between the observed quenched fraction and quenching timescales \citep[e.g.,][]{1402.1498,10.1093/mnras/stv2058,Garrison-Kimmel2019:1903.10515,Akins:2008.02805}. 
In particular, \citet{Akins:2008.02805} suggested that the satellite quenching timescale can vary significantly depending on satellite gas mass and the level of ram pressure stripping.

On the observational side, we have carefully examined potential detection bias against quenched fractions, including biases in the underlying photometric catalogs and in our target selection and completeness. 
Down to our magnitude limit ($M_{r,o} < -12.3$), we found no evidence of missing LSBGs (e.g., $\mueff > 25.5$\,mag\,arcsec$^{-2}$) in our photometric catalogs when compared with deeper HSC catalogs (see \autoref{sec:phot-validation}); this is expected, since both the DES and LS photometric catalogs are much deeper than our magnitude limit. 
For objects that are identified in the photometric catalogs but without a redshift (either because we have not targeted them or we failed to obtain redshift), we model the likelihood of being satellites and quantify our incompleteness based on redshifts we have obtained (see \autoref{sec:incompleteness-correction}). This incompleteness correction is shown as open circles in \autoref{fig:sat_ew_radial} and light green error bars in \autoref{fig:quenched_frac}; and the correction is significant for galaxies below $M_* < 5 \times 10^7 \, \msun$. 
Even with the incompleteness correction, the SAGA quenched fraction is still lower than that of the Local Group; the difference is mainly due to the existence of faint star-forming satellites in the SAGA sample.
We plan to obtain redshifts for these unconfirmed satellites to reduce our error bars on the quenched fraction and to measure star formation rates for each of our satellites, including aperture corrections \citep[e.g.,][]{Brinchmann2004}.

\begin{figure*}[t]
    \centering
    \includegraphics[width=\textwidth,clip,trim=0.1cm 0.2cm 0 0.1cm]{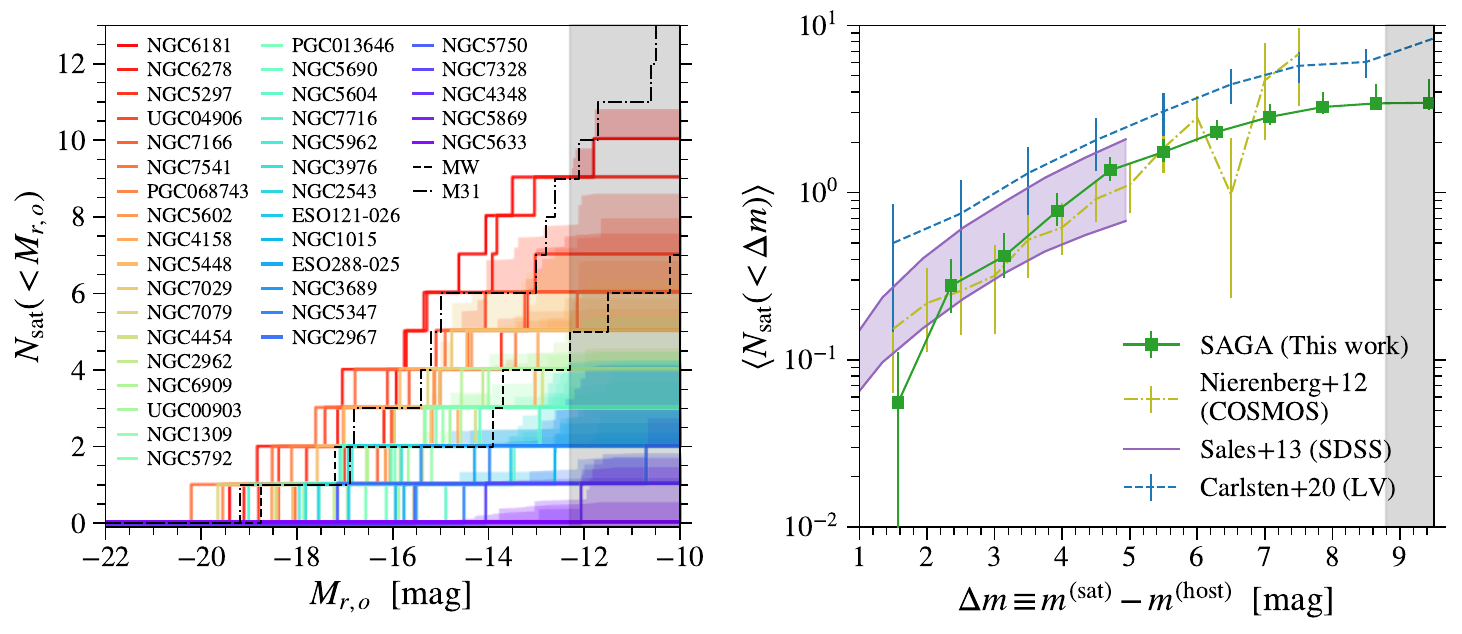}
    \caption{\textit{Left panel}: cumulative satellite luminosity functions for the \nhosts{} complete SAGA hosts; the color gradient follows the number of satellites in each system. Shaded regions represent incompleteness corrections (see \autoref{sec:incompleteness-correction}), and the gray vertical band indicates the region below our survey magnitude limit $M_{r,o} > -12.3$.
    The MW (dashed black line) and M31 (dashed--dotted black line) are shown for comparison.
    \textit{Right panel}: average cumulative satellite luminosity functions in scaled magnitude $\Delta m \equiv m^\mathrm{(sat)} - m^\mathrm{(host)}$ from 
    the SAGA Survey (this work; green squares with error bars showing the combination of Poisson shot noise and incompleteness correction; hosts selected from $0.005 < z < 0.01$ and $10 < \log[M_*/\msun] < 11$), 
    COSMOS (yellow dashed--dotted line with error bars showing Poisson shot noise; taken from Figure~7, top left panel, of \citealt{Nierenberg2012}; hosts selected from $0.1 <z<0.4$ and $10.5 < \log[M_*/\msun] < 11$),
    SDSS (purple band; taken from Figure~3, right panel, of \citealt{Sales2013}; hosts selected from $z<0.055$, and the lower and upper boundaries of the purple band represent the samples of $10 < \log(M_*/\msun)<10.5$ and $10.5 < \log(M_*/\msun)<11$, respectively),
    and the Local Volume (blue dashed line; taken from Figure 8, right panel, of \citealt{Carlsten200602443}) and multiplied by 2 to account for their satellite definition being within only 150\,kpc.
    }
    \label{fig:sat_lf}
\end{figure*}

\begin{figure*}[t]
    \centering
    \includegraphics[width=\textwidth,clip,trim=0.2cm 0.2cm 0cm 0]{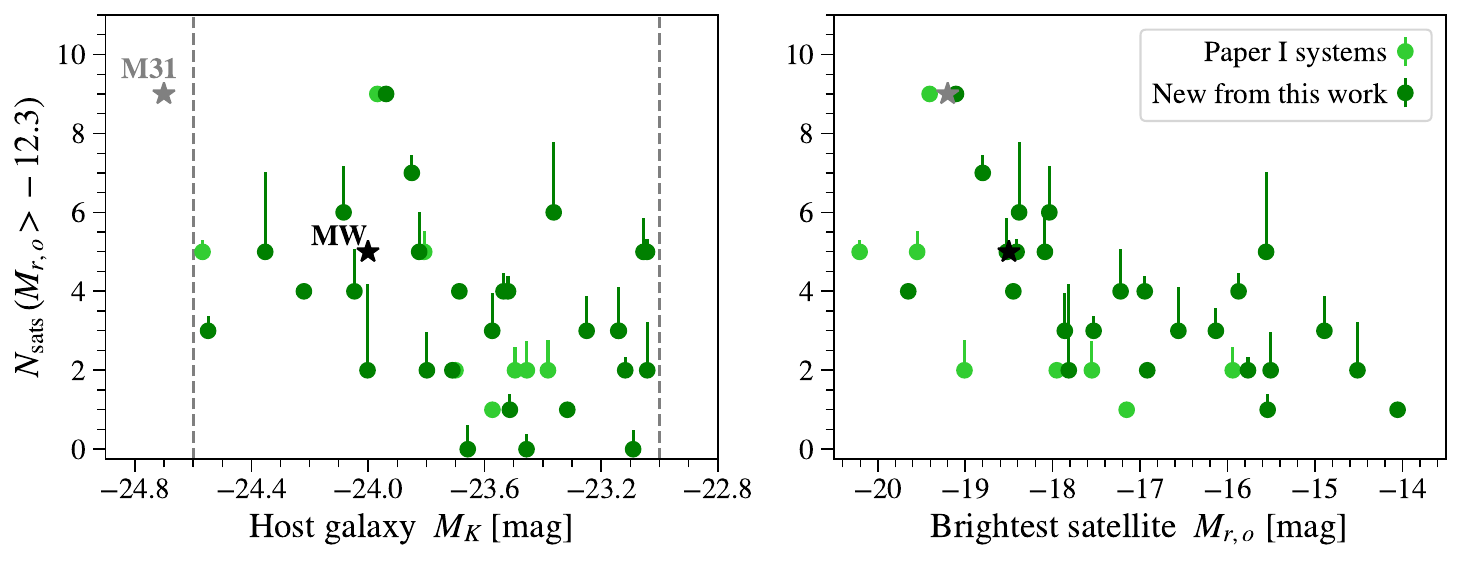}
    \caption{Number of satellites with $M_{r,o}<-12.3$ vs.\ host galaxy $M_K$ (\textit{left}) and brightest satellite $M_{r,o}$ (\textit{right}) for SAGA systems newly presented here (dark green dots) and in \paperone{} (light green dots). The symbols show confirmed satellites counts, and the error bars indicate spectroscopic incompleteness corrections. Total satellite counts for the MW (black star) and M31 (gray star) are shown for comparison. In the left panel, the vertical dashed lines indicate our host luminosity selection criteria. Systems that have no confirmed satellites are not plotted in the right panel. For plotting purposes, we slightly shift the $M_K$ values of hosts that would appear as a single overlapping point.}
    \label{fig:num_sat_host_mk}
\end{figure*}

\section{Properties of Satellite Systems}
\label{sec:sat-system}

We now investigate satellite population-level statistics of SAGA systems, focusing on satellite luminosity functions (Section \ref{sec:LF}), radial distributions (Section \ref{sec:radial_dist}), and satellite groups and planes (Section \ref{sec:sats-plane}).

\subsection{Satellite Luminosity Functions}
\label{sec:LF}

We examine the luminosity functions of the SAGA satellite systems. The left panel of \autoref{fig:sat_lf} shows the cumulative number of satellites around each SAGA host galaxy as a function of absolute magnitude, including our assessment of host-by-host incompleteness as described in \autoref{sec:incompleteness-correction}. This panel demonstrates the diversity of SAGA satellite systems despite our relatively restrictive range of host properties. It further shows that the MW luminosity function is  typical among our statistical sample.

To assess this quantitatively, we compute the Poisson likelihood of observing the MW sample given the set of SAGA systems, marginalizing over the unknown Poisson rate in each absolute magnitude bin following \cite{Nadler180905542}. We find that the likelihood of observing the MW satellite system is nearly identical to that of a model that simply averages the observed count of SAGA satellites across all systems in each absolute magnitude bin (the log-likelihood difference between these scenarios is $\sim$1). 
Thus, the MW satellite luminosity function is consistent with being drawn from the same underlying distribution as the SAGA data.

In contrast, the unnormalized satellite luminosity function of M31 lies in the high satellite number tail of the SAGA luminosity function distribution; two SAGA systems have the same number of satellites as M31. Note that M31 is slightly brighter than our upper limit on host luminosity.
We also find that eight out of \nhosts{} SAGA systems have no confirmed satellites brighter than $M_{r,o} = -12.3$ within 150\,kpc of their hosts. Thus, the M94 satellite system \citep[see][]{Smercina2018} is also not an outlier in the SAGA context.

In \autoref{fig:num_sat_host_mk}, we explore properties that correlate with the amplitude of the satellite luminosity function by showing the number of satellite galaxies down to our absolute magnitude limit of $M_{r,o} = -12.3$ as a function of their host's $K$-band absolute magnitude (left panel) and their brightest satellite's $r$-band absolute magnitude (right panel). We include estimates of host-by-host incompleteness as vertical error bars, based on the model described in \autoref{sec:incompleteness-correction}. Total satellite counts for the MW and M31 are also included. We observe a significant correlation between host $M_K$ and total satellite count, with a Spearman rank correlation coefficient of $\rho_s = -0.4$ ($p\text{-value}=0.02$); this correlation is a factor of $\sim$10 more significant than the trend reported in \paperone, largely due to the increased sample size. 
SAGA systems that host bright/massive satellites also tend to have higher total numbers of satellites, and this correlation is even stronger than that between host magnitude and satellite count.
The Spearman rank correlation coefficient between the $M_{r,o}$ value of the brightest satellite and the total satellite count ($M_{r,o} < -12.3$) is $\rho_s = -0.7$ ($p\text{-value}=4\times10^{-5}$).

The right panel of \autoref{fig:num_sat_host_mk} also reveals that several SAGA hosts have bright/massive satellites that are of similar luminosities or stellar masses as the Large and Small Magellanic Clouds (LMC/SMC). 
About 30\% of the \nhosts{} SAGA systems contain at least one satellite similar to the LMC (more massive than $M_* = 10^9\,\msun$ or $M_{r,o} < -18.3$), and about 25\% of SAGA systems have at least two satellites more massive than $M_* = 3 \times 10^8\,\msun$ (or $M_{r,o} < -17$; similar to the SMC). Considering that the search radius of SAGA is 300\,kpc in projection, our result is consistent with the findings of \citet{Liu2011} and \citet{Tollerud2011}. 

\subsubsection{Comparing Luminosity Functions to Previous Studies}

In the right panel of \autoref{fig:sat_lf}, we inspect the average cumulative satellite luminosity functions in scaled magnitude, $\Delta m \equiv m^\mathrm{(sat)} - m^\mathrm{(host)}$, 
and compare with several cumulative satellite luminosity functions from the literature.
We find that the shape of the SAGA satellite luminosity function is in agreement with previous data, hinting at a universal satellite luminosity function for MW-mass host galaxies across different redshifts. The amplitude is sensitive to the selection of host galaxies.
We caution that this comparison should be considered exploratory rather than conclusive, especially given the different host selections and satellite definition, as discussed below.

The scaled magnitude is calculated in the $r$ band, and we use NSA photometry for 30 SAGA hosts that are in the NSA catalog. For the remaining six hosts we use $M_K + 2.5$ as an approximation. 
The satellite luminosity functions from the literature include COSMOS (\citealt{Nierenberg2012}; hosts selected from $0.1 <z<0.4$ and $10.5 < \log[M_*/\msun] < 11$), SDSS (\citealt{Sales2013}; hosts selected from $z<0.055$ and $10 < \log(M_*/\msun)< 11$), and MW-like hosts in the Local Volume \citep{Carlsten200602443}.
The SAGA hosts are selected from $0.005 < z<0.01$ and $10 < \log(M_*/\msun) < 11$. 
The SAGA result is the first to probe the satellite luminosity functions outside of the Local Volume to this faint-end limit, and we may be seeing a slight hint of flattening at the faint end of the luminosity function. Note that the expected $\sim$0.7 unconfirmed satellites per SAGA host are already included in the figure as part of the green error bars. 

The average satellite luminosity function of MW-like hosts in the Local Volume of \citet[blue dashed line]{Carlsten200602443} is shown after being multiplied by a factor of 2, because the Local Volume luminosity function only includes satellites within 150\,kpc of their host galaxies in projection, while others use some definition of virial radius (close to 300\,kpc for MW-like hosts). Among SAGA satellites, about half are within 150\,kpc of their hosts and the other half are between 150 and 300\,kpc (see \autoref{sec:radial_dist}).
The amplitude difference between the Local Volume systems and the SAGA systems is most likely due to different host stellar masses. The median stellar mass of the SAGA hosts is about $10^{10.4}\,\msun$, while that of the MW-like hosts presented in \citet{Carlsten200602443} is about $10^{10.7}\,\msun$.
Note that the lower and upper boundaries of the purple band show the two host samples in \citet{Sales2013}:  $\log(M_*/\msun) \in [10, 10.5]$ and $[10.5, 11]$ respectively. This 0.5-dex difference in stellar mass results in about a factor of 3 difference in the amplitude of satellite luminosity functions.

Recently, \citet{2008.05479} studied satellite luminosity functions in the COSMOS field using a statistical approach similar to that of \citet{Nierenberg2012} and found a much higher amplitude (by about a factor of 5) than the luminosity functions of SAGA, \citet{Nierenberg2012}, and \citet{Sales2013}. Hence, a closer examination of how the amplitude of satellite luminosity function depends on host selection is warranted.

\begin{figure}[t]
    \centering
    \includegraphics[width=\columnwidth,clip,trim=0.2cm 0.3cm 0 0]{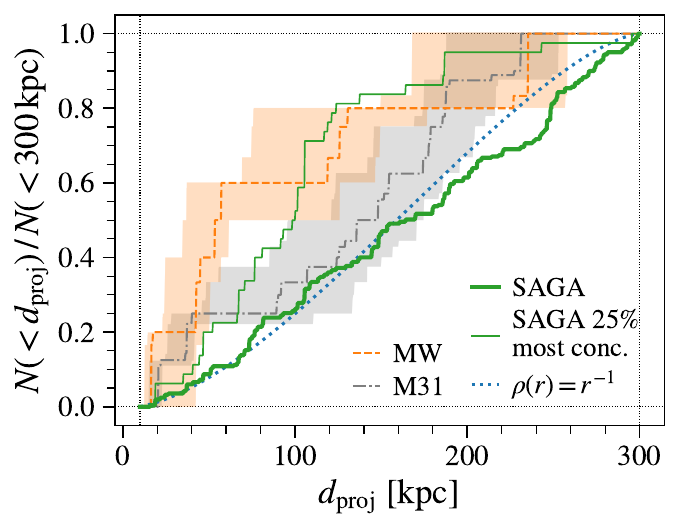}
    \caption{Average cumulative radial distribution of confirmed satellites around \nhosts{} SAGA hosts (thick green line) and Local Group galaxies around MW and M31 (dashed orange and dashed--dotted gray lines). 
    For the latter, we calculated the projected distance using a random set of sight lines, and the 1$\sigma$ range is shown as shaded areas around the median lines. Also shown are the average radial distribution of the eight SAGA hosts that have the most concentrated distribution (thin green line) and a spherical distribution of $r^{-1}$  projected onto 2D (dotted blue line).}
    \label{fig:sat_radial}
\end{figure}

\subsection{Satellite Radial Distributions}
\label{sec:radial_dist}

The average cumulative projected radial distribution of confirmed SAGA satellites is shown in \autoref{fig:sat_radial}. 
Note that we have not attempted to identify satellites within 10\,kpc of the hosts, as the region is dominated by host light; hence, the cumulative function starts at 10\,kpc. 
Since all SAGA satellites are within 300\,kpc by definition, the normalized cumulative radial distribution reaches 1 at 300\,kpc. 
The observed distribution is roughly linear in projected distance, similar to the findings of \citet{Tollerud2011}, and is consistent with a hypothetical case where satellites are distributed according to $1/r_\mathrm{3D}$ within a sphere (shown as a dotted blue line). 
This $1/r_\mathrm{3D}$ profile is less concentrated than a typical NFW profile.

For comparison, we show the radial distributions of galaxies around the MW and M31. We use the galaxies tabulated in \citet{mcconnachie12}, assuming $M_V = M_{r,o} + 0.4$ \citep{jester05} and excluding galaxies that are fainter than the SAGA magnitude limit ($M_{r,o} = -12.3$).
We then produce projected radial distributions by ``mock observing'' these Local Group galaxies with 5000 sight lines that are sampled uniformly at random. We reject any sight line that results in the MW and the M31 being within 300\,kpc in projection of each other and label any Local Group galaxies that are within 300\,kpc in projection of the MW/M31 and within $\pm$500\,kpc in the light-of-sight direction as ``satellites.'' Both these operations are done to mimic the SAGA observations. 
The median distributions for the MW and M31 are shown as dashed orange and dashed--dotted gray lines, respectively, in \autoref{fig:sat_radial}, and the corresponding bands show the $1\sigma$ spread due to random projection. 

We see that the radial distribution of the MW satellites is much more concentrated than the average distribution of SAGA satellites, mostly due to the presence of the LMC and SMC. The radial distribution of the M31 satellites is also slightly more concentrated than that of the SAGA satellites, but they are marginally consistent. 
We note that this difference is most likely due to host-to-host scatter. In \autoref{fig:sat_radial} we also show the radial distribution averaged over only the eight (25\%) of the SAGA hosts that have the most concentrated radial distributions; the resulting radial distribution is almost as concentrated as the MW satellite radial distribution.  

To compare our results with the Local Volume satellite systems presented in \citet{Carlsten200602444}, we calculate $d_\mathrm{proj,half}$, the median projected distance (i.e., radius encompassing half of the satellites), for each SAGA system. 
Among the \nhosts{} SAGA systems, the median and the 16$^\text{th}$/84$^\text{th}$ percentiles of $d_\mathrm{proj,half}$ are $169^{+69}_{-52}$\,kpc.
If we limit our sample to only satellites within 150\,kpc of their hosts to be consistent with \citealt{Carlsten200602444}, the median and the 16$^\text{th}$/84$^\text{th}$ percentiles of $d_\mathrm{proj,half}^{\,(<150\,\mathrm{kpc})}$ become $83^{+48}_{-36}$\,kpc.\footnote{Note that in each SAGA system, there are only $\sim$two satellites within 150\,kpc, on average; hence, the measurement of $d_\mathrm{proj,half}^{\,(<150\,\mathrm{kpc})}$ may be noisy.}
This result can then be directly compared with Figure~4 of \citet{Carlsten200602444}; we find that the distribution of $d_\mathrm{proj,half}$ among SAGA systems is about 1$\sigma$ larger than that of the Local Volume satellite systems (including the MW and M31 satellite systems), but is in full agreement with simulation predictions (see also \autoref{sec:theory-comp}).
Our finding is also consistent with the results of \citet{Samuel2019a}, which showed that the radial distribution of SAGA \paperone{} satellites is in agreement with the Feedback in Realistic Environments (FIRE) baryonic simulations and that the host-to-host variation dominates the scatter. 

Finally, we find no strong dependence between the radial distribution and satellite magnitude. This can been seen in the lower right panel of \autoref{fig:sat_ew_radial} (Spearman rank correlation coefficient $\rho_s\sim0.08$; $p\text{-value} \sim 0.3$). 

\begin{figure}[t]
    \centering
    \includegraphics[width=\columnwidth,clip,trim=0 0.2cm 0 0]{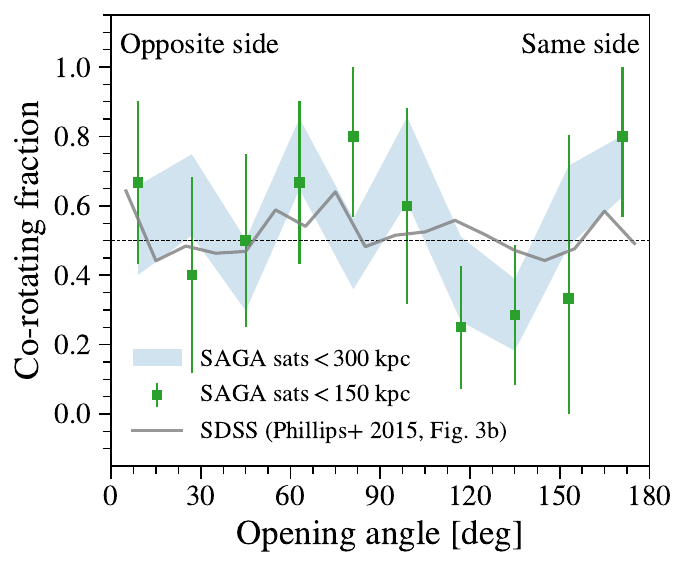}
    \caption{Fraction of corotating pairs in bins of opening angles (a pair of satellites on diametrically opposite sides of a host has an opening angle of 0$^\circ$). The green squares and error bars show the corotating fraction from 36 pairs of satellites within 150 kpc to their hosts.
    The blue band shows the same but including satellites up to 300 kpc from their hosts (189 pairs). 
    The gray line reproduces the SDSS data from Figure 3(b) of \citet{Phillips2015}.
    See \autoref{sec:sats-plane} for details about how the corotating fraction is calculated.} 
    \label{fig:sat_coroating_pairs}
\end{figure}

\subsection{Planes of Satellites}
\label{sec:sats-plane}

Over the past decade, the  ``plane-of-satellites'' question---i.e., whether satellite systems around MW-like galaxies preferentially form a corotating planar structure---has raised much discussion \citep[see review by][]{Pawlowski2018}. Detailed studies have been done on three satellite systems: the MW, M31, and Centaurus\,A \citep[e.g.,][]{1204.5176,1307.6210,1503.05599}. Additionally,
statistical analysis using SDSS data has been done by several groups.  In this work, we present an analysis similar to that in \citet{1407.8178}, \citet{2015MNRAS.449.2576C}, and \citet{Phillips2015}. A more detailed analysis is expected in Stage III of the SAGA Survey.

For each SAGA system, we identify all of the satellite pairs (a pair is any two satellites of the same host) and calculate the ``opening angle'' with respect to their host galaxy. An opening angle of 0$^\circ$ indicates that the two satellites in a pair are on diametrically opposite sides, while an opening angle of 180$^\circ$ indicates that they are on the same side of the host and have exactly the same position angle. For each satellite pair, we check whether their spectroscopic velocities have a corotating signature, that is, whether the two satellites are moving toward the opposite (same) direction with respect to the host when the opening angle is less (greater) than 90$^\circ$. We plot the fraction of pairs that have the corotating signature as a function of opening angle in \autoref{fig:sat_coroating_pairs}. We only consider satellites that have a velocity difference with respect to their respective hosts greater 25\,\kms.

Among the SAGA satellite systems presented here, we have identified 36 satellite pairs for satellites within 150\,kpc of their hosts and with $\Delta V > 25$\,\kms. The number of satellite pairs increases to 189 for satellites between 150 and 300\,kpc. 
The green boxes and blue band in \autoref{fig:sat_coroating_pairs} show the corotating fraction for the inner satellite pairs and the full satellite pairs, respectively, in 10 bins of opening angle. We calculate the fraction using the mean Bayes estimator, $(c+1)/(n+2)$, where $n$ is total number of pairs in that bin, and $c$ is the number of corotating pairs. The error bars are estimated using the Wald method, $\sqrt{p(1-p)/n}$, where $p$ is the estimated fraction. 
The signal is much noisier in this case of inner satellite pairs, but we show it for comparison with the SDSS result, which is computed inside 150\,kpc and presented in Figure 3b of \citet[][shown as a gray line]{Phillips2015}. 

With this analysis, we do not observe a clear excess of high corotating fractions, either for the full satellite sample or using only satellites within 150\,kpc. For pairs that have small opening angles (on nearly diametrically opposed sides), the corotating fraction is close to 0.5. However, we note that the SAGA signal is noisier than the SDSS signal due to the limited number of pairs. 
A possible corotating excess seems to appear in the last bin near 180$^\circ$ opening angle (on the same side of the host). Those ``corotating'' pairs are, in fact, composed of satellites that are close to each other (less than 100\,kpc in projection) and have similar velocities (see \autoref{fig:sat-pos-vel} for an illustration). Hence, it is possible that this excess near the 180$^\circ$ opening angle is due to satellite groups rather than satellite planes. Further analysis is needed to fully distinguish the origin of this excess. 

\begin{figure}[tb!]
    \centering
    \includegraphics[width=1.0\columnwidth,clip,trim=0.55cm 0.6cm 0.4cm 0.3cm]{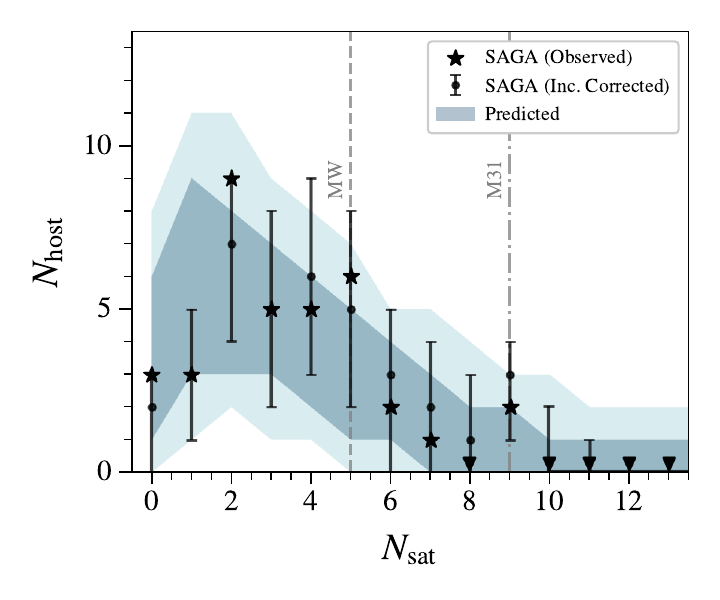}
    \caption{Distribution of satellite number in SAGA hosts. The black stars show the data from our complete sample of~$36$ hosts, and the error bars indicate incompleteness corrections as described in \autoref{sec:incompleteness-correction} (note that the incompleteness-corrected $N_{\mathrm{sat}}$ measurements are highly correlated). Dark blue (light blue) contours indicate the predicted $68\%$ ($95\%$) confidence intervals based on our simulation and galaxy--halo connection model. The MW (M31) is shown as a dashed (dashed--dotted) gray line.
    }
    \label{fig:Nhost_Nsat_pred}
\end{figure}

\begin{figure*}[t]
\centering
\includegraphics[width=\textwidth,clip,trim=0.5cm 0.5cm 0.5cm 0.5cm]{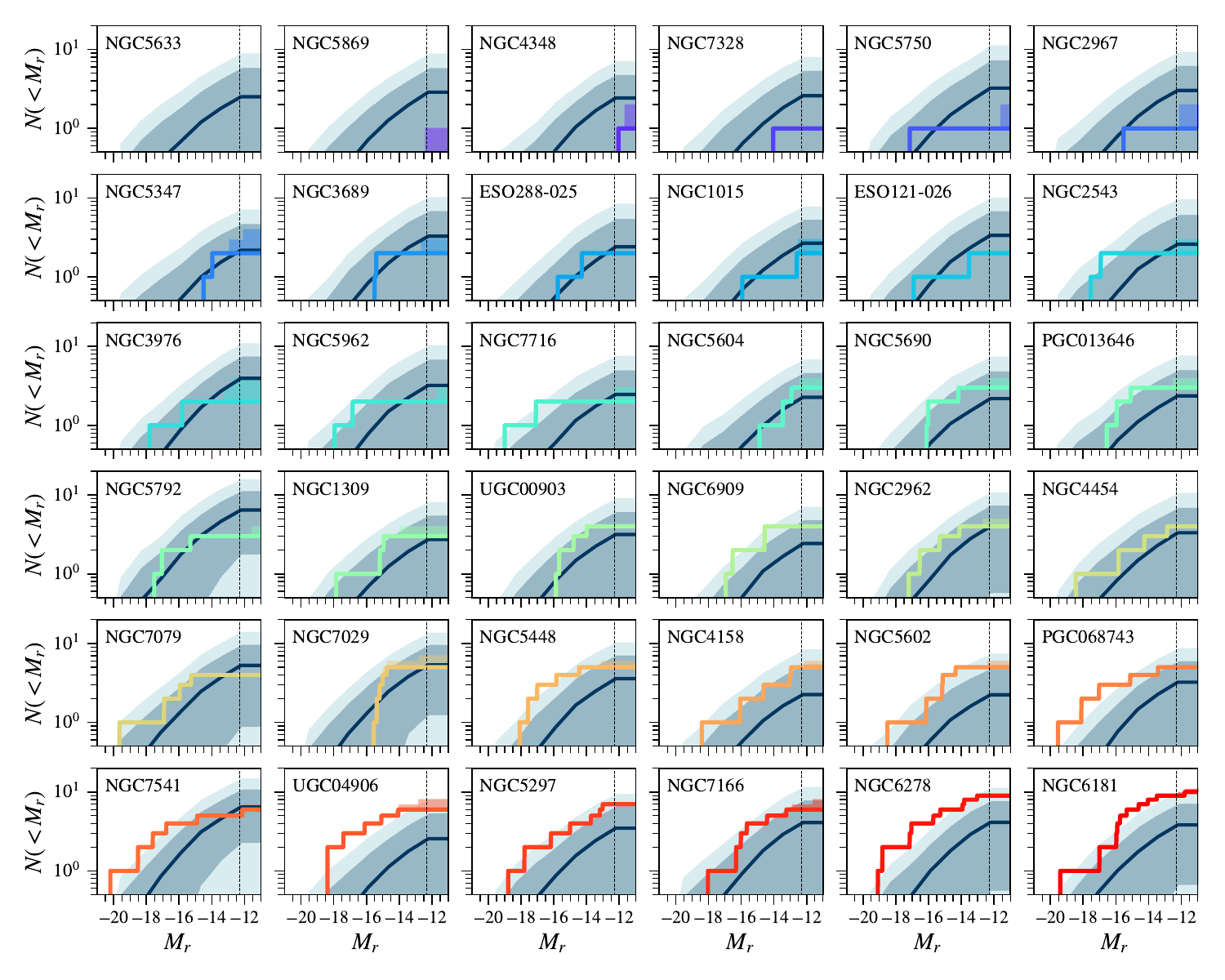}
\caption{SAGA satellite luminosity functions and incompleteness corrections (colored lines and bands) compared to predictions from a cosmological dark matter--only simulation populated with galaxies using the empirical satellite model in \cite{Nadler180905542,Nadler191203303}, which has been fit to the MW satellite population. Dark blue lines indicate the mean prediction for each satellite population, and dark blue (light blue) contours indicate $68\%$ ($95\%$) confidence intervals, which include the effects of host galaxy--halo abundance-matching scatter, uncertainty in our galaxy--halo connection model (see \autoref{sec:theory-model}), and projection effects.}
\label{fig:saga_lf_prediction}
\end{figure*}

\begin{figure*}[t]
\centering
\includegraphics[width=\textwidth,clip,trim=0.5cm 0.5cm 0.5cm 0.5cm]{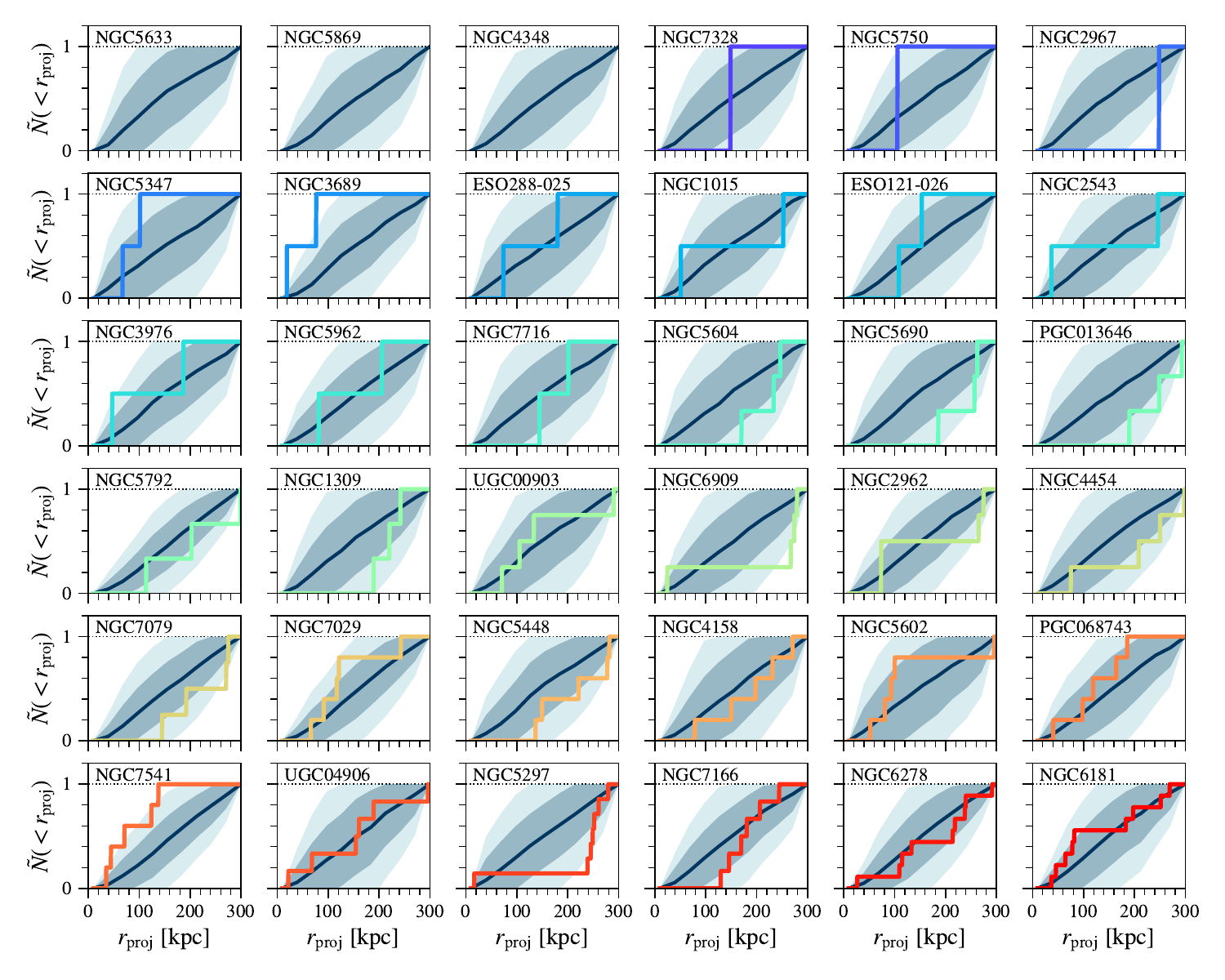}
\caption{Comparison of predicted and observed radial satellite distributions, normalized to the number of satellites within a projected distance of $300\ \mathrm{kpc}$. The predicted mean and confidence interval is identical to \autoref{fig:saga_lf_prediction}, and the observed radial distributions are computed using all satellites above our $M_{r,o}<-12.3$ absolute magnitude limit. 
Hence, no observed data are shown for systems that do not have any satellites with $M_{r,o}<-12.3$. We do not correct the SAGA radial distributions because our incompleteness model predicts that missing satellites should have the same distribution as confirmed ones.}
\label{fig:saga_radial}
\end{figure*}

\section{Implications for the Galaxy--Halo Connection}
\label{sec:theory}

\subsection{The Model}
\label{sec:theory-model}

To place our results in the context of galaxy--halo connection models, we combine the abundance-matching procedure that associates SAGA host galaxies with dark matter halos with the empirical satellite galaxy model from \cite{Nadler180905542,Nadler191203303}. This model populates subhalos with satellite galaxies by extrapolating an abundance-matching relation between luminosity and peak maximum circular velocity (calibrated to the GAMA luminosity function from \citealt{Loveday150501003} for $M_{r,o}<-13$) and a galaxy size--halo size relation (calibrated to \citealt{Kravtsov12122980} for galaxies with half-light radii $r_{1/2}\gtrsim 1\,\mathrm{kpc}$) into the regime of faint satellites. In addition, it includes a model for satellite disruption due to a central disk potential, which is calibrated to the FIRE baryonic simulations. We update the disruption prescription from \cite{Nadler180905542}, which was originally based on the machine-learning algorithm from \cite{Nadler171204467} and the FIRE simulations in \cite{Garrison-Kimmel170103792}, using the analytic fit to subhalo disruption in the FIRE simulations presented in \cite{Samuel2019a}.

As described in \cite{Nadler180905542,Nadler191203303}, this model accurately describes the luminosity, size, and radial distribution of MW satellite galaxies, including the population of satellites accreted with the LMC, when combined with recent observational MW satellite selection functions \citep{Drlica-Wagner191203302}. To incorporate theoretical uncertainties in our prediction for SAGA satellite populations, we sample from the posterior distribution over the galaxy--halo connection and baryonic disruption model parameters presented in \cite{Nadler191203303}, which is derived by fitting the model to most of the known MW satellite population.

To predict the satellite population for each SAGA host, we first follow the procedure in \paperone{} to
map each SAGA host galaxy to a set of potential dark matter host halos in the \code{c125-2048} cosmological simulation \citep[a higher-resolution version of the box used in][]{Mao150302637} using abundance matching. We assume that the scatter in central luminosity at fixed halo properties is 0.15\,dex, and we select all halos mapped to absolute magnitudes within 0.15\,mag of the SAGA host galaxy in question, which roughly corresponds to the quadrature sum of the estimated photometric and distance errors for our hosts. 
We also impose the host $M_K$ cut (\autoref{eq:host-MK-cut}) and the environment cut (\autoref{eq:env-cut-K}) described in \autoref{sec:host-selection}.\footnote{We do not implement the stellar foreground cuts in this mock selection because our simulation does not include stars or a Galactic disk. We do not implement the maximum halo mass cut because the abundance-matching procedure yields very few potential SAGA host halos with present-day virial mass $>10^{13}\,\msun$.}
This procedure yields $\sim$2000 halos per SAGA host galaxy, on average, from which we select a random subset of 300 matched halos for each model realization to capture the probabilistic relation between host galaxy and host halo properties.
In particular, these mock host halos have the cosmologically representative range of masses and formation histories that is expected for each host galaxy luminosity, which allows us to estimate the uncertainty in satellite populations resulting from the scatter in these quantities at a fixed central luminosity. 

For each potential host halo in a given model realization, we then select all other halos above a resolution threshold of $V_\mathrm{peak}=40\,\kms$ that satisfy the $\pm 300$\,kpc and $\pm 275$\,\kms{} projected distance and velocity criteria in various projections of our simulation to mimic the SAGA satellite definition; thus, our model self-consistently includes interloping galaxies. We populate (sub)halos with (satellite) galaxies using the model described above, and we measure the ``observed'' satellite population for each host halo matched to a given SAGA host galaxy for several draws of satellite model parameters from the \cite{Nadler191203303} posterior. We perform three sets of mock observations of galaxies in the $z=0$ snapshot of the simulation by projecting the simulation along perpendicular axes and imposing the $M_{r,o}<-12.3$ SAGA completeness threshold and a surface brightness completeness limit of $\mueff < 26$\,mag\,arcsec$^{-2}$.%
\footnote{The model predicts a very small number of satellites of $\mueff > 26$\,mag\,arcsec$^{-2}$; however, there are no satellites or satellite candidates at this low surface brightness in the SAGA object catalog or in the matched HSC catalog.}
Mock-observed satellites are weighted by their survival probability according to our baryonic subhalo disruption model, and mock-observed interlopers are assumed to have survival probabilities of unity. 

With these procedures, the model predicts that, at 68\% confidence, SAGA hosts inhabit dark matter halos with virial masses in the range $7 \times 10^{11} < M_{\mathrm{vir}}/\msun < 2 \times 10^{12}$ and that SAGA satellites are hosted by halos with peak virial masses in the range $2 \times 10^{10} < M_{\mathrm{peak}} / \msun< 10^{11}$, and typical present-day virial masses a factor of $\sim$1--6 lower.

\subsection{Comparing Model Predictions with SAGA Results}
\label{sec:theory-comp}

\autoref{fig:Nhost_Nsat_pred} shows the comparison between the observed and predicted distributions of the total number of satellites ($M_{r,o} < -12.3$) among the \nhosts{} complete SAGA systems. The error bars on the observed number are generated using the procedure described in Section \ref{sec:incompleteness-correction}. To generate the predicted distribution, we repeatedly draw sets of mock host halos corresponding to the \nhosts{} SAGA systems; thus, the ``predicted'' contour includes both statistical uncertainties (due to the limited host sample size) and systematic uncertainties (due to scatter in the host galaxy--halo connection and our satellite model). Since this plot shows predicted and incompleteness-corrected realizations of the SAGA sample, adjacent data points are anticorrelated. We find good agreement between the total predicted $N_{\mathrm{sat}}$ distribution and SAGA observations. 

To compare our predictions to the data in more detail, \autoref{fig:saga_lf_prediction} shows predicted luminosity functions, and \autoref{fig:saga_radial} shows predicted radial distributions (normalized to the number of satellites with $M_{r,o} < -12.3$ within a projected distance of 300\,kpc) compared to the observed radial distribution for each complete SAGA system. Again, contours indicate 68\% and 95\% confidence intervals due to scatter in the host galaxy--halo connection, draws from our satellite galaxy model parameters, and projections of our simulation. Our predictions are largely consistent with the observed luminosity functions, and they are in excellent agreement with the observed normalized radial distributions. However, for SAGA hosts with the largest numbers of observed satellites, our model underpredicts both the total number of satellites, particularly the number of \emph{bright} ($M_{r,o} < -15$) satellites, for which SAGA observations are highly complete (e.g., see the right panel of \autoref{fig:host_completeness_def}).

To quantify this bright-end tension, we estimate that $0.13$ ($0.09$) predicted satellites with $M_{r,o} < -15$ must be added to our fiducial prediction per host to bring it into agreement with the data at 68\% (95\%) confidence assuming Poisson errors on the observed counts, which corresponds to an $\sim$1$\sigma$--2$\sigma$ discrepancy. This is consistent with \autoref{fig:saga_lf_prediction}, which shows that $\sim$five SAGA hosts have $\sim$one additional bright satellite relative to our predicted 95\% confidence interval.
There is also a hint that the model overpredicts the number of dim satellites, although it is formally consistent with the data given our current incompleteness estimates.

Overall, it is encouraging that the model predictions are in broad agreement with SAGA data, reinforcing our finding that the MW satellite population is not highly atypical.

\subsection{Implications}

Before we discuss the implications of the bright-end tension noted above, we emphasize several caveats associated with our current predictions. The satellite model we employed has specifically been fit to the MW satellite population; thus, our predictions for subhalo and satellite galaxy disruption do not account for the varying masses, morphological properties, and host halo density profiles of SAGA host galaxies relative to the MW. In addition, we have not included ``orphan'' satellites in the model\footnote{In $N$-body simulations, a dark matter subhalo may be disrupted by tidal stripping earlier than in reality due to the lack of concentrated baryonic content or spurious numerical effects. Modeling this effect is commonly referred to as including ``orphan'' satellites.}, nor have we explored exactly how our predictions depend on galaxy--halo connection parameters within the region of parameter space allowed by the MW satellite population.

Nonetheless, the potential bright-end discrepancy is reminiscent of similar tensions noted for Local Volume field \citep{Neuzil191204307} and satellite \citep{Carlsten200602443} galaxies when compared with models that differ from ours in detail, hinting at a more systematic issue that may exist in various theoretical predictions, including those from hydrodynamical simulations. The flexibility of our model allows us to quantify the possible sources of the tension and study potential solutions. We therefore briefly describe possible solutions in the context of our model, leaving a more thorough investigation to future work.

\begin{enumerate}
    \item \emph{Stellar mass--halo mass relation}. Forcing the halos that host the richest SAGA satellite systems to be a factor of $\sim 3$ more massive resolves the bright-end discrepancy. However, this will significantly alter the well-validated stellar mass--halo mass relation in the MW-mass regime \citep{WechslerTinker}.\footnote{We note that varying the 0.15\,dex scatter by $\pm 0.1$\,dex in our host abundance-matching relation does not significantly affect the bright-end tension.} Alternatively, this halo mass shift can be achieved if SAGA host magnitudes are systematically biased dim by $\sim\,0.5$\,mag, which we also regard as unlikely.
    \item \emph{Disruption model}. Removing subhalo disruption from our model reduces the bright-end tension to less than~$\sim\,1\sigma$. It is unlikely that subhalos that host bright satellites undergo no disruption; however, disruption prescriptions calibrated to hydrodynamical zoom-in simulations are dominated by low-mass subhalos and therefore might overpredict disruption efficiencies for more massive subhalos.
    \item \emph{Global luminosity function}. Our abundance-matching prediction is calibrated to the GAMA luminosity function down to $M_{r,o} = -13$; this luminosity function contains both statistical errors (captured by uncertainties in the Schechter function fit to GAMA data) and potential systematic errors (e.g., due to survey incompleteness). While there is no evidence that the GAMA survey is incomplete down to $r = -19.8$, we find that varying the GAMA luminosity function amplitude from \cite{Loveday150501003} within its quoted $2\sigma$ error can fully resolve the bright-end tension.
\end{enumerate}

In future work, we plan to address these questions in detail and to explicitly test for consistency between the MW and SAGA satellite populations by refitting the satellite model to the SAGA data. We also plan to compare SAGA results with additional models and hydrodynamical simulations.

\bigskip

\section{Summary and Outlook}
\label{sec:summary}

In this work, we present the Stage II results from the ongoing SAGA Survey, including \nhosts{} spectroscopically complete satellite systems around MW analogs. This release marks the completion of just over one-third of our planned total (\autoref{tab:roadmap}). 
The full redshift data set around these hosts includes \nzsaga{} SAGA-obtained redshifts, peaking around $z=0.2$, and  \nzother{} redshifts from the literature and preexisting surveys. For galaxies in $17.5 < r_o < 20.75$, 77\% of those redshifts are first obtained by SAGA.
Full SAGA redshift data will be made publicly available with our Stage III publication.

Several improvements to the observing strategy and analysis have been made compared to SAGA Stage I (\paperone{}). 
We discussed these survey improvements in \autoref{sec:survey}--\ref{sec:systematic} and summarize selected highlights here.
\begin{enumerate}
    \item We have updated our list of MW analog hosts, and the Grand List from which these hosts are selected, incorporating updates in HyperLEDA and EDD. We now identify \nhostsinclnoimage{} hosts that meet our criteria, \nhoststotal{} of which have full photometric coverage at present (\autoref{fig:hosts_coverage}). 
    \item Our photometric sources now include SDSS DR14, DES DR1, and LS DR6/DR7.  We have developed a FoF merging procedure to combine overlapping photometric coverage (\autoref{fig:fof_illustration}). Checking against the HSC-SSP, we have verified that we do not miss low surface brightness objects in our photometric catalogs.
    \item Using redshifts collected from \paperone{}, we have improved our target selection strategy to efficiently select very low redshift galaxy candidates  (Figures \ref{fig:z_distribution} and \ref{fig:target_sel}). 
    \item We have developed a robust method to evaluate our survey completeness (\autoref{fig:host_completeness_def}), based on both spectroscopic coverage and an accurate model that characterizes of the rate of satellite discovery as a function of photometric properties (Figures \ref{fig:model_demo} and \ref{fig:sat_prob_model}). This satellite rate model is used to correct for survey incompleteness. 
\end{enumerate}

Our primary science results are presented in \autoref{sec:sat-properties}--\ref{sec:theory}, focused on the properties of satellites and satellite systems. We summarize them as follows.
\begin{enumerate}
    \item We identify \nsats{} satellites around \nhosts{} SAGA hosts (\autoref{fig:all-sats}; Tables \ref{tab:complete-hosts} and \ref{tab:satellites}). This is the first time that dozens of complete satellite luminosity functions (down to $M_{r,o} = -12.3$) of MW-like hosts have been measured.
    \item SAGA satellites follow a surface brightness vs.\ magnitude relation that is consistent with that of the MW (\autoref{fig:sat_properties}). We find a small number of UDGs in our satellite sample.
    \item Consistent with \paperone{}, the satellite quenched fraction among SAGA systems is lower than that in the Local Group (\autoref{fig:quenched_frac}), even when incompleteness is fully accounted for (see \autoref{sec:quenching}). The quenched fraction increases with decreasing stellar mass; 2/50 bright satellites are quenched ($M_{r,o} < -16$, similar to the LMC/SMC) while 16/55  faint satellites ($-16 < M_{r,o} < -12.3$) are quenched.   We also see a slight trend of increasing quenched fraction as the projected radius decreases (\autoref{fig:sat_ew_radial}).
    \item The satellite luminosity function of the MW is consistent with being drawn from the same distribution as the SAGA systems (\autoref{fig:sat_lf}).
    The average SAGA luminosity function is in agreement with past studies but probing into fainter regimes. The amplitude of the satellite luminosity function is sensitive to host mass (\autoref{fig:sat_lf}).
    \item The total number of satellites ($M_{r,o} < -12.3$) per host ranges from zero to nine. This number modestly correlates with the host galaxy $K$-band luminosity ($\rho_s = -0.4$, $p\text{-value}=0.02$; \autoref{fig:num_sat_host_mk}, left), and more strongly correlates with the brightest satellite magnitude $M_{r,o}$ ($\rho_s = -0.7$, $p\text{-value}=4\times10^{-5}$; \autoref{fig:num_sat_host_mk}, right).
    \item The satellite radial distribution of SAGA systems is much less concentrated than that of the MW and, to a lesser extent, M31, but the difference can be largely accounted for by host-to-host scatter (\autoref{fig:sat_radial}). Among SAGA satellite systems, we find no evidence for corotating planes of satellites (\autoref{fig:sat_coroating_pairs}).
    \item Our measured total satellite number, satellite luminosity functions, and radial distributions are all largely consistent with predictions based on a $\text{\LCDM} + \text{SHAM}$ fit to the MW satellite population and, in particular, inform the galaxy--halo connection in the halo mass regime of 10$^9$--10$^{10}\,\msun$  (Figures \ref{fig:Nhost_Nsat_pred}--\ref{fig:saga_radial}).
\end{enumerate}

As more satellite systems are observed and characterized, both by the SAGA Survey and by others, we will be equipped with a powerful tool to test the \LCDM{} model and galaxy formation theory.
In this work, we have touched on several topics and already found intriguing results, such as the wide range of radial distributions and the low quenched fraction of faint satellites, but much remains to be explored.
As we make concrete progress toward the completion of SAGA Survey Stage III, we plan to follow up on these questions. In addition, we are publishing our current satellite data in machine-readable format (see footnote \ref{fn:saga}), hoping to foster independent studies of the SAGA satellite systems.

Perhaps the most remarkable aspect of the SAGA results is how this exhibit of \nhosts{} satellite systems around MW analogs solidifies the idea that our very own satellite system of the MW is just one ``realization'' from a diverse distribution. 
While we can now say that the MW seems to sit comfortably among these SAGA systems in terms of number of satellites, it is interesting to imagine if the MW were like ``Odyssey'' (NGC\,6181; nine satellites) or ``DonQuixote'' (NGC\,5633; no satellites); the development of galaxy formation theory might have proceeded quite differently than it has in the past few decades.
The fascination of seeing how our knowledge evolves as we build a statistical sample is at the heart of the SAGA Survey. 

\medskip

{This work was supported by NSF collaborative grants AST-1517148 and AST-1517422 awarded to M.G. and R.H.W. and by Heising-Simons Foundation grant 2019-1402. 
The authors thank Peter Behroozi, Alyson Brooks, Scott Carlsten, Elise Darragh-Ford, Alex Drlica-Wagner, Jenny Greene, William Hernandez, Andrey Kravtsov, Philip Mansfield, Marta Nowotka, Ekta Patel, Annika Peter, Richie (Yunchong) Wang, and Andrew Wetzel for helpful discussions and feedback that have improved this manuscript; 
Rebecca Bernstein, Yu Lu, Phil Marshall, and Emily Sandford for contributions to the early stages of the survey; 
Boris Leistedt and Li-Cheng Tsai for inputs on statistical methods used in the survey; 
Dustin Lang for developing and maintaining the Legacy Surveys Viewer; 
Chris Lidman for guidance and support of AAT observing; 
and the Center for Computational Astrophysics at the Flatiron Institute for hosting several SAGA team meetings.
Our gratitude also goes to all essential workers that support our lives and work, especially during the COVID-19 pandemic.

Observations reported here were obtained in part at the MMT Observatory, a joint facility of the University of Arizona and the Smithsonian Institution. 
Data were also acquired at the Anglo-Australian Telescope (AAT) under programs A/3000 and NOAO\,0144/0267. We acknowledge the traditional owners of the land on which the AAT stands, the Gamilaraay people, and pay our respects to elders past and present. 

This research made use of computational resources at SLAC National Accelerator Laboratory, a U.S.\ Department of Energy Office, and at the Sherlock
cluster at the Stanford Research Computing Center (SRCC); Y.-Y.M., R.H.W., and E.O.N.\ are thankful for the support of the SLAC and SRCC computational teams. 
This research used the resources of the National Energy Research Scientific Computing Center (NERSC), a U.S. Department of Energy Office of Science User Facility operated under contract No.\ DE-AC02-05CH11231. 

Support for Y.-Y.M.\ was provided by the Pittsburgh Particle Physics, Astrophysics and Cosmology Center through the Samuel P.\ Langley PITT PACC Postdoctoral Fellowship and by NASA through the NASA Hubble Fellowship grant no.\ HST-HF2-51441.001 awarded by the Space Telescope Science Institute, which is operated by the Association of Universities for Research in Astronomy, Incorporated, under NASA contract NAS5-26555. 
This research received support from the National Science Foundation (NSF) under grant no.\ NSF DGE-1656518 through the NSF Graduate Research Fellowship received by E.O.N. 
Part of this work was performed by Y.-Y.M. and R.H.W. at the Aspen Center for Physics, which is supported by National Science Foundation grant PHY-1607611. 
N.K.\ is supported by NSF CAREER award 1455260.

This project used public data from the Sloan Digital Sky Survey (SDSS).
Funding for the Sloan Digital Sky Survey IV has been provided by the Alfred P. Sloan Foundation, the U.S.\ Department of Energy Office of Science, and the Participating Institutions. SDSS-IV acknowledges support and resources from the Center for High-Performance Computing at the University of Utah. The SDSS web site is \http{www.sdss.org}.

SDSS-IV is managed by the Astrophysical Research Consortium for the 
Participating Institutions of the SDSS Collaboration including the 
Brazilian Participation Group, the Carnegie Institution for Science, 
Carnegie Mellon University, the Chilean Participation Group, the French Participation Group, Harvard-Smithsonian Center for Astrophysics, 
Instituto de Astrof\'isica de Canarias, The Johns Hopkins University, Kavli Institute for the Physics and Mathematics of the Universe (IPMU) / 
University of Tokyo, the Korean Participation Group, Lawrence Berkeley National Laboratory, 
Leibniz Institut f\"ur Astrophysik Potsdam (AIP),  
Max-Planck-Institut f\"ur Astronomie (MPIA Heidelberg), 
Max-Planck-Institut f\"ur Astrophysik (MPA Garching), 
Max-Planck-Institut f\"ur Extraterrestrische Physik (MPE), 
National Astronomical Observatories of China, New Mexico State University, 
New York University, University of Notre Dame, 
Observat\'ario Nacional / MCTI, The Ohio State University, 
Pennsylvania State University, Shanghai Astronomical Observatory, 
United Kingdom Participation Group,
Universidad Nacional Aut\'onoma de M\'exico, University of Arizona, 
University of Colorado Boulder, University of Oxford, University of Portsmouth, 
University of Utah, University of Virginia, University of Washington, University of Wisconsin, 
Vanderbilt University, and Yale University.

This project used public data from the Legacy Surveys.
The Legacy Surveys consist of three individual and complementary projects: the Dark Energy Camera Legacy Survey (DECaLS; NOAO Proposal ID \# 2014B-0404; PIs: David Schlegel and Arjun Dey), the Beijing-Arizona Sky Survey (BASS; NOAO Proposal ID \# 2015A-0801; PIs: Zhou Xu and Xiaohui Fan), and the Mayall z-band Legacy Survey (MzLS; NOAO Proposal ID \# 2016A-0453; PI: Arjun Dey). DECaLS, BASS and MzLS together include data obtained, respectively, at the Blanco telescope, Cerro Tololo Inter-American Observatory, National Optical Astronomy Observatory (NOAO); the Bok telescope, Steward Observatory, University of Arizona; and the Mayall telescope, Kitt Peak National Observatory, NOAO. The Legacy Surveys project is honored to be permitted to conduct astronomical research on Iolkam Du'ag (Kitt Peak), a mountain with particular significance to the Tohono O'odham Nation.

NOAO is operated by the Association of Universities for Research in Astronomy (AURA) under a cooperative agreement with the National Science Foundation.

BASS is a key project of the Telescope Access Program (TAP), which has been funded by the National Astronomical Observatories of China, the Chinese Academy of Sciences (the Strategic Priority Research Program "The Emergence of Cosmological Structures" Grant \# XDB09000000), and the Special Fund for Astronomy from the Ministry of Finance. The BASS is also supported by the External Cooperation Program of Chinese Academy of Sciences (Grant \# 114A11KYSB20160057), and Chinese National Natural Science Foundation (Grant \# 11433005).

The Legacy Survey team makes use of data products from the Near-Earth Object Wide-field Infrared Survey Explorer (NEOWISE), which is a project of the Jet Propulsion Laboratory/California Institute of Technology. NEOWISE is funded by the National Aeronautics and Space Administration.

The Legacy Surveys imaging of the DESI footprint is supported by the Director, Office of Science, Office of High Energy Physics of the U.S. Department of Energy under Contract No.\ DE-AC02-05CH1123, by the National Energy Research Scientific Computing Center, a DOE Office of Science User Facility under the same contract; and by the U.S. National Science Foundation, Division of Astronomical Sciences under Contract No.\ AST-0950945 to NOAO.

This project used public archival data from the Dark Energy Survey (DES). 
Funding for the DES Projects has been provided by the U.S. Department of Energy, the U.S. National Science Foundation, the Ministry of Science and Education of Spain, the Science and Technology Facilities Council of the United Kingdom, the Higher Education Funding Council for England, the National Center for Supercomputing Applications at the University of Illinois at Urbana-Champaign, the Kavli Institute of Cosmological Physics at the University of Chicago, the Center for Cosmology and Astro-Particle Physics at the Ohio State University, the Mitchell Institute for Fundamental Physics and Astronomy at Texas A\&M University, Financiadora de Estudos e Projetos, Funda{\c c}{\~a}o Carlos Chagas Filho de Amparo {\`a} Pesquisa do Estado do Rio de Janeiro, Conselho Nacional de Desenvolvimento Cient{\'i}fico e Tecnol{\'o}gico and the Minist{\'e}rio da Ci{\^e}ncia, Tecnologia e Inova{\c c}{\~a}o, the Deutsche Forschungsgemeinschaft, and the Collaborating Institutions in the Dark Energy Survey.

The Collaborating Institutions in the Dark Energy Survey are Argonne National Laboratory, the University of California at Santa Cruz, the University of Cambridge, Centro de Investigaciones Energ{\'e}ticas, Medioambientales y Tecnol{\'o}gicas-Madrid, the University of Chicago, University College London, the DES-Brazil Consortium, the University of Edinburgh, the Eidgen{\"o}ssische Technische Hochschule (ETH) Z{\"u}rich,  Fermi National Accelerator Laboratory, the University of Illinois at Urbana-Champaign, the Institut de Ci{\`e}ncies de l'Espai (IEEC/CSIC), the Institut de F{\'i}sica d'Altes Energies, Lawrence Berkeley National Laboratory, the Ludwig-Maximilians Universit{\"a}t M{\"u}nchen and the associated Excellence Cluster Universe, the University of Michigan, the National Optical Astronomy Observatory, the University of Nottingham, The Ohio State University, the OzDES Membership Consortium, the University of Pennsylvania, the University of Portsmouth, SLAC National Accelerator Laboratory, Stanford University, the University of Sussex, and Texas A\&M University.

The public archival data from the DES is based in part on observations at Cerro Tololo Inter-American Observatory, National Optical Astronomy Observatory, which is operated by the Association of Universities for Research in Astronomy (AURA) under a cooperative agreement with the National Science Foundation.

This project used public data from the GAMA Survey.
GAMA is a joint European-Australasian project based around a spectroscopic campaign using the Anglo-Australian Telescope. The GAMA input catalogue is based on data taken from the Sloan Digital Sky Survey and the UKIRT Infrared Deep Sky Survey. Complementary imaging of the GAMA regions is being obtained by a number of independent survey programmes including GALEX MIS, VST KiDS, VISTA VIKING, WISE, Herschel-ATLAS, GMRT and ASKAP providing UV to radio coverage. GAMA is funded by the STFC (UK), the ARC (Australia), the AAO, and the participating institutions. The GAMA website is \http{www.gama-survey.org}.

This work made use of public data collected at the Subaru Telescope and retrieved from the Hyper Suprime-Cam (HSC) data archive system, which is operated by Subaru Telescope and Astronomy Data Center (ADC) at National Astronomical Observatory of Japan.
The Hyper Suprime-Cam (HSC) collaboration includes the astronomical communities of Japan and Taiwan, and Princeton University. The HSC instrumentation and software were developed by the National Astronomical Observatory of Japan (NAOJ), the Kavli Institute for the Physics and Mathematics of the Universe (Kavli IPMU), the University of Tokyo, the High Energy Accelerator Research Organization (KEK), the Academia Sinica Institute for Astronomy and Astrophysics in Taiwan (ASIAA), and Princeton University. Funding was contributed by the FIRST program from the Japanese Cabinet Office, the Ministry of Education, Culture, Sports, Science and Technology (MEXT), the Japan Society for the Promotion of Science (JSPS), Japan Science and Technology Agency (JST), the Toray Science Foundation, NAOJ, Kavli IPMU, KEK, ASIAA, and Princeton University. 

This work made use of images from the Digitized Sky Surveys (DSS).
The Digitized Sky Surveys were produced at the Space Telescope Science Institute (STScI) under U.S.\ Government grant NAG W-2166. The images of these surveys are based on photographic data obtained using the Oschin Schmidt Telescope on Palomar Mountain and the UK Schmidt Telescope. The plates were processed into the present compressed digital form with the permission of these institutions.
The National Geographic Society - Palomar Observatory Sky Atlas (POSS-I) was made by the California Institute of Technology with grants from the National Geographic Society.
The Second Palomar Observatory Sky Survey (POSS-II) was made by the California Institute of Technology with funds from the National Science Foundation, the National Geographic Society, the Sloan Foundation, the Samuel Oschin Foundation, and the Eastman Kodak Corporation.
The Oschin Schmidt Telescope is operated by the California Institute of Technology and Palomar Observatory.
The UK Schmidt Telescope was operated by the Royal Observatory Edinburgh, with funding from the UK Science and Engineering Research Council (later the UK Particle Physics and Astronomy Research Council), until 1988 June, and thereafter by the Anglo-Australian Observatory. The blue plates of the southern Sky Atlas and its Equatorial Extension (together known as the SERC-J), as well as the Equatorial Red (ER), and the Second Epoch [red] Survey (SES) were all taken with the UK Schmidt.
All DSS data are subject to the copyright given in the copyright summary. Copyright information specific to individual plates is provided in the downloaded FITS headers.
Supplemental funding for sky-survey work at the STScI is provided by the European Southern Observatory.

We acknowledge the usage of the HyperLeda database (\http{leda.univ-lyon1.fr}) and the Extragalactic Distance Database (EDD; \http{edd.ifa.hawaii.edu}). Support for the development of content for the EDD is provided by the National Science Foundation under Grant No.\ AST09-08846.
This research has made use of NASA's Astrophysics Data System. 
}

\facilities{AAT (2dF), MMT (Hectospec), Hale (DBSP), NERSC} 

\software{
    Numpy \citep{numpy, 2020NumPy-Array},
    SciPy \citep{scipy, 2020SciPy-NMeth},
    numexpr \citep{numexpr},
    Matplotlib \citep{matplotlib},
    IPython \citep{ipython},
    Jupyter \citep{jupyter},
    Astropy \citep{astropy},
    statsmodels \citep{statsmodels},
    astroplan \citep{astroplan2018},
    healpy \citep{2005ApJ...622..759G, Zonca2019},
    easyquery (\https{github.com/yymao/easyquery}),
    adstex (\https{github.com/yymao/adstex}),
    xfitfibs \citep{10.1117/12.316837},
    HSRED \citep{10.1086/589642},
    qplot (\https{github.com/bjweiner/qplot}),
    2dfdr \citep{2015ascl.soft05015A},
    Marz \citep{10.1016/j.ascom.2016.03.001}}

\appendix
\counterwithin{figure}{section}

\section{Distribution of SAGA Redshifts in Photometric Space}
\label{app:redshift-stats}

\begin{figure*}[hbt]
    \centering
    \includegraphics[width=\textwidth,clip,trim=0 0.2cm 0 0]{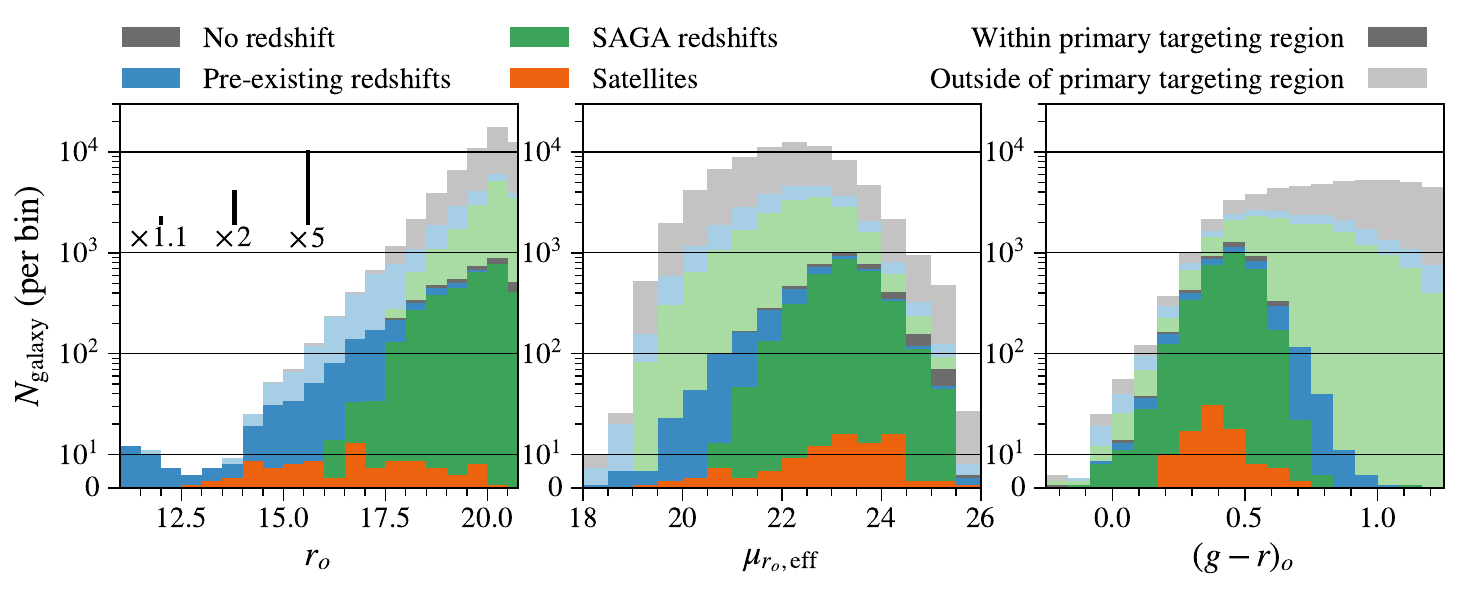}
    \caption{Stacked histograms of galaxies down to $r_o = 20.75$ within 28 complete SAGA systems that are also in the SDSS footprint (``complete systems'' are defined in \autoref{sec:completeness-def}), in bins of magnitude $r_o$ (\textit{left}), effective surface brightness $\mueff$ (\textit{middle}), and color $(g-r)_o$ (\textit{right})..
    Color indicates whether the galaxy has no redshift (gray), has a preexisting redshift (blue), has a SAGA redshift (green), or is a satellite (orange). For each of these categories/colors, brightness indicates whether the galaxy is in our primary targeting region (dark) or not (light); the primary targeting region is defined in \autoref{eq:targeting-cuts}. Note that the $y$-axis scale is linear between zero and 10, and logarithmic above 10. Hence, above 10, the visual height difference between histograms should be interpreted as multiples; the vertical bars in the left panel demonstrate the height differences corresponding to a few common multiples.}
    \label{fig:target_spec_counts}
\end{figure*}

Here we examine in detail how SAGA redshifts and targets distribute in the photometric space.   \autoref{fig:target_spec_counts} provides a close look at the numbers of galaxies that are within the footprints of both SAGA and SDSS and whether they have only SAGA redshifts, preexisting redshifts (e.g., from SDSS), or no redshift information. We plot these numbers with respect to three photometric quantities: apparent magnitude ($r_o$), effective surface brightness ($\mueff$), and color ($(g-r)_o$). 

In the left panel of \autoref{fig:target_spec_counts}, galaxies within the SAGA footprint and brighter than $r_o = 16$ basically all have preexisting redshifts. In the magnitude range of  $ 16 < r_o < 17.5 $, about half of the galaxies in our primary targeting region (dark colors in the figure) do not have preexisting redshifts, and SAGA has obtained redshifts for these galaxies. 
This result implies that even in the magnitude range of  $ 16 < r_o < 17.5 $, very low redshift galaxies ($z<0.03$) do not have complete spectroscopic coverage. 
The incomplete spectroscopic coverage in the very low-redshift universe is consistent with the findings of \cite{2018RNAAS...2..234L} and \cite{2020ApJ...895...32F}.
Below the magnitude of $r_o= 18$, most redshifts were, unsurprisingly, obtained by SAGA; similar information is also shown in the right panel of \autoref{fig:z_distribution}.

In the middle and right panels of \autoref{fig:target_spec_counts}, we see more clearly that SAGA has obtained thousands of redshifts for galaxies outside of the primary targeting region, covering about 10--20\% of the total objects. These redshifts provide us with the quantitative evidence that our primary targeting region does not exclude very low redshift galaxies or potential satellites. 
In these panels, we also see how the satellites (orange histogram) distribute in these photometric quantities. 
We note that in the photometric region where satellites are more populated, the majority of redshifts were obtained by SAGA.

\section{Modeling Satellite Rate in the Photometric Space}
\label{app:completeness-model}

\begin{figure*}[t]
    \centering
    \includegraphics[width=\textwidth,clip,trim=0 0.2cm 0 0.1cm]{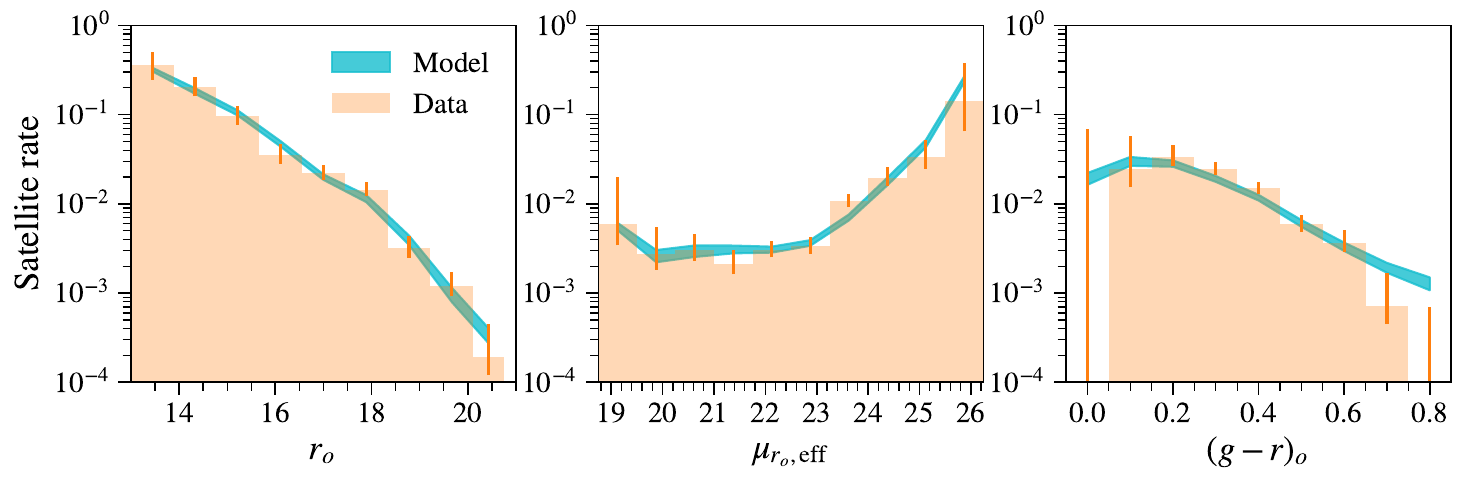}
    \caption{Satellite rates (ratio of number of satellites to number of targets with redshifts; orange histograms) in bins of magnitude $r_o$ (\textit{left}), effective surface brightness $\mueff$ (\textit{middle}), and color $(g-r)_o$ (\textit{right}). The orange error bars are estimated using the Wald method for a binomial distribution. 
    The cyan bands show the satellite rates produced by the model, where the upper and lower edges of the bands show 84\% and 16\% confidence levels.}
    \label{fig:model_demo}
\end{figure*}

\begin{figure*}[t!]
    \centering
    \includegraphics[width=\textwidth,clip,trim=0.3cm 0.6cm 1.2cm 0.55cm]{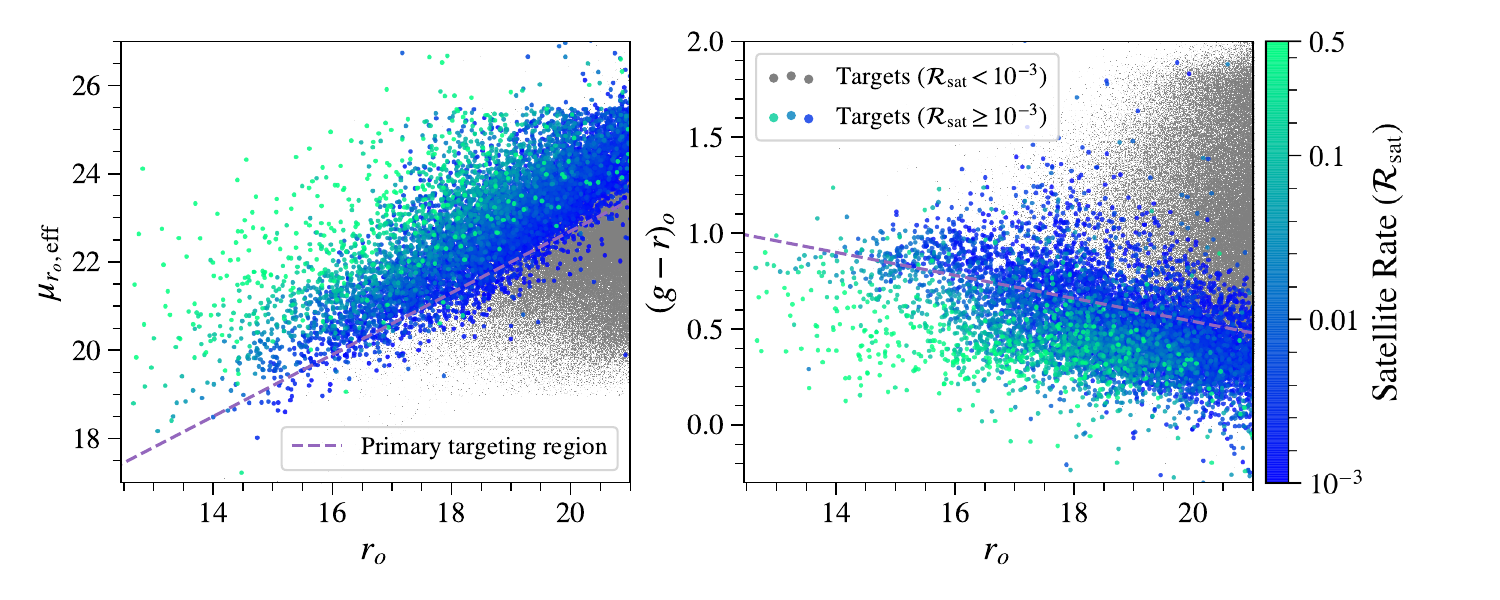}
    \caption{Distribution of galaxies in the surface brightness--magnitude (\emph{left}) and color--magnitude (\emph{right}) planes, color-coded by the satellite rate ($\mathcal{R}_\mathrm{sat}$) produced by our model. The $\mathcal{R}_\mathrm{sat}$ value can be read off from the rightmost color bar. Galaxies whose $\mathcal{R}_\mathrm{sat}<10^{-3}$ are shown as small gray points to avoid crowding. The boundaries of our primary targeting region are shown as purple dashed lines.}
    \label{fig:sat_prob_model}
\end{figure*}


In \autoref{sec:incompleteness-correction}, we calculated the ``satellite rate,'' the fraction of satellites among the target with redshifts, as a function of photometric properties.   We detail our construction of this satellite rate here.  
The construction involves a modified logistic regression method, behind which the core ideas are straightforward. 

Our goal is to build a smooth function in photometric space where the value of the function represents the satellite rate. We first choose to work in the photometric space of $(r_o, \mueff, (g-r)_o)$. We then assume that (1) there is a latent variable $\ell$ that takes the form of a linear combination of the photometric properties,
\begin{align}
\ell = \beta_0 + \beta_1 r_o + \beta_2 \mueff + \beta_3 (g-r)_o,
\end{align}
and (2) the satellite rate (probability) is a monotonic function of the latent variable $\ell$. 
In other words, we aim to identify a specific direction in the photometric space, determined by $(\beta_1, \beta_2, \beta_3)$, for which the satellite rate varies monotonically only along that direction. 
The monotonic function we choose to model the satellite rate is a logistic function, 
\begin{align}
\mathcal{R}_\mathrm{sat}(\ell) = \frac{\mathcal{R}_\mathrm{max}}{1+\exp(-\ell)},
\label{eq:p_sat_formula}
\end{align}
where the satellite rate $\mathcal{R}_\mathrm{sat}(\ell \rightarrow -\infty) = 0$ and $\mathcal{R}_\mathrm{sat}(\ell \rightarrow \infty) = \mathcal{R}_\mathrm{max}$. 
With this setup, the remaining task is to fit for the parameters $(\beta_0, \beta_1, \beta_2, \beta_3, \mathcal{R}_\mathrm{max})$.
We use the maximum-likelihood method, and the log-likelihood  can be calculated exactly as
\begin{align}
\log L = \sum_{i \in \mathrm{sats}} \log \mathcal{R}_\mathrm{sat}(\ell_i) + \sum_{i \notin \mathrm{sats}} \log \left[1-\mathcal{R}_\mathrm{sat}(\ell_i)\right].
\label{eq:model_log_likelihood}
\end{align}

This model construction is almost identical to logistic regression, except that we also fit for $\mathcal{R}_\mathrm{max}$, rather than fixing it to unity. This is because it is impossible to find a region in the photometric space where one would only find satellites; hence, the maximum satellite rate should be capped at some $\mathcal{R}_\mathrm{max} < 1$.
Operationally, we modify the Logit model in the \textsc{statsmodels} \textsc{Python} package\footnote{\https{www.statsmodels.org}} \citep{statsmodels} to carry out the maximum-likelihood fit. 

We fit this model to all galaxy redshifts we have collected among \nhosts{} complete SAGA systems.  
The best-fit parameter values are 
$(\beta_0 = 0.303, \beta_1=-1.96, \beta_2=1.507, \beta_3=-5.498, \text{ and } \mathcal{R}_\mathrm{max}=0.487)$.
To estimate the errors, we bootstrap the satellite population and refit the parameters 1000 times. We choose to not bootstrap the full population because the satellites are too rare in the full population. We also fix the $\mathcal{R}_\mathrm{max}$ value in the error estimation process.
 
We note that the model we built here aims to be descriptive rather than predictive, as there are multiple complications if one desires to build a predictive model for rare events \citep[e.g.,][]{King2001}. 
The main goal here is to construct a model to reproduce the ``satellite rate'' in different regions of the photometric space and smooth out the noise due to limited data points. 

\autoref{fig:model_demo} demonstrates that our model can indeed reproduce the input data set. 
The satellite rates our model produces (cyan bands) agree very well with the data in different slices of the photometric space. 
In the left panel, the satellite rate decreases with magnitude, which is a simple result of the relation between distance, volume, and magnitude. 
In the middle and left panels, we observe that the model can accurately capture the satellite rate as a function of surface brightness and color, even if the function is not monotonic. 

With this model, we can then calculate the satellite rate ($\mathcal{R}_\mathrm{sat}$) at any given point in the three-dimensional photometric space. We demonstrate our calculation in \autoref{fig:sat_prob_model}, where each point represents a target and its green--blue color corresponds to our estimate of its satellite rate $\mathcal{R}_\mathrm{sat}$. 
These values are then used to correct for incompleteness, as discussed in \autoref{sec:incompleteness-correction}.

\section{Satellite Positions and Velocities}
\label{app:sat-pos-vel}

\begin{figure*}[p]
    \centering
    \includegraphics[width=\textwidth]{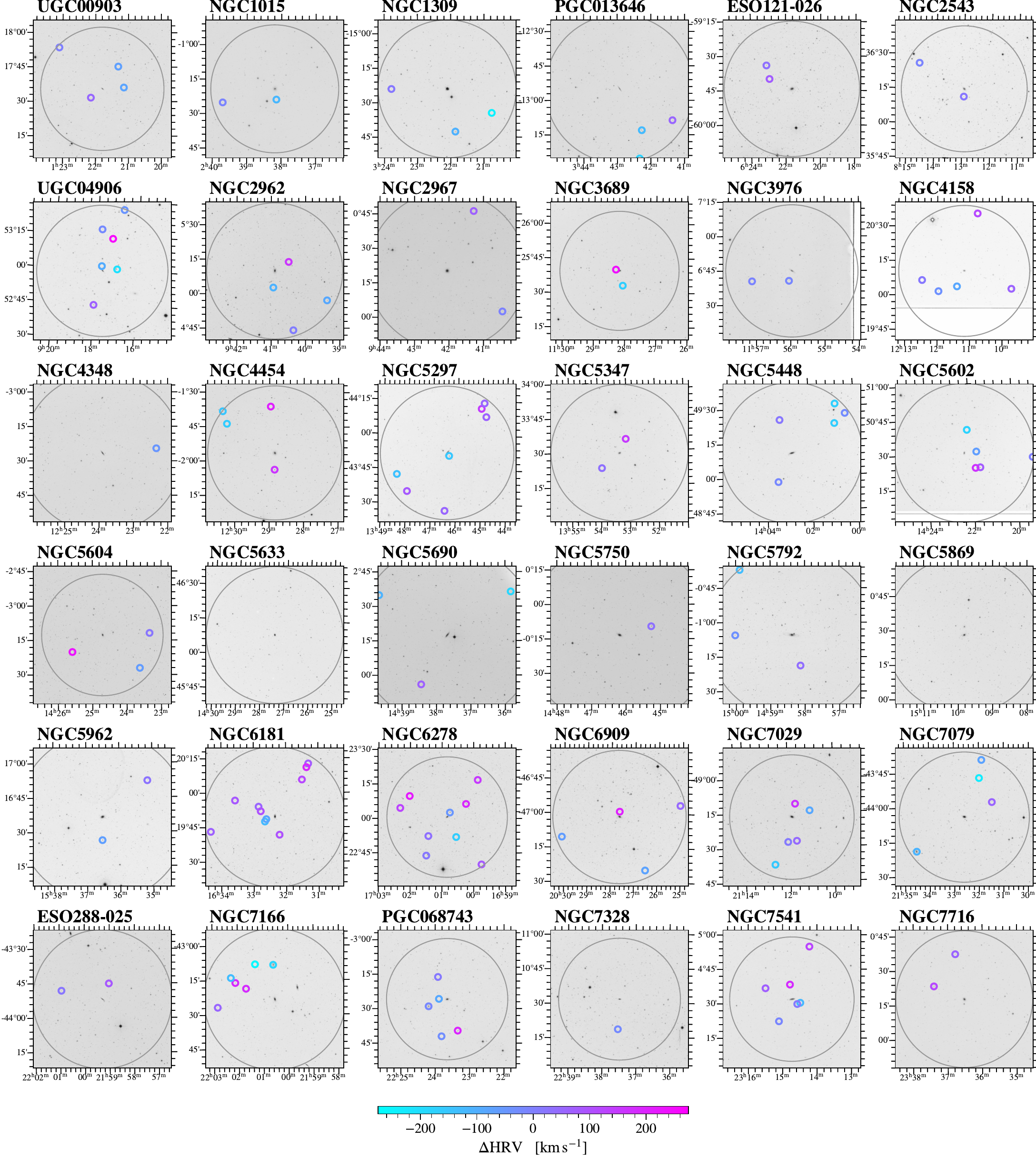}
    \caption{Positions and velocities of the SAGA Satellites. Each panel shows one SAGA system, as listed in \autoref{tab:complete-hosts}, with a background image from the Digitized Sky Survey (DSS-1) and 1$^\circ$ on a side. The $x$- and $y$-axes show R.A.\ and decl., respectively. The large gray circle indicates a radius of 300\,kpc (virial radius) to each host. Satellites are plotted as small circles, color-coded by the difference in heliocentric velocity with respect to their hosts.}
    \label{fig:sat-pos-vel}
\end{figure*}

For completeness, we plot the positions and velocities (with respect to their respective hosts) of all SAGA satellites in \autoref{fig:sat-pos-vel}.
It is noticeable that some close satellite pairs also have similar velocities to each other and contribute to the excess near 180$^\circ$ in \autoref{fig:sat_coroating_pairs}.

\renewcommand*{\arraystretch}{0.98}

\centerwidetable
\begin{deluxetable*}{llrRccCCcrlc}
\tablehead{\colhead{Common} & \colhead{SAGA} & \colhead{R.A.} & \colhead{Decl.} & \colhead{HRV} & \colhead{$D$} & \colhead{$M_{K,o}$} & \colhead{Phot.} & \colhead{Spec.} & \multicolumn{2}{c}{Confirmed} & \colhead{Potential}\\[-6pt]
\colhead{Name} & \colhead{Name} & \colhead{[deg]} & \colhead{[deg]} & \colhead{[\kms]} & \colhead{[Mpc]} & \colhead{[mag]} & \colhead{Coverage} & \colhead{Cover. (\%)} & \multicolumn{2}{c}{$N_\text{sat}$} & \colhead{$N_\text{sat}$}\\[-6pt]
\colhead{(1)} & \colhead{(2)} & \colhead{(3)} & \colhead{(4)} & \colhead{(5)} & \colhead{(6)} & \colhead{(7)} & \colhead{(8)} & \colhead{(9)} & \multicolumn{2}{c}{(10)} & \colhead{(11)}}
\tablecaption{List of host (primary) galaxies in the \nhosts{} complete SAGA systems. \label{tab:complete-hosts}}
\startdata
UGC00903 & HarryPotter & 20.449 & 17.592 & 2516 & 38.4 & -23.54 & S-L & \phn92.4 & 4 &  & 0.46 \\
NGC1015 & Narnia & 39.548 & -1.319 & 2625 & 37.0 & -23.50 & SDL & \phn95.9 & 2 &  & 0.60 \\
NGC1309 & Hiccup & 50.527 & -15.400 & 2136 & 34.3 & -23.57 & -DL & \phn89.7 & 3 &  & 0.97 \\
PGC013646 & Genji & 55.734 & -12.916 & 2163 & 31.9 & -23.14 & -DL & \phn82.9 & 3 &  & 1.15 \\
ESO121-026 & Mulan & 95.412 & -59.740 & 2266 & 35.0 & -23.71 & -D- & \phn85.6 & 2 &  & 0.12 \\
NGC2543 & StarTrek & 123.241 & 36.255 & 2470 & 37.6 & -23.46 & S-L & \phn92.2 & 2 &  & 1.00 \\
UGC04906 & Okonkwo & 139.416 & 52.993 & 2273 & 36.1 & -23.36 & S-L & \phn81.1 & 6 &  & 2.54 \\
NGC2962 & Aeneid & 145.225 & 5.166 & 1958 & 34.8 & -24.05 & S-L & \phn91.9 & 4 &  & 1.33 \\
NGC2967 & Skywalker & 145.514 & 0.336 & 1892 & 29.9 & -23.51 & S-L & \phn88.8 & 1 & (1) & 0.69 \\
NGC3689 & Chihiro & 172.046 & 25.661 & 2737 & 39.8 & -23.80 & S-L & \phn93.8 & 2 &  & 0.98 \\
NGC3976 & Gaukur & 178.989 & 6.750 & 2496 & 35.9 & -24.00 & S-L & \phn82.2 & 2 &  & 2.21 \\
NGC4158 & ScoobyDoo & 182.792 & 20.176 & 2450 & 36.2 & -23.04 & S-L & \phn90.5 & 5 &  & 0.85 \\
NGC4348 & Macondo & 185.975 & -3.443 & 2005 & 29.7 & -23.46 & S-L & \phn98.3 & 0 & (1) & 0.65 \\
NGC4454 & Metamorphoses & 187.211 & -1.939 & 2329 & 35.3 & -23.69 & S-L & \phn95.5 & 4 &  & 0.55 \\
NGC5297 & Rand & 206.599 & 43.872 & 2405 & 35.5 & -23.85 & S-L & \phn83.3 & 7 &  & 0.50 \\
NGC5347 & Trisolaris & 208.324 & 33.491 & 2371 & 34.6 & -23.04 & S-L & \phn92.2 & 2 &  & 2.02 \\
NGC5448 & Essun & 210.708 & 49.173 & 2013 & 33.5 & -23.82 & S-L & \phn82.8 & 5 &  & 1.28 \\
NGC5602 & Pippi & 215.578 & 50.501 & 2221 & 34.0 & -23.05 & S-L & \phn86.9 & 5 &  & 1.13 \\
NGC5604 & Beloved & 216.178 & -3.212 & 2749 & 39.0 & -23.25 & S-L & \phn93.8 & 3 &  & 0.89 \\
NGC5633 & DonQuixote & 216.868 & 46.147 & 2325 & 34.6 & -23.09 & S-L & \phn80.2 & 0 &  & 0.51 \\
NGC5690 & SunWukong & 219.421 & 2.291 & 1756 & 26.3 & -23.14 & S-L & \phn97.4 & 3 &  & 1.15 \\
NGC5750 & Dune & 221.546 & -0.223 & 1659 & 25.3 & -23.57 & S-L & \phn95.7 & 1 &  & 0.80 \\
NGC5792 & Othello & 224.594 & -1.091 & 1924 & 28.3 & -24.55 & S-L & \phn94.7 & 3 &  & 0.64 \\
NGC5869 & Ynglinga & 227.456 & 0.470 & 2074 & 30.1 & -23.66 & S-L & \phn91.8 & 0 &  & 1.40 \\
NGC5962 & Gilgamesh & 234.132 & 16.608 & 1963 & 27.9 & -23.70 & S-L & \phn94.9 & 2 &  & 0.85 \\
NGC6181 & Odyssey & 248.088 & 19.824 & 2370 & 33.6 & -23.97 & S-L & \phn97.4 & 9 & (1) & 0.77 \\
NGC6278 & Arya & 255.210 & 23.011 & 2795 & 39.3 & -23.98 & S-L & 100.0 & 9 &  & 0.07 \\
NGC6909 & Moana & 306.912 & -47.027 & 2778 & 35.6 & -23.52 & -D- & \phn85.3 & 4 &  & 0.39 \\
NGC7029 & Ozymandias & 317.967 & -49.284 & 2783 & 38.1 & -24.35 & -D- & \phn89.5 & 5 &  & 2.04 \\
NGC7079 & Middlemarch & 323.147 & -44.068 & 2653 & 36.4 & -24.22 & -D- & \phn96.3 & 4 &  & 0.06 \\
ESO288-025 & Bilbo & 329.824 & -43.867 & 2493 & 34.1 & -23.12 & -D- & \phn95.0 & 2 &  & 0.33 \\
NGC7166 & Frodo & 330.137 & -43.390 & 2458 & 33.5 & -24.08 & -D- & \phn81.8 & 6 &  & 1.43 \\
PGC068743 & OBrother & 335.913 & -3.432 & 2865 & 39.1 & -23.81 & S-L & \phn98.0 & 5 &  & 0.53 \\
NGC7328 & PiPatel & 339.372 & 10.532 & 2824 & 38.9 & -23.32 & S-L & \phn84.2 & 1 &  & 0.14 \\
NGC7541 & Catch22 & 348.683 & 4.534 & 2680 & 38.1 & -24.57 & S-L & \phn94.9 & 5 & (1) & 0.79 \\
NGC7716 & AnaK & 354.131 & 0.297 & 2558 & 34.6 & -23.38 & SDL & \phn93.3 & 2 &  & 1.02
\enddata
\tablecomments{Columns (1) and (3)-(7): host galaxy properties taken from HyperLEDA, EDD, and 2MRS; distances (6) for some hosts were taken from the NSA catalog (see \autoref{sec:grand-list} for details). 
Column (2): SAGA name given to each galaxy for ease of reference.  
Column (8): existing coverage by photometric surveys, $S$ for SDSS, $D$ for DES, and $L$ for LS. 
Column (9): spectroscopic coverage within the SAGA primary targeting region. 
Column (10): number of confirmed satellites down to $M_{r,o}=-12.3$; number in parentheses are confirmed satellites below that magnitude limit. 
Column (11): expected number of unconfirmed potential satellites down to $M_{r,o}=-12.3$ (see \autoref{sec:incompleteness-correction} for detail). A machine-readable version of this table is available on the SAGA website (see footnote \ref{fn:saga}).}
\end{deluxetable*}

\clearpage

\startlongtable
\centerwidetable
\begin{deluxetable*}{llrRrRrRRrrcl}
\tablehead{%
\colhead{Host} & \colhead{Object} & \colhead{R.A.} & \colhead{Decl.} & \colhead{$d_\text{proj}$} & \colhead{$\Delta$\,HRV} & \colhead{$r_o$} & \colhead{$M_{r,o}$} & \colhead{$(g-r)_o$} & \colhead{$\mueff$} & \colhead{$M_*$} & \colhead{H$\alpha$} & \colhead{Redshift}\\[-8pt]
\colhead{Name} & \colhead{ID} & \colhead{[deg]} & \colhead{[deg]} & \colhead{[kpc]} & \colhead{[\kms]} & \colhead{[mag]} & \colhead{[mag]} & \colhead{[mag]} & \colhead{~} & \colhead{[dex]} & \colhead{~} & \colhead{Source}\\[-6pt]
\colhead{(1)} & \colhead{(2)} & \colhead{(3)} & \colhead{(4)} & \colhead{(5)} & \colhead{(6)} & \colhead{(7)} & \colhead{(8)} & \colhead{(9)} & \colhead{(10)} & \colhead{(11)} & \colhead{(12)} & \colhead{(13)}
}
\tablecaption{List of \nsats{} satellites in \nhostswsats{} complete SAGA systems (\nhostsnosats{} complete systems have no satellites) \label{tab:satellites}}
\startdata
UGC00903 & LS-432563-224 & 20.7772 & 17.8916 & 290 & -1 & 17.06 & -15.9 & 0.26 & 23.57 & 7.89 & Y & ALFALF \\
UGC00903 & LS-429811-3398 & 20.2850 & 17.6022 & 105 & -78 & 17.26 & -15.7 & 0.45 & 24.04 & 8.02 & Y & MMT \\
UGC00903 & LS-431187-1672 & 20.3280 & 17.7539 & 133 & -70 & 18.13 & -14.8 & 0.33 & 24.24 & 7.54 & Y & MMT \\
UGC00903 & LS-429812-2469 & 20.5362 & 17.5279 & 70 & 50 & 18.93 & -14.0 & 0.21 & 23.90 & 7.08 & Y & MMT \\ \hline
NGC1015 & DES-313240666 & 39.9254 & -1.4187 & 252 & -9 & 16.91 & -15.9 & 0.42 & 21.04 & 8.09 & Y & MMT \\
NGC1015 & DES-310691517 & 39.5360 & -1.3965 & 50 & -119 & 20.24 & -12.6 & 0.19 & 23.82 & 6.51 & Y & AAT \\ \hline
NGC1309 & DES-353757883 & 50.4652 & -15.7104 & 189 & -106 & 14.83 & -17.9 & 0.36 & 22.76 & 8.80 & Y & 6dF \\
NGC1309 & DES-350665706 & 50.1913 & -15.5749 & 220 & -244 & 17.50 & -15.2 & 0.37 & 22.03 & 7.74 & Y & AAT \\
NGC1309 & DES-353742769 & 50.9464 & -15.4004 & 242 & 16 & 17.74 & -15.0 & 0.51 & 22.33 & 7.79 & N & AAT \\ \hline
PGC013646 & DES-371747881 & 55.3397 & -13.1446 & 248 & 55 & 15.97 & -16.6 & 0.33 & 20.75 & 8.24 & Y & AAT \\
PGC013646 & DES-373383928 & 55.5682 & -13.2170 & 189 & -127 & 16.57 & -16.0 & 0.31 & 21.14 & 7.97 & Y & AAT \\
PGC013646 & DES-373393030 & 55.5841 & -13.4218 & 292 & -175 & 17.44 & -15.1 & 0.37 & 21.95 & 7.69 & Y & AAT \\ \hline
ESO121-026 & DES-467669052 & 95.7756 & -59.5691 & 153 & 32 & 15.81 & -16.9 & 0.22 & 21.76 & 8.26 & Y & 6dF \\
ESO121-026 & DES-467510764 & 95.7327 & -59.6675 & 108 & 74 & 19.20 & -13.5 & 0.25 & 23.19 & 6.94 & Y & AAT \\ \hline
NGC2543 & NSA-162577 & 123.2432 & 36.1984 & 37 & 18 & 15.34 & -17.6 & 0.30 & 21.42 & 8.61 & Y & SDSS \\
NGC2543 & NSA-162596 & 123.6499 & 36.4344 & 246 & -8 & 15.96 & -16.9 & 0.33 & 24.96 & 8.39 & Y & SDSS \\ \hline
UGC04906 & NSA-78947 & 139.4972 & 52.7426 & 160 & 65 & 14.41 & -18.4 & 0.42 & 22.07 & 9.06 & Y & SDSS \\
UGC04906 & NSA-648311 & 139.1897 & 53.4429 & 296 & -45 & 14.42 & -18.4 & 0.21 & 19.47 & 8.84 & Y & SDSS \\
UGC04906 & NSA-78956 & 139.4444 & 53.2935 & 189 & -22 & 15.39 & -17.4 & 0.29 & 22.33 & 8.54 & Y & SDSS \\
UGC04906 & LS-595052-1698 & 139.2389 & 53.0101 & 68 & -209 & 16.69 & -16.1 & 0.43 & 22.66 & 8.17 & Y & MMT \\
UGC04906 & LS-595052-1940 & 139.4253 & 53.0271 & 21 & -96 & 17.70 & -15.1 & 0.42 & 23.15 & 7.75 & Y & MMT \\
UGC04906 & LS-595921-1395 & 139.3098 & 53.2298 & 154 & 259 & 18.74 & -14.1 & 0.42 & 24.03 & 7.34 & Y & MMT+ \\ \hline
NGC2962 & NSA-192008 & 144.8430 & 4.9524 & 264 & -87 & 15.50 & -17.2 & 0.34 & 20.69 & 8.51 & Y & SDSS+ \\
NGC2962 & NSA-191990 & 145.2345 & 5.0449 & 73 & -108 & 16.17 & -16.5 & 0.25 & 23.64 & 8.15 & Y & ALFALF \\
NGC2962 & NSA-82770 & 145.0880 & 4.7351 & 274 & 5 & 17.38 & -15.3 & 0.20 & 22.66 & 7.61 & Y & SDSS+ \\
NGC2962 & LS-361164-2394 & 145.1227 & 5.2311 & 73 & 160 & 18.60 & -14.1 & 0.48 & 24.47 & 7.42 & Y & PAL \\ \hline
NGC2967 & NSA-195 & 145.3209 & 0.7711 & 248 & 65 & 16.85 & -15.5 & 0.33 & 20.64 & 7.83 & Y & SDSS \\
NGC2967 & LS-330948-4542 & 145.1122 & 0.0426 & 259 & -5 & 21.68 & -10.7 & 0.10 & 24.86 & 5.65 & Y & ALFALF \\ \hline
NGC3689 & LS-474605-827 & 172.0746 & 25.6697 & 18 & 240 & 17.50 & -15.5 & 0.60 & 23.73 & 8.11 & N & MMT \\
NGC3689 & LS-473304-3777 & 172.0184 & 25.5538 & 76 & -168 & 17.60 & -15.4 & 0.47 & 24.48 & 7.93 & N & PAL \\ \hline
NGC3976 & NSA-328500 & 179.0103 & 6.6779 & 46 & -16 & 14.97 & -17.8 & 0.48 & 21.71 & 8.91 & N & SDSS \\
NGC3976 & NSA-328502 & 179.2801 & 6.6755 & 186 & -10 & 16.97 & -15.8 & 0.40 & 23.26 & 8.01 & Y & SDSS+ \\ \hline
NGC4158 & NSA-542307 & 182.9907 & 20.0279 & 150 & -37 & 14.39 & -18.4 & 0.36 & 20.20 & 9.02 & Y & SDSS+ \\
NGC4158 & 1237668298203070473 & 182.4280 & 20.0469 & 230 & 60 & 16.75 & -16.1 & 0.37 & 23.31 & 8.09 & Y & ALFALF \\
NGC4158 & LS-446799-3923 & 182.6898 & 20.5927 & 270 & 118 & 18.16 & -14.6 & 0.25 & 23.80 & 7.39 & Y & MMT \\
NGC4158 & LS-444092-4124 & 182.8481 & 20.0633 & 78 & -84 & 19.78 & -13.0 & 0.29 & 23.25 & 6.78 & Y & MMT \\
NGC4158 & LS-444093-4832 & 183.1177 & 20.1081 & 197 & 16 & 19.84 & -13.0 & 0.32 & 23.99 & 6.79 & Y & MMT \\ \hline
NGC4348 & LS-310953-4341 & 185.5829 & -3.4094 & 203 & -48 & 20.31 & -12.1 & 0.13 & 23.78 & 6.23 & Y & AAT \\ \hline
NGC4454 & NSA-19694 & 187.2117 & -2.0609 & 74 & 160 & 14.30 & -18.4 & 0.62 & 20.70 & 9.31 & Y & SDSS+ \\
NGC4454 & LS-321038-4238 & 187.5895 & -1.6383 & 297 & -152 & 16.91 & -15.8 & 0.23 & 24.25 & 7.84 & Y & 2dF \\
NGC4454 & NSA-628407 & 187.2415 & -1.6030 & 208 & 204 & 18.48 & -14.3 & 0.61 & 23.42 & 7.63 & Y & NSA \\
NGC4454 & LS-321038-2612 & 187.5592 & -1.7291 & 250 & -163 & 19.88 & -12.9 & 0.35 & 23.93 & 6.78 & Y & AAT \\ \hline
NGC5297 & NSA-677369 & 206.5777 & 43.8514 & 15 & -154 & 13.96 & -18.8 & 0.50 & 20.80 & 9.32 & Y & SDSS \\
NGC5297 & NSA-265851 & 207.0967 & 43.7087 & 245 & -139 & 14.96 & -17.8 & 0.31 & 21.73 & 8.72 & Y & SDSS \\
NGC5297 & LS-558703-937 & 206.6083 & 43.4521 & 260 & 54 & 16.58 & -16.2 & 0.33 & 23.89 & 8.09 & Y & MMT \\
NGC5297 & 1237661852013166857 & 206.9900 & 43.5872 & 248 & 82 & 17.75 & -15.0 & 0.53 & 23.00 & 7.84 & N & PAL \\
NGC5297 & LS-561826-172 & 206.2129 & 44.1393 & 238 & 63 & 19.03 & -13.7 & 0.50 & 22.96 & 7.30 & N & MMT \\
NGC5297 & LS-561826-1312 & 206.2337 & 44.2385 & 279 & 55 & 19.58 & -13.2 & 0.30 & 25.48 & 6.85 & Y & MMT \\
NGC5297 & LS-561826-847 & 206.2622 & 44.1987 & 251 & 144 & 19.74 & -13.0 & 0.37 & 22.82 & 6.87 & N & MMT \\ \hline
NGC5347 & LS-513578-3277 & 208.2657 & 33.5912 & 67 & 149 & 18.19 & -14.5 & 0.41 & 22.59 & 7.51 & Y & MMT \\
NGC5347 & LS-513579-159 & 208.4808 & 33.3850 & 101 & 33 & 18.72 & -14.0 & 0.42 & 23.72 & 7.31 & Y & MMT \\ \hline
NGC5448 & NSA-339206 & 210.2074 & 49.5161 & 277 & -169 & 14.55 & -18.1 & 0.39 & 24.33 & 8.92 & Y & SDSS \\
NGC5448 & NSA-630990 & 210.2220 & 49.3748 & 220 & -165 & 15.06 & -17.6 & 0.42 & 23.43 & 8.74 & Y & NSA \\
NGC5448 & NSA-339202 & 210.1000 & 49.4443 & 281 & -26 & 15.62 & -17.0 & 0.44 & 22.28 & 8.55 & Y & SDSS \\
NGC5448 & NSA-339134 & 210.8700 & 48.9655 & 136 & -18 & 16.79 & -15.8 & 0.47 & 23.45 & 8.10 & Y & SDSS \\
NGC5448 & LS-582565-481 & 210.8297 & 49.4151 & 149 & 17 & 18.18 & -14.5 & 0.33 & 25.29 & 7.39 & Y & MMT \\ \hline
NGC5602 & NSA-208848 & 215.4049 & 50.3903 & 93 & 75 & 14.14 & -18.5 & 0.69 & 20.35 & 9.43 & Y & SDSS \\
NGC5602 & NSA-208508 & 215.4590 & 50.3884 & 80 & 169 & 16.52 & -16.2 & 0.36 & 22.34 & 8.10 & Y & SDSS \\
NGC5602 & NSA-208516 & 214.8035 & 50.4409 & 295 & 0 & 17.50 & -15.2 & 0.39 & 23.26 & 7.75 & Y & SDSS \\
NGC5602 & 1237659119323119642 & 215.4408 & 50.5062 & 52 & -65 & 17.53 & -15.1 & 0.38 & 22.91 & 7.73 & Y & MMT \\
NGC5602 & LS-587209-635 & 215.5340 & 50.6674 & 100 & -180 & 18.32 & -14.4 & 0.52 & 25.73 & 7.56 & N & MMT \\ \hline
NGC5604 & LS-311074-3415 & 215.9066 & -3.4513 & 246 & -60 & 18.08 & -14.9 & 0.24 & 25.46 & 7.48 & Y & AAT \\
NGC5604 & LS-312512-3375 & 215.8357 & -3.1983 & 233 & 20 & 19.50 & -13.5 & 0.32 & 23.39 & 6.99 & Y & AAT \\
NGC5604 & LS-312514-628 & 216.3962 & -3.3346 & 170 & 233 & 20.04 & -12.9 & 0.32 & 24.82 & 6.78 & Y & AAT \\ \hline
NGC5690 & NSA-74334 & 219.9139 & 2.5818 & 262 & -111 & 15.97 & -16.1 & 0.37 & 22.77 & 8.12 & Y & SDSS+ \\
NGC5690 & NSA-74161 & 218.9589 & 2.6061 & 256 & -188 & 16.08 & -16.0 & 0.27 & 23.74 & 7.96 & Y & SDSS+ \\
NGC5690 & LS-342766-1926 & 219.6106 & 1.9336 & 185 & 66 & 17.93 & -14.2 & 0.65 & 23.30 & 7.64 & N & GAMA \\ \hline
NGC5750 & NSA-14784 & 221.3160 & -0.1599 & 105 & 30 & 14.87 & -17.2 & 0.42 & 21.78 & 8.57 & Y & SDSS+ \\ \hline
NGC5792 & NSA-171561 & 225.0054 & -1.0913 & 202 & -36 & 14.73 & -17.5 & 0.26 & 20.74 & 8.56 & Y & SDSS+ \\
NGC5792 & NSA-171385 & 224.5327 & -1.3126 & 113 & 38 & 15.21 & -17.1 & 0.57 & 21.68 & 8.70 & Y & SDSS+ \\
NGC5792 & LS-328387-73 & 224.9713 & -0.6169 & 298 & -132 & 16.94 & -15.3 & 0.34 & 24.20 & 7.76 & Y & AAT \\ \hline
NGC5962 & NSA-685595 & 234.1329 & 16.4405 & 81 & -68 & 14.28 & -18.0 & 0.51 & 22.05 & 8.99 & Y & SDSS+ \\
NGC5962 & NSA-571923 & 233.7870 & 16.8704 & 205 & 27 & 15.39 & -16.8 & 0.37 & 22.66 & 8.40 & Y & SDSS+ \\ \hline
NGC6181 & NSA-337507 & 247.8399 & 20.1841 & 251 & 193 & 13.24 & -19.4 & 0.57 & 21.35 & 9.65 & Y & NSA \\
NGC6181 & NSA-337485 & 248.3932 & 19.9461 & 183 & 84 & 15.63 & -17.0 & 0.40 & 20.02 & 8.49 & Y & SDSS \\
NGC6181 & 1237662698115432544 & 248.0513 & 19.6957 & 77 & 100 & 16.66 & -16.0 & 0.38 & 22.22 & 8.07 & Y & AAT+ \\
NGC6181 & LS-444337-4335 & 247.8755 & 20.0948 & 197 & 119 & 16.72 & -15.9 & 0.31 & 24.29 & 7.97 & Y & SDSS \\
NGC6181 & NSA-633453 & 247.8259 & 20.2109 & 269 & 78 & 16.89 & -15.8 & 0.34 & 23.44 & 7.93 & Y & NSA \\
NGC6181 & LS-442982-4210 & 248.1526 & 19.8095 & 36 & -84 & 17.30 & -15.3 & 0.38 & 24.34 & 7.81 & Y & MMT \\
NGC6181 & LS-442983-1849 & 248.5809 & 19.7198 & 279 & 66 & 18.03 & -14.6 & 0.34 & 24.79 & 7.48 & Y & AAT+ \\
NGC6181 & 1237662224092496776 & 248.1954 & 19.8670 & 64 & 99 & 18.59 & -14.1 & 0.23 & 23.04 & 7.13 & Y & MMT \\
NGC6181 & LS-444338-475 & 248.2119 & 19.9001 & 81 & 57 & 19.14 & -13.5 & 0.45 & 24.37 & 7.15 & N & PAL \\
NGC6181 & LS-442982-3970 & 248.1636 & 19.7923 & 45 & -124 & 20.84 & -11.8 & 0.10 & 23.92 & 6.08 & Y & MMT \\ \hline
NGC6278 & NSA-343657 & 255.1879 & 23.0440 & 26 & -48 & 13.88 & -19.1 & 0.39 & 19.96 & 9.33 & Y & SDSS \\
NGC6278 & NSA-687367 & 255.0608 & 23.1063 & 114 & 189 & 14.16 & -18.8 & 0.25 & 22.22 & 9.06 & Y & SDSS \\
NGC6278 & NSA-343463 & 255.5795 & 23.0765 & 237 & 128 & 15.83 & -17.2 & 0.27 & 20.66 & 8.41 & Y & SDSS \\
NGC6278 & NSA-343648 & 255.5052 & 23.1634 & 213 & 218 & 15.93 & -17.1 & 0.23 & 22.87 & 8.32 & Y & SDSS \\
NGC6278 & 1237662301379166288 & 255.3735 & 22.7311 & 218 & 15 & 17.27 & -15.7 & 0.37 & 22.03 & 7.95 & Y & SDSS \\
NGC6278 & NSA-633932 & 255.1382 & 22.8660 & 109 & -167 & 17.70 & -15.3 & 0.30 & 22.62 & 7.70 & Y & NSA \\
NGC6278 & LS-459131-5790 & 255.3576 & 22.8730 & 133 & 32 & 19.06 & -13.9 & 0.39 & 24.43 & 7.26 & Y & MMT \\
NGC6278 & LS-461785-3431 & 254.9681 & 23.2804 & 239 & 150 & 19.16 & -13.8 & 0.44 & 23.51 & 7.27 & Y & MMT \\
NGC6278 & LS-459129-1060 & 254.9390 & 22.6673 & 291 & 82 & 19.95 & -13.0 & 0.35 & 24.29 & 6.85 & Y & MMT \\ \hline
NGC6909 & DES-168457190 & 306.9106 & -46.9880 & 24 & 205 & 15.82 & -16.9 & 0.63 & 21.48 & 8.72 & N & AAT \\
NGC6909 & DES-170146699 & 307.5227 & -47.1748 & 273 & -63 & 16.24 & -16.5 & 0.48 & 21.26 & 8.39 & Y & AAT \\
NGC6909 & DES-168455364 & 306.2666 & -46.9370 & 279 & 50 & 18.19 & -14.6 & 0.52 & 23.87 & 7.66 & N & AAT \\
NGC6909 & DES-168615265 & 306.6276 & -47.4109 & 266 & -75 & 18.20 & -14.6 & 0.39 & 22.06 & 7.51 & Y & AAT \\ \hline
NGC7029 & DES-188054728 & 317.9267 & -49.1875 & 66 & 162 & 17.35 & -15.6 & 0.62 & 22.84 & 8.15 & N & AAT \\
NGC7029 & DES-188072275 & 318.0170 & -49.4633 & 121 & 12 & 17.53 & -15.4 & 0.63 & 23.64 & 8.10 & N & AAT \\
NGC7029 & DES-188072825 & 317.9191 & -49.4568 & 116 & 45 & 17.71 & -15.2 & 0.52 & 21.97 & 7.91 & Y & 2dF \\
NGC7029 & DES-191931631 & 318.1685 & -49.6251 & 242 & -155 & 17.90 & -15.0 & 0.42 & 22.31 & 7.73 & Y & AAT \\
NGC7029 & DES-188058282 & 317.7678 & -49.2384 & 91 & -86 & 18.14 & -14.8 & 0.49 & 22.37 & 7.70 & Y & AAT \\ \hline
NGC7079 & DES-206747419 & 323.6315 & -44.3144 & 270 & -113 & 13.16 & -19.7 & 0.39 & 21.95 & 9.54 & Y & 6dF \\
NGC7079 & DES-201237392 & 322.8677 & -43.9600 & 144 & 57 & 15.89 & -16.9 & 0.51 & 22.28 & 8.58 & Y & AAT \\
NGC7079 & DES-203016260 & 322.9709 & -43.6539 & 274 & -77 & 16.85 & -16.0 & 0.36 & 23.52 & 8.04 & N & AAT \\
NGC7079 & DES-203024920 & 322.9955 & -43.7851 & 192 & -232 & 17.55 & -15.3 & 0.46 & 23.42 & 7.87 & Y & AAT \\ \hline
ESO288-025 & DES-247129294 & 330.2345 & -43.8028 & 179 & 21 & 16.90 & -15.8 & 0.44 & 23.19 & 8.05 & Y & AAT \\
ESO288-025 & DES-247126789 & 329.7579 & -43.7531 & 73 & 84 & 18.38 & -14.3 & 0.31 & 22.65 & 7.31 & Y & AAT \\ \hline
NGC7166 & DES-219806824 & 330.1490 & -43.1405 & 145 & -187 & 14.59 & -18.0 & 0.47 & 22.27 & 8.98 & Y & 6dF \\
NGC7166 & DES-247094045 & 330.5277 & -43.2685 & 180 & 185 & 16.32 & -16.3 & 0.26 & 22.03 & 8.06 & Y & 6dF \\
NGC7166 & DES-72187492 & 330.7073 & -43.4457 & 244 & 55 & 16.61 & -16.0 & 0.39 & 23.98 & 8.09 & Y & AAT \\
NGC7166 & DES-247096912 & 330.4219 & -43.3113 & 129 & 181 & 16.98 & -15.7 & 0.56 & 25.00 & 8.13 & Y & AAT \\
NGC7166 & DES-247092888 & 330.5717 & -43.2322 & 206 & -132 & 18.21 & -14.4 & 0.45 & 23.52 & 7.52 & Y & AAT \\
NGC7166 & DES-219810437 & 330.3291 & -43.1353 & 169 & -259 & 19.39 & -13.2 & 0.15 & 23.46 & 6.71 & Y & AAT \\ \hline
PGC068743 & NSA-636065 & 336.0479 & -3.4834 & 98 & -17 & 13.42 & -19.5 & 0.40 & 22.55 & 9.51 & Y & 6dF \\
PGC068743 & NSA-636047 & 335.8363 & -3.6598 & 164 & 195 & 14.86 & -18.1 & 0.57 & 21.65 & 9.12 & Y & 6dF \\
PGC068743 & LS-312992-1567 & 335.9799 & -3.2706 & 118 & 14 & 15.94 & -17.0 & 0.59 & 22.20 & 8.72 & Y & AAT \\
PGC068743 & LS-311554-3218 & 335.9729 & -3.4296 & 40 & -85 & 17.84 & -15.1 & 0.63 & 23.28 & 8.00 & N & MMT \\
PGC068743 & LS-310115-2825 & 335.9540 & -3.7010 & 185 & -30 & 19.54 & -13.4 & 0.22 & 24.25 & 6.87 & Y & AAT \\ \hline
NGC7328 & LS-390486-3250 & 339.3833 & 10.3132 & 148 & -21 & 18.90 & -14.1 & 0.38 & 24.01 & 7.29 & N & PAL \\ \hline
NGC7541 & NSA-637123 & 348.6438 & 4.4984 & 34 & 11 & 12.71 & -20.2 & 0.67 & 19.85 & 10.07 & Y & ALFALF \\
NGC7541 & LS-359104-131 & 348.6965 & 4.6396 & 70 & 182 & 14.40 & -18.5 & 0.25 & 23.86 & 8.93 & Y & ALFALF \\
NGC7541 & LS-356233-3798 & 348.7769 & 4.3732 & 123 & -13 & 15.31 & -17.6 & 0.52 & 23.21 & 8.86 & Y & MMT \\
NGC7541 & LS-357669-3767 & 348.8745 & 4.6131 & 137 & 68 & 16.13 & -16.8 & 0.37 & 24.18 & 8.37 & Y & ALFALF \\
NGC7541 & LS-357668-2728 & 348.6214 & 4.5073 & 44 & -121 & 18.01 & -14.9 & 0.48 & 23.47 & 7.74 & N & AAT \\
NGC7541 & LS-360540-737 & 348.5546 & 4.9151 & 267 & 125 & 20.77 & -12.1 & 0.06 & 24.29 & 6.18 & Y & MMT \\ \hline
NGC7716 & NSA-31702 & 354.3508 & 0.3910 & 144 & 119 & 13.70 & -19.0 & 0.39 & 23.33 & 9.29 & Y & SDSS+ \\
NGC7716 & NSA-31683 & 354.1952 & 0.6234 & 200 & 59 & 15.61 & -17.1 & 0.41 & 22.84 & 8.55 & Y & SDSS+
\enddata
\tablecomments{Column (1): corresponding host galaxy name. Column (2): object ID; the prefix indicates the primary photometric survey used for the object (a lack of prefix refers to the SDSS). 
Columns (3), (4), (7), (9), and (10): photometric properties taken from the primary survey; effective surface brightness \mueff{} (10) is in mag\,arcsec$^{-2}$. 
Column (5): projected distance to the host galaxy in kiloparsecs.
Column (6): difference in the heliocentric velocity with respect to the host galaxy, in kilometers per second.
Column (8): $k$-corrected absolute $r$-band magnitude, assuming the object is at the same physical distance as the host galaxy. 
Column (11): stellar mass inferred from $M_{r,o}$ and $(g-r)_o$ color (see \autoref{sec:phot-quantities}), values reported as $\log [M_*/\msun]$. 
Column (12): presence of the H$\alpha$ line (with an EW larger than 2\,\text{\AA{}}, indicating star forming; see \autoref{sec:quenching}).
Column (13): redshift source; the AAT/MMT/PAL labels indicate that the redshift was first obtained by SAGA. 
A machine-readable version of this table is available on the SAGA website (see footnote \ref{fn:saga}).}
\end{deluxetable*}

~ 
\clearpage

\bibliographystyle{aasjournal}
\bibliography{main}

\begin{thebibliography}{}
\expandafter\ifx\csname natexlab\endcsname\relax\def\natexlab#1{#1}\fi
\providecommand{\url}[1]{\href{#1}{#1}}
\providecommand{\dodoi}[1]{doi:~\href{http://doi.org/#1}{\nolinkurl{#1}}}
\providecommand{\doeprint}[1]{\href{http://ascl.net/#1}{\nolinkurl{http://ascl.net/#1}}}
\providecommand{\doarXiv}[1]{\href{https://arxiv.org/abs/#1}{\nolinkurl{https://arxiv.org/abs/#1}}}

\bibitem[{{AAO Software Team}(2015)}]{2015ascl.soft05015A}
{AAO Software Team}. 2015, {2dfdr: Data reduction software}.
\newblock \doeprint{1505.015}

\bibitem[{{Abbott} {et~al.}(2018){Abbott}, {Abdalla}, {Allam}, {Amara},
  {Annis}, {Asorey}, {Avila}, {Ballester}, {Banerji}, {Barkhouse}, \&
  et~al.}]{DES_DR1}
{Abbott}, T.~M.~C., {Abdalla}, F.~B., {Allam}, S., {et~al.} 2018,
  \href{http://dx.doi.org/10.3847/1538-4365/aae9f0}{\apjs, 239, 18}

\bibitem[{{Abolfathi} {et~al.}(2018){Abolfathi}, {Aguado}, {Aguilar}, {Allende
  Prieto}, {Almeida}, {Ananna}, {Anders}, {Anderson}, {Andrews}, {Anguiano}, \&
  et~al.}]{SDSS_DR14}
{Abolfathi}, B., {Aguado}, D.~S., {Aguilar}, G., {et~al.} 2018,
  \href{http://dx.doi.org/10.3847/1538-4365/aa9e8a}{\apjs, 235, 42}

\bibitem[{{Adhikari} {et~al.}(2014){Adhikari}, {Dalal}, \&
  {Chamberlain}}]{Adhikari14094482}
{Adhikari}, S., {Dalal}, N., \& {Chamberlain}, R.~T. 2014,
  \href{http://dx.doi.org/10.1088/1475-7516/2014/11/019}{\jcap, 2014, 019}

\bibitem[{{Aihara} {et~al.}(2019){Aihara}, {AlSayyad}, {Ando}, {Armstrong},
  {Bosch}, {Egami}, {Furusawa}, {Furusawa}, {Goulding}, {Harikane}, {Hikage},
  {Ho}, {Hsieh}, {Huang}, {Ikeda}, {Imanishi}, {Ito}, {Iwata}, {Jaelani},
  {Kakuma}, {Kawana}, {Kikuta}, {Kobayashi}, {Koike}, {Komiyama}, {Li},
  {Liang}, {Lin}, {Luo}, {Lupton}, {Lust}, {MacArthur}, {Matsuoka}, {Mineo},
  {Miyatake}, {Miyazaki}, {More}, {Murata}, {Namiki}, {Nishizawa}, {Oguri},
  {Okabe}, {Okamoto}, {Okura}, {Ono}, {Onodera}, {Onoue}, {Osato}, {Ouchi},
  {Shibuya}, {Strauss}, {Sugiyama}, {Suto}, {Takada}, {Takagi}, {Takata},
  {Takita}, {Tanaka}, {Terai}, {Toba}, {Uchiyama}, {Utsumi}, {Wang}, {Wang}, \&
  {Yamada}}]{hsc-pdr2:1905.12221}
{Aihara}, H., {AlSayyad}, Y., {Ando}, M., {et~al.} 2019,
  \href{http://dx.doi.org/10.1093/pasj/psz103}{\pasj, 71, 114}

\bibitem[{{Akins} {et~al.}(2020){Akins}, {Christensen}, {Brooks}, {Munshi},
  {Applebaum}, {Angelhardt}, \& {Chamberland}}]{Akins:2008.02805}
{Akins}, H.~B., {Christensen}, C.~R., {Brooks}, A.~M., {et~al.} 2020,
  \href{http://arxiv.org/abs/2008.02805}{arXiv:2008.02805}

\bibitem[{{An} {et~al.}(2019){An}, {Kim}, {Moon}, \& {Yoon}}]{An191111782}
{An}, S.-H., {Kim}, J., {Moon}, J.-S., \& {Yoon}, S.-J. 2019,
  \href{http://dx.doi.org/10.3847/1538-4357/ab535f}{\apj, 887, 59}

\bibitem[{{Applebaum} {et~al.}(2020){Applebaum}, {Brooks}, {Christensen},
  {Munshi}, {Quinn}, {Shen}, \& {Tremmel}}]{2008.11207}
{Applebaum}, E., {Brooks}, A.~M., {Christensen}, C.~R., {et~al.} 2020,
  \href{http://arxiv.org/abs/2008.11207}{arXiv:2008.11207}

\bibitem[{{Astropy Collaboration} {et~al.}(2013){Astropy Collaboration},
  {Robitaille}, {Tollerud}, {Greenfield}, {Droettboom}, {Bray}, {Aldcroft},
  {Davis}, {Ginsburg}, {Price-Whelan}, {Kerzendorf}, {Conley}, {Crighton},
  {Barbary}, {Muna}, {Ferguson}, {Grollier}, {Parikh}, {Nair}, {Unther},
  {Deil}, {Woillez}, {Conseil}, {Kramer}, {Turner}, {Singer}, {Fox}, {Weaver},
  {Zabalza}, {Edwards}, {Azalee Bostroem}, {Burke}, {Casey}, {Crawford},
  {Dencheva}, {Ely}, {Jenness}, {Labrie}, {Lim}, {Pierfederici}, {Pontzen},
  {Ptak}, {Refsdal}, {Servillat}, \& {Streicher}}]{astropy}
{Astropy Collaboration}, {Robitaille}, T.~P., {Tollerud}, E.~J., {et~al.} 2013,
  \href{http://dx.doi.org/10.1051/0004-6361/201322068}{\aap, 558, A33}

\bibitem[{{Baldry} {et~al.}(2018){Baldry}, {Liske}, {Brown}, {Robotham},
  {Driver}, {Dunne}, {Alpaslan}, {Brough}, {Cluver}, {Eardley}, {Farrow},
  {Heymans}, {Hildebrandt}, {Hopkins}, {Kelvin}, {Loveday}, {Moffett},
  {Norberg}, {Owers}, {Taylor}, {Wright}, {Bamford}, {Bland -Hawthorn},
  {Bourne}, {Bremer}, {Colless}, {Conselice}, {Croom}, {Davies}, {Foster},
  {Grootes}, {Holwerda}, {Jones}, {Kafle}, {Kuijken}, {Lara-Lopez},
  {L{\'o}pez-S{\'a}nchez}, {Meyer}, {Phillipps}, {Sutherland}, {van Kampen}, \&
  {Wilkins}}]{Baldry2018:GAMA:DR3}
{Baldry}, I.~K., {Liske}, J., {Brown}, M.~J.~I., {et~al.} 2018,
  \href{http://dx.doi.org/10.1093/mnras/stx3042}{\mnras, 474, 3875}

\bibitem[{{Bell} {et~al.}(2003){Bell}, {McIntosh}, {Katz}, \&
  {Weinberg}}]{bell2003}
{Bell}, E.~F., {McIntosh}, D.~H., {Katz}, N., \& {Weinberg}, M.~D. 2003,
  \href{http://dx.doi.org/10.1086/378847}{\apjs, 149, 289}

\bibitem[{{Bennet} {et~al.}(2019){Bennet}, {Sand}, {Crnojevi{\'c}}, {Spekkens},
  {Karunakaran}, {Zaritsky}, \& {Mutlu-Pakdil}}]{Bennet2019a}
{Bennet}, P., {Sand}, D.~J., {Crnojevi{\'c}}, D., {et~al.} 2019,
  \href{http://dx.doi.org/10.3847/1538-4357/ab46ab}{\apj, 885, 153}

\bibitem[{{Blake} {et~al.}(2016){Blake}, {Amon}, {Childress}, {Erben},
  {Glazebrook}, {Harnois-Deraps}, {Heymans}, {Hildebrandt}, {Hinton},
  {Janssens}, {Johnson}, {Joudaki}, {Klaes}, {Kuijken}, {Lidman}, {Marin},
  {Parkinson}, {Poole}, \& {Wolf}}]{2dFlens}
{Blake}, C., {Amon}, A., {Childress}, M., {et~al.} 2016,
  \href{http://dx.doi.org/10.1093/mnras/stw1990}{\mnras, 462, 4240}

\bibitem[{{Blanton} {et~al.}(2011){Blanton}, {Kazin}, {Muna}, {Weaver}, \&
  {Price-Whelan}}]{Blanton2011}
{Blanton}, M.~R., {Kazin}, E., {Muna}, D., {Weaver}, B.~A., \& {Price-Whelan},
  A. 2011, \href{http://dx.doi.org/10.1088/0004-6256/142/1/31}{\aj, 142, 31}

\bibitem[{{Bolton} {et~al.}(2012){Bolton}, {Schlegel}, {Aubourg}, {Bailey},
  {Bhardwaj}, {Brownstein}, {Burles}, {Chen}, {Dawson}, {Eisenstein}, {Gunn},
  {Knapp}, {Loomis}, {Lupton}, {Maraston}, {Muna}, {Myers}, {Olmstead},
  {Padmanabhan}, {P{\^a}ris}, {Percival}, {Petitjean}, {Rockosi}, {Ross},
  {Schneider}, {Shu}, {Strauss}, {Thomas}, {Tremonti}, {Wake}, {Weaver}, \&
  {Wood-Vasey}}]{2012AJ....144..144B}
{Bolton}, A.~S., {Schlegel}, D.~J., {Aubourg}, {\'E}., {et~al.} 2012,
  \href{http://dx.doi.org/10.1088/0004-6256/144/5/144}{\aj, 144, 144}

\bibitem[{{Brinchmann} {et~al.}(2004){Brinchmann}, {Charlot}, {White},
  {Tremonti}, {Kauffmann}, {Heckman}, \& {Brinkmann}}]{Brinchmann2004}
{Brinchmann}, J., {Charlot}, S., {White}, S.~D.~M., {et~al.} 2004,
  \href{http://dx.doi.org/10.1111/j.1365-2966.2004.07881.x}{\mnras, 351, 1151}

\bibitem[{{Brooks} {et~al.}(2017){Brooks}, {Papastergis}, {Christensen},
  {Governato}, {Stilp}, {Quinn}, \& {Wadsley}}]{Brooks2017}
{Brooks}, A.~M., {Papastergis}, E., {Christensen}, C.~R., {et~al.} 2017,
  \href{http://dx.doi.org/10.3847/1538-4357/aa9576}{\apj, 850, 97}

\bibitem[{{Buck} {et~al.}(2019){Buck}, {Macci{\`o}}, {Dutton}, {Obreja}, \&
  {Frings}}]{buck:2019MNRAS.483.1314B}
{Buck}, T., {Macci{\`o}}, A.~V., {Dutton}, A.~A., {Obreja}, A., \& {Frings}, J.
  2019, \href{http://dx.doi.org/10.1093/mnras/sty2913}{\mnras, 483, 1314}

\bibitem[{{Bullock} \& {Boylan-Kolchin}(2017)}]{Bullock2017}
{Bullock}, J.~S. \& {Boylan-Kolchin}, M. 2017,
  \href{http://dx.doi.org/10.1146/annurev-astro-091916-055313}{\araa, 55, 343}

\bibitem[{{Cao} {et~al.}(2020){Cao}, {Tinker}, {Mao}, \& {Wechsler}}]{Cao2019}
{Cao}, J.-z., {Tinker}, J.~L., {Mao}, Y.-Y., \& {Wechsler}, R.~H. 2020,
  \href{http://dx.doi.org/10.1093/mnras/staa2644}{\mnras, 498, 5080}

\bibitem[{{Carlsten} {et~al.}(2019){Carlsten}, {Beaton}, {Greco}, \&
  {Greene}}]{Carlsten:2019ApJ...878L..16C}
{Carlsten}, S.~G., {Beaton}, R.~L., {Greco}, J.~P., \& {Greene}, J.~E. 2019,
  \href{http://dx.doi.org/10.3847/2041-8213/ab24d2}{\apjl, 878, L16}

\bibitem[{{Carlsten} {et~al.}(2020{\natexlab{a}}){Carlsten}, {Greco}, {Beaton},
  \& {Greene}}]{Carlsten2019}
{Carlsten}, S.~G., {Greco}, J.~P., {Beaton}, R.~L., \& {Greene}, J.~E.
  2020{\natexlab{a}}, \href{http://dx.doi.org/10.3847/1538-4357/ab7758}{\apj,
  891, 144}

\bibitem[{{Carlsten} {et~al.}(2020{\natexlab{b}}){Carlsten}, {Greene}, {Peter},
  {Beaton}, \& {Greco}}]{Carlsten200602443}
{Carlsten}, S.~G., {Greene}, J.~E., {Peter}, A. H.~G., {Beaton}, R.~L., \&
  {Greco}, J.~P. 2020{\natexlab{b}},
  \href{http://arxiv.org/abs/2006.02443}{arXiv:2006.02443}

\bibitem[{{Carlsten} {et~al.}(2020{\natexlab{c}}){Carlsten}, {Greene}, {Peter},
  {Greco}, \& {Beaton}}]{Carlsten200602444}
{Carlsten}, S.~G., {Greene}, J.~E., {Peter}, A. H.~G., {Greco}, J.~P., \&
  {Beaton}, R.~L. 2020{\natexlab{c}},
  \href{http://dx.doi.org/10.3847/1538-4357/abb60b}{\apj, 902, 124}

\bibitem[{{Cautun} {et~al.}(2015){Cautun}, {Wang}, {Frenk}, \&
  {Sawala}}]{2015MNRAS.449.2576C}
{Cautun}, M., {Wang}, W., {Frenk}, C.~S., \& {Sawala}, T. 2015,
  \href{http://dx.doi.org/10.1093/mnras/stv490}{\mnras, 449, 2576}

\bibitem[{{Chiboucas} {et~al.}(2013){Chiboucas}, {Jacobs}, {Tully}, \&
  {Karachentsev}}]{Chiboucas2013}
{Chiboucas}, K., {Jacobs}, B.~A., {Tully}, R.~B., \& {Karachentsev}, I.~D.
  2013, \href{http://dx.doi.org/10.1088/0004-6256/146/5/126}{\aj, 146, 126}

\bibitem[{{Chilingarian} {et~al.}(2010){Chilingarian}, {Melchior}, \&
  {Zolotukhin}}]{Chilingarian2010}
{Chilingarian}, I.~V., {Melchior}, A.-L., \& {Zolotukhin}, I.~Y. 2010,
  \href{http://dx.doi.org/10.1111/j.1365-2966.2010.16506.x}{\mnras, 405, 1409}

\bibitem[{{Colless} {et~al.}(2001){Colless}, {Dalton}, {Maddox}, {Sutherland },
  {Norberg}, {Cole}, {Bland -Hawthorn}, {Bridges}, {Cannon}, {Collins},
  {Couch}, {Cross}, {Deeley}, {De Propris}, {Driver}, {Efstathiou}, {Ellis},
  {Frenk}, {Glazebrook}, {Jackson}, {Lahav}, {Lewis}, {Lumsden}, {Madgwick},
  {Peacock}, {Peterson}, {Price}, {Seaborne}, \& {Taylor}}]{2dFGRS}
{Colless}, M., {Dalton}, G., {Maddox}, S., {et~al.} 2001,
  \href{http://dx.doi.org/10.1046/j.1365-8711.2001.04902.x}{\mnras, 328, 1039}

\bibitem[{{Collins} {et~al.}(2020){Collins}, {Tollerud}, {Rich}, {Ibata},
  {Martin}, {Chapman}, {Gilbert}, \& {Preston}}]{Collins2020}
{Collins}, M. L.~M., {Tollerud}, E.~J., {Rich}, R.~M., {et~al.} 2020,
  \href{http://dx.doi.org/10.1093/mnras/stz3252}{\mnras, 491, 3496}

\bibitem[{{Cool} {et~al.}(2008){Cool}, {Eisenstein}, {Fan}, {Fukugita},
  {Jiang}, {Maraston}, {Meiksin}, {Schneider}, \& {Wake}}]{10.1086/589642}
{Cool}, R.~J., {Eisenstein}, D.~J., {Fan}, X., {et~al.} 2008,
  \href{http://dx.doi.org/10.1086/589642}{\apj, 682, 919}

\bibitem[{{Crnojevi{\'c}} {et~al.}(2019){Crnojevi{\'c}}, {Sand}, {Bennet},
  {Pasetto}, {Spekkens}, {Caldwell}, {Guhathakurta}, {McLeod}, {Seth}, {Simon},
  {Strader}, \& {Toloba}}]{Crnojevic2019}
{Crnojevi{\'c}}, D., {Sand}, D.~J., {Bennet}, P., {et~al.} 2019,
  \href{http://dx.doi.org/10.3847/1538-4357/aafbe7}{\apj, 872, 80}

\bibitem[{{Danieli} {et~al.}(2018){Danieli}, {van Dokkum}, \&
  {Conroy}}]{Danieli2018}
{Danieli}, S., {van Dokkum}, P., \& {Conroy}, C. 2018,
  \href{http://dx.doi.org/10.3847/1538-4357/aaadfb}{\apj, 856, 69}

\bibitem[{{Danieli} {et~al.}(2017){Danieli}, {van Dokkum}, {Merritt},
  {Abraham}, {Zhang}, {Karachentsev}, \&
  {Makarova}}]{Danieli:2017ApJ...837..136D}
{Danieli}, S., {van Dokkum}, P., {Merritt}, A., {et~al.} 2017,
  \href{http://dx.doi.org/10.3847/1538-4357/aa615b}{\apj, 837, 136}

\bibitem[{{Davis} {et~al.}(2021){Davis}, {Nierenberg}, {Peter}, {Garling},
  {Greco}, {Kochanek}, {Utomo}, {Casey}, {Pogge}, {Roberts}, {Sand}, \&
  {Sardone}}]{LBT-SONG:2003.08352}
{Davis}, A.~B., {Nierenberg}, A.~M., {Peter}, A. H.~G., {et~al.} 2021,
  \href{http://dx.doi.org/10.1093/mnras/staa3246}{\mnras, 500, 3854}

\bibitem[{{de Jong} {et~al.}(2012){de Jong}, {Bellido-Tirado}, {Chiappini},
  {Depagne}, {Haynes}, {Johl}, {Schnurr}, {Schwope}, {Walcher}, {Dionies},
  {Haynes}, {Kelz}, {Kitaura}, {Lamer}, {Minchev}, {M{\"u}ller}, {Nuza},
  {Olaya}, {Piffl}, {Popow}, {Steinmetz}, {Ural}, {Williams}, {Winkler},
  {Wisotzki}, {Ansorge}, {Banerji}, {Gonzalez Solares}, {Irwin}, {Kennicutt},
  {King}, {McMahon}, {Koposov}, {Parry}, {Sun}, {Walton}, {Finger}, {Iwert},
  {Krumpe}, {Lizon}, {Vincenzo}, {Amans}, {Bonifacio}, {Cohen}, {Francois},
  {Jagourel}, {Mignot}, {Royer}, {Sartoretti}, {Bender}, {Grupp}, {Hess},
  {Lang-Bardl}, {Muschielok}, {B{\"o}hringer}, {Boller}, {Bongiorno}, {Brusa},
  {Dwelly}, {Merloni}, {Nandra}, {Salvato}, {Pragt}, {Navarro}, {Gerlofsma},
  {Roelfsema}, {Dalton}, {Middleton}, {Tosh}, {Boeche}, {Caffau}, {Christlieb},
  {Grebel}, {Hansen}, {Koch}, {Ludwig}, {Quirrenbach}, {Sbordone}, {Seifert},
  {Thimm}, {Trifonov}, {Helmi}, {Trager}, {Feltzing}, {Korn}, \&
  {Boland}}]{2012SPIE.8446E..0TD}
{de Jong}, R.~S., {Bellido-Tirado}, O., {Chiappini}, C., {et~al.} 2012, in
  Society of Photo-Optical Instrumentation Engineers (SPIE) Conference Series,
  Vol. 8446, Ground-based and Airborne Instrumentation for Astronomy IV,
  84460T, \dodoi{10.1117/12.926239}

\bibitem[{{DESI Collaboration} {et~al.}(2016){DESI Collaboration}, {Aghamousa},
  {Aguilar}, {Ahlen}, {Alam}, {Allen}, {Allende Prieto}, {Annis}, {Bailey},
  {Balland}, \& et~al.}]{1611.00036}
{DESI Collaboration}, {Aghamousa}, A., {Aguilar}, J., {et~al.} 2016,
  \href{http://arxiv.org/abs/1611.00036}{arXiv:1611.00036}

\bibitem[{{Dey} {et~al.}(2019){Dey}, {Schlegel}, {Lang}, {Blum}, {Burleigh},
  {Fan}, {Findlay}, {Finkbeiner}, {Herrera}, {Juneau}, {Landriau}, {Levi},
  {McGreer}, {Meisner}, {Myers}, {Moustakas}, {Nugent}, {Patej}, {Schlafly},
  {Walker}, {Valdes}, {Weaver}, {Y{\`e}che}, {Zou}, {Zhou}, {Abareshi},
  {Abbott}, {Abolfathi}, {Aguilera}, {Alam}, {Allen}, {Alvarez}, {Annis},
  {Ansarinejad}, {Aubert}, {Beechert}, {Bell}, {BenZvi}, {Beutler}, {Bielby},
  {Bolton}, {Brice{\~n}o}, {Buckley-Geer}, {Butler}, {Calamida}, {Carlberg},
  {Carter}, {Casas}, {Castander}, {Choi}, {Comparat}, {Cukanovaite}, {Delubac},
  {DeVries}, {Dey}, {Dhungana}, {Dickinson}, {Ding}, {Donaldson}, {Duan},
  {Duckworth}, {Eftekharzadeh}, {Eisenstein}, {Etourneau}, {Fagrelius},
  {Farihi}, {Fitzpatrick}, {Font-Ribera}, {Fulmer}, {G{\"a}nsicke},
  {Gaztanaga}, {George}, {Gerdes}, {Gontcho}, {Gorgoni}, {Green}, {Guy},
  {Harmer}, {Hernand ez}, {Honscheid}, {Huang}, {James}, {Jannuzi}, {Jiang},
  {Joyce}, {Karcher}, {Karkar}, {Kehoe}, {Kneib}, {Kueter-Young}, {Lan},
  {Lauer}, {Le Guillou}, {Le Van Suu}, {Lee}, {Lesser}, {Perreault Levasseur},
  {Li}, {Mann}, {Marshall}, {Mart{\'\i}nez-V{\'a}zquez}, {Martini}, {du Mas des
  Bourboux}, {McManus}, {Meier}, {M{\'e}nard}, {Metcalfe},
  {Mu{\~n}oz-Guti{\'e}rrez}, {Najita}, {Napier}, {Narayan}, {Newman}, {Nie},
  {Nord}, {Norman}, {Olsen}, {Paat}, {Palanque-Delabrouille}, {Peng},
  {Poppett}, {Poremba}, {Prakash}, {Rabinowitz}, {Raichoor}, {Rezaie},
  {Robertson}, {Roe}, {Ross}, {Ross}, {Rudnick}, {Safonova}, {Saha},
  {S{\'a}nchez}, {Savary}, {Schweiker}, {Scott}, {Seo}, {Shan}, {Silva},
  {Slepian}, {Soto}, {Sprayberry}, {Staten}, {Stillman}, {Stupak}, {Summers},
  {Sien Tie}, {Tirado}, {Vargas-Maga{\~n}a}, {Vivas}, {Wechsler}, {Williams},
  {Yang}, {Yang}, {Yapici}, {Zaritsky}, {Zenteno}, {Zhang}, {Zhang}, {Zhou}, \&
  {Zhou}}]{Dey2019}
{Dey}, A., {Schlegel}, D.~J., {Lang}, D., {et~al.} 2019,
  \href{http://dx.doi.org/10.3847/1538-3881/ab089d}{\aj, 157, 168}

\bibitem[{{Diemer} \& {Kravtsov}(2014)}]{Diemer:2014xya}
{Diemer}, B. \& {Kravtsov}, A.~V. 2014,
  \href{http://dx.doi.org/10.1088/0004-637X/789/1/1}{\apj, 789, 1}

\bibitem[{{Drinkwater} {et~al.}(2018){Drinkwater}, {Byrne}, {Blake},
  {Glazebrook}, {Brough}, {Colless}, {Couch}, {Croton}, {Croom}, {Davis},
  {Forster}, {Gilbank}, {Hinton}, {Jelliffe}, {Jurek}, {Li}, {Martin},
  {Pimbblet}, {Poole}, {Pracy}, {Sharp}, {Smillie}, {Spolaor}, {Wisnioski},
  {Woods}, {Wyder}, \& {Yee}}]{WiggleZ}
{Drinkwater}, M.~J., {Byrne}, Z.~J., {Blake}, C., {et~al.} 2018,
  \href{http://dx.doi.org/10.1093/mnras/stx2963}{\mnras, 474, 4151}

\bibitem[{{Drlica-Wagner} {et~al.}(2020){Drlica-Wagner}, {Bechtol}, {Mau},
  {McNanna}, {Nadler}, {Pace}, {Li}, {Pieres}, {Rozo}, {Simon}, {Walker},
  {Wechsler}, {Abbott}, {Allam}, {Annis}, {Bertin}, {Brooks}, {Burke},
  {Rosell}, {Carrasco Kind}, {Carretero}, {Costanzi}, {da Costa}, {De Vicente},
  {Desai}, {Diehl}, {Doel}, {Eifler}, {Everett}, {Flaugher}, {Frieman},
  {Garc{\'\i}a-Bellido}, {Gaztanaga}, {Gruen}, {Gruendl}, {Gschwend},
  {Gutierrez}, {Honscheid}, {James}, {Krause}, {Kuehn}, {Kuropatkin}, {Lahav},
  {Maia}, {Marshall}, {Melchior}, {Menanteau}, {Miquel}, {Palmese}, {Plazas},
  {Sanchez}, {Scarpine}, {Schubnell}, {Serrano}, {Sevilla-Noarbe}, {Smith},
  {Suchyta}, {Tarle}, \& {DES Collaboration}}]{Drlica-Wagner191203302}
{Drlica-Wagner}, A., {Bechtol}, K., {Mau}, S., {et~al.} 2020,
  \href{http://dx.doi.org/10.3847/1538-4357/ab7eb9}{\apj, 893, 47}

\bibitem[{{Fabricant} {et~al.}(2005){Fabricant}, {Fata}, {Roll}, {Hertz},
  {Caldwell}, {Gauron}, {Geary}, {McLeod}, {Szentgyorgyi}, {Zajac}, {Kurtz},
  {Barberis}, {Bergner}, {Brown}, {Conroy}, {Eng}, {Geller}, {Goddard},
  {Honsa}, {Mueller}, {Mink}, {Ordway}, {Tokarz}, {Woods}, {Wyatt}, {Epps}, \&
  {Dell'Antonio}}]{2005PASP..117.1411F}
{Fabricant}, D., {Fata}, R., {Roll}, J., {et~al.} 2005,
  \href{http://dx.doi.org/10.1086/497385}{\pasp, 117, 1411}

\bibitem[{{Fielder} {et~al.}(2019){Fielder}, {Mao}, {Newman}, {Zentner}, \&
  {Licquia}}]{1807.05180}
{Fielder}, C.~E., {Mao}, Y.-Y., {Newman}, J.~A., {Zentner}, A.~R., \&
  {Licquia}, T.~C. 2019, \href{http://dx.doi.org/10.1093/mnras/stz1098}{\mnras,
  486, 4545}

\bibitem[{{Fillingham} {et~al.}(2015){Fillingham}, {Cooper}, {Wheeler},
  {Garrison-Kimmel}, {Boylan-Kolchin}, \& {Bullock}}]{10.1093/mnras/stv2058}
{Fillingham}, S.~P., {Cooper}, M.~C., {Wheeler}, C., {et~al.} 2015,
  \href{http://dx.doi.org/10.1093/mnras/stv2058}{\mnras, 454, 2039}

\bibitem[{{Flores Vel{\'a}zquez} {et~al.}(2020){Flores Vel{\'a}zquez},
  {Gurvich}, {Faucher-Gigu{\`e}re}, {Bullock}, {Starkenburg}, {Moreno},
  {Lazar}, {Mercado}, {Stern}, {Sparre}, {Hayward}, {Wetzel}, \&
  {El-Badry}}]{2008.08582}
{Flores Vel{\'a}zquez}, J.~A., {Gurvich}, A.~B., {Faucher-Gigu{\`e}re}, C.-A.,
  {et~al.} 2020, \href{http://arxiv.org/abs/2008.08582}{arXiv:2008.08582}

\bibitem[{{Fremling} {et~al.}(2020){Fremling}, {Miller}, {Sharma}, {Dugas},
  {Perley}, {Taggart}, {Sollerman}, {Goobar}, {Graham}, {Neill}, {Nordin},
  {Rigault}, {Walters}, {Andreoni}, {Bagdasaryan}, {Belicki}, {Cannella},
  {Bellm}, {Cenko}, {De}, {Dekany}, {Frederick}, {Golkhou}, {Graham}, {Helou},
  {Ho}, {Kasliwal}, {Kupfer}, {Laher}, {Mahabal}, {Masci}, {Riddle},
  {Rusholme}, {Schulze}, {Shupe}, {Smith}, {van Velzen}, {Yan}, {Yao},
  {Zhuang}, \& {Kulkarni}}]{2020ApJ...895...32F}
{Fremling}, C., {Miller}, A.~A., {Sharma}, Y., {et~al.} 2020,
  \href{http://dx.doi.org/10.3847/1538-4357/ab8943}{\apj, 895, 32}

\bibitem[{{Garrison-Kimmel} {et~al.}(2014){Garrison-Kimmel}, {Boylan-Kolchin},
  {Bullock}, \& {Kirby}}]{GK2014}
{Garrison-Kimmel}, S., {Boylan-Kolchin}, M., {Bullock}, J.~S., \& {Kirby},
  E.~N. 2014, \href{http://dx.doi.org/10.1093/mnras/stu1477}{\mnras, 444, 222}

\bibitem[{{Garrison-Kimmel} {et~al.}(2017){Garrison-Kimmel}, {Wetzel},
  {Bullock}, {Hopkins}, {Boylan-Kolchin}, {Faucher-Gigu{\`e}re}, {Kere{\v{s}}},
  {Quataert}, {Sanderson}, {Graus}, \& {Kelley}}]{Garrison-Kimmel170103792}
{Garrison-Kimmel}, S., {Wetzel}, A., {Bullock}, J.~S., {et~al.} 2017,
  \href{http://dx.doi.org/10.1093/mnras/stx1710}{\mnras, 471, 1709}

\bibitem[{{Garrison-Kimmel} {et~al.}(2019{\natexlab{a}}){Garrison-Kimmel},
  {Hopkins}, {Wetzel}, {Bullock}, {Boylan-Kolchin}, {Kere{\v{s}}},
  {Faucher-Gigu{\`e}re}, {El-Badry}, {Lamberts}, {Quataert}, \& {Sand
  erson}}]{Garrison-Kimmel2019:1806.04143}
{Garrison-Kimmel}, S., {Hopkins}, P.~F., {Wetzel}, A., {et~al.}
  2019{\natexlab{a}}, \href{http://dx.doi.org/10.1093/mnras/stz1317}{\mnras,
  487, 1380}

\bibitem[{{Garrison-Kimmel} {et~al.}(2019{\natexlab{b}}){Garrison-Kimmel},
  {Wetzel}, {Hopkins}, {Sanderson}, {El-Badry}, {Graus}, {Chan}, {Feldmann},
  {Boylan-Kolchin}, {Hayward}, {Bullock}, {Fitts}, {Samuel}, {Wheeler},
  {Kere{\v{s}}}, \& {Faucher-Gigu{\`e}re}}]{Garrison-Kimmel2019:1903.10515}
{Garrison-Kimmel}, S., {Wetzel}, A., {Hopkins}, P.~F., {et~al.}
  2019{\natexlab{b}}, \href{http://dx.doi.org/10.1093/mnras/stz2507}{\mnras,
  489, 4574}

\bibitem[{{Geha} {et~al.}(2012){Geha}, {Blanton}, {Yan}, \&
  {Tinker}}]{geha2012}
{Geha}, M., {Blanton}, M.~R., {Yan}, R., \& {Tinker}, J.~L. 2012,
  \href{http://dx.doi.org/10.1088/0004-637X/757/1/85}{\apj, 757, 85}

\bibitem[{{Geha} {et~al.}(2017){Geha}, {Wechsler}, {Mao}, {Tollerud}, {Weiner},
  {Bernstein}, {Hoyle}, {Marchi}, {Marshall}, {Mu{\~n}oz}, \& {Lu}}]{Geha2017}
{Geha}, M., {Wechsler}, R.~H., {Mao}, Y.-Y., {et~al.} 2017,
  \href{http://dx.doi.org/10.3847/1538-4357/aa8626}{\apj, 847, 4}

\bibitem[{{G{\'o}rski} {et~al.}(2005){G{\'o}rski}, {Hivon}, {Banday},
  {Wandelt}, {Hansen}, {Reinecke}, \& {Bartelmann}}]{2005ApJ...622..759G}
{G{\'o}rski}, K.~M., {Hivon}, E., {Banday}, A.~J., {et~al.} 2005,
  \href{http://dx.doi.org/10.1086/427976}{\apj, 622, 759}

\bibitem[{{Graham} \& {Driver}(2005)}]{Graham2005}
{Graham}, A.~W. \& {Driver}, S.~P. 2005,
  \href{http://dx.doi.org/10.1071/AS05001}{\pasa, 22, 118}

\bibitem[{{Greco} {et~al.}(2020){Greco}, {van Dokkum}, {Danieli}, {Carlsten},
  \& {Conroy}}]{Greco:2004.07273}
{Greco}, J.~P., {van Dokkum}, P., {Danieli}, S., {Carlsten}, S.~G., \&
  {Conroy}, C. 2020, \href{http://arxiv.org/abs/2004.07273}{arXiv:2004.07273}

\bibitem[{{Greco} {et~al.}(2018){Greco}, {Greene}, {Strauss}, {Macarthur},
  {Flowers}, {Goulding}, {Huang}, {Kim}, {Komiyama}, {Leauthaud}, {Leisman},
  {Lupton}, {Sif{\'o}n}, \& {Wang}}]{Greco18:1709.04474}
{Greco}, J.~P., {Greene}, J.~E., {Strauss}, M.~A., {et~al.} 2018,
  \href{http://dx.doi.org/10.3847/1538-4357/aab842}{\apj, 857, 104}

\bibitem[{Harris {et~al.}(2020)Harris, Millman, van~der Walt, Gommers,
  Virtanen, Cournapeau, Wieser, Taylor, Berg, Smith, Kern, Picus, Hoyer, van
  Kerkwijk, Brett, Haldane, Fernández~del Río, Wiebe, Peterson,
  Gérard-Marchant, Sheppard, Reddy, Weckesser, Abbasi, Gohlke, \&
  Oliphant}]{2020NumPy-Array}
Harris, C.~R., Millman, K.~J., van~der Walt, S.~J., {et~al.} 2020,
  \href{http://dx.doi.org/10.1038/s41586-020-2649-2}{Nature, 585, 357–362}

\bibitem[{{Hausammann} {et~al.}(2019){Hausammann}, {Revaz}, \&
  {Jablonka}}]{Hausammann:2019A&A...624A..11H}
{Hausammann}, L., {Revaz}, Y., \& {Jablonka}, P. 2019,
  \href{http://dx.doi.org/10.1051/0004-6361/201834871}{\aap, 624, A11}

\bibitem[{{Haynes} {et~al.}(2018){Haynes}, {Giovanelli}, {Kent}, {Adams},
  {Balonek}, {Craig}, {Fertig}, {Finn}, {Giovanardi}, {Hallenbeck}, {Hess},
  {Hoffman}, {Huang}, {Jones}, {Koopmann}, {Kornreich}, {Leisman}, {Miller},
  {Moorman}, {O'Connor}, {O'Donoghue}, {Papastergis}, {Troischt}, {Stark}, \&
  {Xiao}}]{ALFALFA100}
{Haynes}, M.~P., {Giovanelli}, R., {Kent}, B.~R., {et~al.} 2018,
  \href{http://dx.doi.org/10.3847/1538-4357/aac956}{\apj, 861, 49}

\bibitem[{{Hinton} {et~al.}(2016){Hinton}, {Davis}, {Lidman}, {Glazebrook}, \&
  {Lewis}}]{10.1016/j.ascom.2016.03.001}
{Hinton}, S.~R., {Davis}, T.~M., {Lidman}, C., {Glazebrook}, K., \& {Lewis},
  G.~F. 2016, \href{http://dx.doi.org/10.1016/j.ascom.2016.03.001}{Astronomy
  and Computing, 15, 61}

\bibitem[{{Huchra} {et~al.}(2012){Huchra}, {Macri}, {Masters}, {Jarrett},
  {Berlind}, {Calkins}, {Crook}, {Cutri}, {Erdo{\v{g}}du}, {Falco}, {George},
  {Hutcheson}, {Lahav}, {Mader}, {Mink}, {Martimbeau}, {Schneider},
  {Skrutskie}, {Tokarz}, \& {Westover}}]{2012ApJS..199...26H}
{Huchra}, J.~P., {Macri}, L.~M., {Masters}, K.~L., {et~al.} 2012,
  \href{http://dx.doi.org/10.1088/0067-0049/199/2/26}{\apjs, 199, 26}

\bibitem[{{Hunter}(2007)}]{matplotlib}
{Hunter}, J.~D. 2007, \href{http://dx.doi.org/10.1109/MCSE.2007.55}{Computing
  in Science and Engineering, 9, 90}

\bibitem[{{Ibata} {et~al.}(2014){Ibata}, {Ibata}, {Famaey}, \&
  {Lewis}}]{1407.8178}
{Ibata}, N.~G., {Ibata}, R.~A., {Famaey}, B., \& {Lewis}, G.~F. 2014,
  \href{http://dx.doi.org/10.1038/nature13481}{\nat, 511, 563}

\bibitem[{{Jester} {et~al.}(2005){Jester}, {Schneider}, {Richards}, {Green},
  {Schmidt}, {Hall}, {Strauss}, {Vand en Berk}, {Stoughton}, {Gunn},
  {Brinkmann}, {Kent}, {Smith}, {Tucker}, \& {Yanny}}]{jester05}
{Jester}, S., {Schneider}, D.~P., {Richards}, G.~T., {et~al.} 2005,
  \href{http://dx.doi.org/10.1086/432466}{\aj, 130, 873}

\bibitem[{{Jiang} {et~al.}(2020){Jiang}, {Dekel}, {Freundlich}, {van den
  Bosch}, {Green}, {Hopkins}, {Benson}, \& {Du}}]{Jiang:2005.05974}
{Jiang}, F., {Dekel}, A., {Freundlich}, J., {et~al.} 2020,
  \href{http://arxiv.org/abs/2005.05974}{arXiv:2005.05974}

\bibitem[{{Jones} {et~al.}(2009){Jones}, {Read}, {Saunders}, {Colless},
  {Jarrett}, {Parker}, {Fairall}, {Mauch}, {Sadler}, {Watson}, {Burton},
  {Campbell}, {Cass}, {Croom}, {Dawe}, {Fiegert}, {Frankcombe}, {Hartley},
  {Huchra}, {James}, {Kirby}, {Lahav}, {Lucey}, {Mamon}, {Moore}, {Peterson},
  {Prior}, {Proust}, {Russell}, {Safouris}, {Wakamatsu}, {Westra}, \&
  {Williams}}]{6dF}
{Jones}, D.~H., {Read}, M.~A., {Saunders}, W., {et~al.} 2009,
  \href{http://dx.doi.org/10.1111/j.1365-2966.2009.15338.x}{\mnras, 399, 683}

\bibitem[{Jones {et~al.}(2001)Jones, Oliphant, Peterson, {et~al.}}]{scipy}
Jones, E., Oliphant, T., Peterson, P., {et~al.} 2001, {SciPy}: Open source
  scientific tools for {Python}.
\newblock \url{http://www.scipy.org/}

\bibitem[{King \& Zeng(2001)}]{King2001}
King, G. \& Zeng, L. 2001,
  \href{http://dx.doi.org/10.1093/oxfordjournals.pan.a004868}{Political
  Analysis, 9, 137}

\bibitem[{{Kirby} {et~al.}(2013){Kirby}, {Cohen}, {Guhathakurta}, {Cheng},
  {Bullock}, \& {Gallazzi}}]{Kirby2013}
{Kirby}, E.~N., {Cohen}, J.~G., {Guhathakurta}, P., {et~al.} 2013,
  \href{http://dx.doi.org/10.1088/0004-637X/779/2/102}{\apj, 779, 102}

\bibitem[{Kluyver {et~al.}(2016)Kluyver, Ragan-Kelley, P{\'e}rez, Granger,
  Bussonnier, Frederic, Kelley, Hamrick, Grout, Corlay, Ivanov, Avila, Abdalla,
  Willing, \& development team}]{jupyter}
Kluyver, T., Ragan-Kelley, B., P{\'e}rez, F., {et~al.} 2016, in Positioning and
  Power in Academic Publishing: Players, Agents and Agendas, ed. F.~Loizides \&
  B.~Scmidt (Netherlands: IOS Press), 87--90.
\newblock \url{https://eprints.soton.ac.uk/403913/}

\bibitem[{{Kondapally} {et~al.}(2018){Kondapally}, {Russell}, {Conselice}, \&
  {Penny}}]{Kondapally2018}
{Kondapally}, R., {Russell}, G.~A., {Conselice}, C.~J., \& {Penny}, S.~J. 2018,
  \href{http://dx.doi.org/10.1093/mnras/sty2333}{\mnras, 481, 1759}

\bibitem[{{Kravtsov}(2013)}]{Kravtsov12122980}
{Kravtsov}, A.~V. 2013,
  \href{http://dx.doi.org/10.1088/2041-8205/764/2/L31}{\apjl, 764, L31}

\bibitem[{{Lazo} {et~al.}(2018){Lazo}, {Zahid}, {Sohn}, \&
  {Geller}}]{2018RNAAS...2..234L}
{Lazo}, B., {Zahid}, H.~J., {Sohn}, J., \& {Geller}, M.~J. 2018,
  \href{http://dx.doi.org/10.3847/2515-5172/aaf8b1}{Research Notes of the
  American Astronomical Society, 2, 234}

\bibitem[{{Leroy} {et~al.}(2019){Leroy}, {Sandstrom}, {Lang}, {Lewis}, {Salim},
  {Behrens}, {Chastenet}, {Chiang}, {Gallagher}, {Kessler}, \&
  {Utomo}}]{2019ApJS..244...24L}
{Leroy}, A.~K., {Sandstrom}, K.~M., {Lang}, D., {et~al.} 2019,
  \href{http://dx.doi.org/10.3847/1538-4365/ab3925}{\apjs, 244, 24}

\bibitem[{{Li} {et~al.}(2019){Li}, {Koposov}, {Zucker}, {Lewis}, {Kuehn},
  {Simpson}, {Ji}, {Shipp}, {Mao}, {Geha}, {Pace}, {Mackey}, {Allam}, {Tucker},
  {Da Costa}, {Erkal}, {Simon}, {Mould}, {Martell}, {Wan}, {De Silva},
  {Bechtol}, {Balbinot}, {Belokurov}, {Bland-Hawthorn}, {Casey}, {Cullinane},
  {Drlica-Wagner}, {Sharma}, {Vivas}, {Wechsler}, {Yanny}, \& {S5
  Collaboration}}]{10.1093/mnras/stz2731}
{Li}, T.~S., {Koposov}, S.~E., {Zucker}, D.~B., {et~al.} 2019,
  \href{http://dx.doi.org/10.1093/mnras/stz2731}{\mnras, 490, 3508}

\bibitem[{{Lidman} {et~al.}(2020){Lidman}, {Tucker}, {Davis}, {Uddin},
  {Asorey}, {Bolejko}, {Brout}, {Calcino}, {Carollo}, {Carr}, {Childress},
  {Hoormann}, {Foley}, {Galbany}, {Glazebrook}, {Hinton}, {Kessler}, {Kim},
  {King}, {Kremin}, {Kuehn}, {Lagattuta}, {Lewis}, {Macaulay}, {Malik},
  {March}, {Martini}, {M{\"o}ller}, {Mudd}, {Nichol}, {Panther}, {Parkinson},
  {Pursiainen}, {Sako}, {Swann}, {Scalzo}, {Scolnic}, {Sharp}, {Smith},
  {Sommer}, {Sullivan}, {Webb}, {Wiseman}, {Yu}, {Yuan}, {Zhang}, {Abbott},
  {Aguena}, {Allam}, {Annis}, {Avila}, {Bertin}, {Bhargava}, {Brooks}, {Carnero
  Rosell}, {Carrasco Kind}, {Carretero}, {Castander}, {Costanzi}, {da Costa},
  {De Vicente}, {Doel}, {Eifler}, {Everett}, {Fosalba}, {Frieman},
  {Garc{\'\i}a-Bellido}, {Gaztanaga}, {Gruen}, {Gruendl}, {Gschwend},
  {Gutierrez}, {Hartley}, {Hollowood}, {Honscheid}, {James}, {Kuropatkin},
  {Li}, {Lima}, {Lin}, {Maia}, {Marshall}, {Melchior}, {Menanteau}, {Miquel},
  {Palmese}, {Paz-Chinch{\'o}n}, {Plazas}, {Roodman}, {Rykoff}, {Sanchez},
  {Santiago}, {Scarpine}, {Schubnell}, {Serrano}, {Sevilla-Noarbe}, {Suchyta},
  {Swanson}, {Tarle}, {Tucker}, {Varga}, {Walker}, {Wester}, {Wilkinson}, \&
  {DES Collaboration}}]{OzDES}
{Lidman}, C., {Tucker}, B.~E., {Davis}, T.~M., {et~al.} 2020,
  \href{http://dx.doi.org/10.1093/mnras/staa1341}{\mnras, 496, 19}

\bibitem[{{Lim} {et~al.}(2017){Lim}, {Mo}, {Lu}, {Wang}, \&
  {Yang}}]{2017MNRAS.470.2982L}
{Lim}, S.~H., {Mo}, H.~J., {Lu}, Y., {Wang}, H., \& {Yang}, X. 2017,
  \href{http://dx.doi.org/10.1093/mnras/stx1462}{\mnras, 470, 2982}

\bibitem[{{Liu} {et~al.}(2011){Liu}, {Gerke}, {Wechsler}, {Behroozi}, \&
  {Busha}}]{Liu2011}
{Liu}, L., {Gerke}, B.~F., {Wechsler}, R.~H., {Behroozi}, P.~S., \& {Busha},
  M.~T. 2011, \href{http://dx.doi.org/10.1088/0004-637X/733/1/62}{\apj, 733,
  62}

\bibitem[{{Loveday} {et~al.}(2015){Loveday}, {Norberg}, {Baldry}, {Bland
  -Hawthorn}, {Brough}, {Brown}, {Driver}, {Kelvin}, \&
  {Phillipps}}]{Loveday150501003}
{Loveday}, J., {Norberg}, P., {Baldry}, I.~K., {et~al.} 2015,
  \href{http://dx.doi.org/10.1093/mnras/stv1013}{\mnras, 451, 1540}

\bibitem[{{Macci{\`o}} {et~al.}(2020){Macci{\`o}}, {Courteau}, {Ouellette}, \&
  {Dutton}}]{Maccio200600818}
{Macci{\`o}}, A.~V., {Courteau}, S., {Ouellette}, N. N.~Q., \& {Dutton}, A.~A.
  2020, \href{http://dx.doi.org/10.1093/mnrasl/slaa094}{\mnras, 496, L101}

\bibitem[{{Makarov} {et~al.}(2014){Makarov}, {Prugniel}, {Terekhova},
  {Courtois}, \& {Vauglin}}]{Makarov2014}
{Makarov}, D., {Prugniel}, P., {Terekhova}, N., {Courtois}, H., \& {Vauglin},
  I. 2014, \href{http://dx.doi.org/10.1051/0004-6361/201423496}{\aap, 570, A13}

\bibitem[{{Mao} {et~al.}(2015){Mao}, {Williamson}, \&
  {Wechsler}}]{Mao150302637}
{Mao}, Y.-Y., {Williamson}, M., \& {Wechsler}, R.~H. 2015,
  \href{http://dx.doi.org/10.1088/0004-637X/810/1/21}{\apj, 810, 21}

\bibitem[{{McConnachie}(2012)}]{mcconnachie12}
{McConnachie}, A.~W. 2012,
  \href{http://dx.doi.org/10.1088/0004-6256/144/1/4}{\aj, 144, 4}

\bibitem[{{McConnachie} {et~al.}(2016){McConnachie}, {Babusiaux}, {Balogh},
  {Caffau}, {C{\^o}t{\'e}}, {Driver}, {Robotham}, {Starkenburg}, {Venn},
  {Walker}, {Bauman}, {Flagey}, {Ho}, {Isani}, {Laychak}, {Mignot},
  {Murowinski}, {Salmon}, {Simons}, {Szeto}, {Vermeulen}, \&
  {Withington}}]{2016arXiv160600060M}
{McConnachie}, A.~W., {Babusiaux}, C., {Balogh}, M., {et~al.} 2016,
  \href{http://arxiv.org/abs/1606.00060}{arXiv:1606.00060}

\bibitem[{McLeod {et~al.}(2018)McLeod, Alted, Valentino, de~Menten, Wiebe,
  cgohlke, Bedini, mamrehn, anatoly techtonik, Erb, Shadchin, Bunin, Kooij,
  Pavlyk, Jelloul, Garrison, Hurtado, Carey, Sarahan, Cox, Plesivčak,
  Borgdorff, Courbet, Dickinson, Leitao, de~Laat, Pitrou, Portnoy, Ortega, \&
  Böhn}]{numexpr}
McLeod, R., Alted, F., Valentino, A., {et~al.} 2018, pydata/numexpr: NumExpr
  v2.6.9, v2.6.9,  Zenodo, \dodoi{10.5281/zenodo.2483274}

\bibitem[{{More} {et~al.}(2015){More}, {Diemer}, \& {Kravtsov}}]{More150405591}
{More}, S., {Diemer}, B., \& {Kravtsov}, A.~V. 2015,
  \href{http://dx.doi.org/10.1088/0004-637X/810/1/36}{\apj, 810, 36}

\bibitem[{{Morris} {et~al.}(2018){Morris}, {Tollerud}, {Sip{\H o}cz}, {Deil},
  {Douglas}, {Berlanga Medina}, {Vyhmeister}, {Smith}, {Littlefair},
  {Price-Whelan}, {Gee}, \& {Jeschke}}]{astroplan2018}
{Morris}, B.~M., {Tollerud}, E., {Sip{\H o}cz}, B., {et~al.} 2018,
  \href{http://dx.doi.org/10.3847/1538-3881/aaa47e}{\aj, 155, 128}

\bibitem[{{M{\"u}ller} {et~al.}(2019){M{\"u}ller}, {Rejkuba}, {Pawlowski},
  {Ibata}, {Lelli}, {Hilker}, \& {Jerjen}}]{Muller:1907.02012}
{M{\"u}ller}, O., {Rejkuba}, M., {Pawlowski}, M.~S., {et~al.} 2019,
  \href{http://dx.doi.org/10.1051/0004-6361/201935807}{\aap, 629, A18}

\bibitem[{{Munshi} {et~al.}(2019){Munshi}, {Brooks}, {Christensen},
  {Applebaum}, {Holley-Bockelmann}, {Quinn}, \& {Wadsley}}]{Munshi2019}
{Munshi}, F., {Brooks}, A.~M., {Christensen}, C., {et~al.} 2019,
  \href{http://dx.doi.org/10.3847/1538-4357/ab0085}{\apj, 874, 40}

\bibitem[{{Nadler} {et~al.}(2019{\natexlab{a}}){Nadler}, {Gluscevic}, {Boddy},
  \& {Wechsler}}]{Nadler190410000}
{Nadler}, E.~O., {Gluscevic}, V., {Boddy}, K.~K., \& {Wechsler}, R.~H.
  2019{\natexlab{a}}, \href{http://dx.doi.org/10.3847/2041-8213/ab1eb2}{\apjl,
  878, L32}

\bibitem[{{Nadler} {et~al.}(2019{\natexlab{b}}){Nadler}, {Mao}, {Green}, \&
  {Wechsler}}]{Nadler180905542}
{Nadler}, E.~O., {Mao}, Y.-Y., {Green}, G.~M., \& {Wechsler}, R.~H.
  2019{\natexlab{b}}, \href{http://dx.doi.org/10.3847/1538-4357/ab040e}{\apj,
  873, 34}

\bibitem[{{Nadler} {et~al.}(2018){Nadler}, {Mao}, {Wechsler},
  {Garrison-Kimmel}, \& {Wetzel}}]{Nadler171204467}
{Nadler}, E.~O., {Mao}, Y.-Y., {Wechsler}, R.~H., {Garrison-Kimmel}, S., \&
  {Wetzel}, A. 2018, \href{http://dx.doi.org/10.3847/1538-4357/aac266}{\apj,
  859, 129}

\bibitem[{{Nadler} {et~al.}(2020{\natexlab{a}}){Nadler}, {Wechsler}, {Bechtol},
  {Mao}, {Green}, {Drlica-Wagner}, {McNanna}, {Mau}, {Pace}, {Simon},
  {Kravtsov}, {Dodelson}, {Li}, {Riley}, {Wang}, {Abbott}, {Aguena}, {Allam},
  {Annis}, {Avila}, {Bernstein}, {Bertin}, {Brooks}, {Burke}, {Rosell}, {Kind},
  {Carretero}, {Costanzi}, {da Costa}, {De Vicente}, {Desai}, {Evrard},
  {Flaugher}, {Fosalba}, {Frieman}, {Garc{\'\i}a-Bellido}, {Gaztanaga},
  {Gerdes}, {Gruen}, {Gschwend}, {Gutierrez}, {Hartley}, {Hinton}, {Honscheid},
  {Krause}, {Kuehn}, {Kuropatkin}, {Lahav}, {Maia}, {Marshall}, {Menanteau},
  {Miquel}, {Palmese}, {Paz-Chinch{\'o}n}, {Plazas}, {Romer}, {Sanchez},
  {Santiago}, {Scarpine}, {Serrano}, {Smith}, {Soares-Santos}, {Suchyta},
  {Tarle}, {Thomas}, {Varga}, {Walker}, \& {DES
  Collaboration}}]{Nadler191203303}
{Nadler}, E.~O., {Wechsler}, R.~H., {Bechtol}, K., {et~al.} 2020{\natexlab{a}},
  \href{http://dx.doi.org/10.3847/1538-4357/ab846a}{\apj, 893, 48}

\bibitem[{{Nadler} {et~al.}(2020{\natexlab{b}}){Nadler}, {Drlica-Wagner},
  {Bechtol}, {Mau}, {Wechsler}, {Gluscevic}, {Boddy}, {Pace}, {Li}, {McNanna},
  {Riley}, {Garc{\'\i}a-Bellido}, {Mao}, {Green}, {Burke}, {Peter}, {Jain},
  {Abbott}, {Aguena}, {Allam}, {Annis}, {Avila}, {Brooks}, {Carrasco Kind},
  {Carretero}, {Costanzi}, {da Costa}, {De Vicente}, {Desai}, {Diehl}, {Doel},
  {Everett}, {Evrard}, {Flaugher}, {Frieman}, {Gerdes}, {Gruen}, {Gruendl},
  {Gschwend}, {Gutierrez}, {Hinton}, {Honscheid}, {Huterer}, {James}, {Krause},
  {Kuehn}, {Kuropatkin}, {Lahav}, {Maia}, {Marshall}, {Menanteau}, {Miquel},
  {Palmese}, {Paz-Chinch{\'o}n}, {Plazas}, {Romer}, {Sanchez}, {Scarpine},
  {Serrano}, {Sevilla-Noarbe}, {Smith}, {Soares-Santos}, {Suchyta}, {Swanson},
  {Tarle}, {Tucker}, {Walker}, \& {Wester}}]{Nadler:2008.00022}
{Nadler}, E.~O., {Drlica-Wagner}, A., {Bechtol}, K., {et~al.}
  2020{\natexlab{b}}, \href{http://arxiv.org/abs/2008.00022}{arXiv:2008.00022}

\bibitem[{{Navarro} {et~al.}(1996){Navarro}, {Frenk}, \& {White}}]{nfw1996}
{Navarro}, J.~F., {Frenk}, C.~S., \& {White}, S. D.~M. 1996,
  \href{http://dx.doi.org/10.1086/177173}{\apj, 462, 563}

\bibitem[{{Navarro} {et~al.}(1997){Navarro}, {Frenk}, \& {White}}]{nfw1997}
---. 1997, \href{http://dx.doi.org/10.1086/304888}{\apj, 490, 493}

\bibitem[{{Neuzil} {et~al.}(2020){Neuzil}, {Mansfield}, \&
  {Kravtsov}}]{Neuzil191204307}
{Neuzil}, M.~K., {Mansfield}, P., \& {Kravtsov}, A.~V. 2020,
  \href{http://dx.doi.org/10.1093/mnras/staa898}{\mnras, 494, 2600}

\bibitem[{{Newman} {et~al.}(2013){Newman}, {Cooper}, {Davis}, {Faber}, {Coil},
  {Guhathakurta}, {Koo}, {Phillips}, {Conroy}, {Dutton}, {Finkbeiner}, {Gerke},
  {Rosario}, {Weiner}, {Willmer}, {Yan}, {Harker}, {Kassin}, {Konidaris},
  {Lai}, {Madgwick}, {Noeske}, {Wirth}, {Connolly}, {Kaiser}, {Kirby},
  {Lemaux}, {Lin}, {Lotz}, {Luppino}, {Marinoni}, {Matthews}, {Metevier}, \&
  {Schiavon}}]{Newman2013}
{Newman}, J.~A., {Cooper}, M.~C., {Davis}, M., {et~al.} 2013,
  \href{http://dx.doi.org/10.1088/0067-0049/208/1/5}{\apjs, 208, 5}

\bibitem[{{Nierenberg} {et~al.}(2012){Nierenberg}, {Auger}, {Treu}, {Marshall},
  {Fassnacht}, \& {Busha}}]{Nierenberg2012}
{Nierenberg}, A.~M., {Auger}, M.~W., {Treu}, T., {et~al.} 2012,
  \href{http://dx.doi.org/10.1088/0004-637X/752/2/99}{\apj, 752, 99}

\bibitem[{{Oke} \& {Gunn}(1982)}]{oke1982}
{Oke}, J.~B. \& {Gunn}, J.~E. 1982,
  \href{http://dx.doi.org/10.1086/131027}{\pasp, 94, 586}

\bibitem[{{Pawlowski}(2018)}]{Pawlowski2018}
{Pawlowski}, M.~S. 2018,
  \href{http://dx.doi.org/10.1142/S0217732318300045}{Modern Physics Letters A,
  33, 1830004}

\bibitem[{{Pawlowski} {et~al.}(2013){Pawlowski}, {Kroupa}, \&
  {Jerjen}}]{1307.6210}
{Pawlowski}, M.~S., {Kroupa}, P., \& {Jerjen}, H. 2013,
  \href{http://dx.doi.org/10.1093/mnras/stt1384}{\mnras, 435, 1928}

\bibitem[{{Pawlowski} {et~al.}(2012){Pawlowski}, {Pflamm-Altenburg}, \&
  {Kroupa}}]{1204.5176}
{Pawlowski}, M.~S., {Pflamm-Altenburg}, J., \& {Kroupa}, P. 2012,
  \href{http://dx.doi.org/10.1111/j.1365-2966.2012.20937.x}{\mnras, 423, 1109}

\bibitem[{{Perez} \& {Granger}(2007)}]{ipython}
{Perez}, F. \& {Granger}, B.~E. 2007,
  \href{http://dx.doi.org/10.1109/MCSE.2007.53}{Computing in Science and
  Engineering, 9, 21}

\bibitem[{{Phillips} {et~al.}(2015){Phillips}, {Cooper}, {Bullock}, \&
  {Boylan-Kolchin}}]{Phillips2015}
{Phillips}, J.~I., {Cooper}, M.~C., {Bullock}, J.~S., \& {Boylan-Kolchin}, M.
  2015, \href{http://dx.doi.org/10.1093/mnras/stv1770}{\mnras, 453, 3839}

\bibitem[{{Pillepich} {et~al.}(2018){Pillepich}, {Springel}, {Nelson}, {Genel},
  {Naiman}, {Pakmor}, {Hernquist}, {Torrey}, {Vogelsberger}, {Weinberger}, \&
  {Marinacci}}]{Pillepich2018}
{Pillepich}, A., {Springel}, V., {Nelson}, D., {et~al.} 2018,
  \href{http://dx.doi.org/10.1093/mnras/stx2656}{\mnras, 473, 4077}

\bibitem[{{Polisensky} \& {Ricotti}(2014)}]{Polisensky2014}
{Polisensky}, E. \& {Ricotti}, M. 2014,
  \href{http://dx.doi.org/10.1093/mnras/stt2105}{\mnras, 437, 2922}

\bibitem[{{Read} \& {Erkal}(2019)}]{Read2019}
{Read}, J.~I. \& {Erkal}, D. 2019,
  \href{http://dx.doi.org/10.1093/mnras/stz1320}{\mnras, 487, 5799}

\bibitem[{{Reddick} {et~al.}(2013){Reddick}, {Wechsler}, {Tinker}, \&
  {Behroozi}}]{Reddick2013}
{Reddick}, R.~M., {Wechsler}, R.~H., {Tinker}, J.~L., \& {Behroozi}, P.~S.
  2013, \href{http://dx.doi.org/10.1088/0004-637X/771/1/30}{\apj, 771, 30}

\bibitem[{{Roberts} {et~al.}(2020){Roberts}, {Nierenberg}, \&
  {Peter}}]{2008.05479}
{Roberts}, D.~M., {Nierenberg}, A.~M., \& {Peter}, A. H.~G. 2020,
  \href{http://arxiv.org/abs/2008.05479}{arXiv:2008.05479}

\bibitem[{{Roll} {et~al.}(1998){Roll}, {Fabricant}, \&
  {McLeod}}]{10.1117/12.316837}
{Roll}, J.~B., {Fabricant}, D.~G., \& {McLeod}, B.~A. 1998, in Society of
  Photo-Optical Instrumentation Engineers (SPIE) Conference Series, Vol. 3355,
  Optical Astronomical Instrumentation, ed. S.~{D'Odorico}, 324--332,
  \dodoi{10.1117/12.316837}

\bibitem[{{Sales} {et~al.}(2013){Sales}, {Wang}, {White}, \&
  {Navarro}}]{Sales2013}
{Sales}, L.~V., {Wang}, W., {White}, S. D.~M., \& {Navarro}, J.~F. 2013,
  \href{http://dx.doi.org/10.1093/mnras/sts054}{\mnras, 428, 573}

\bibitem[{{Samuel} {et~al.}(2020){Samuel}, {Wetzel}, {Tollerud},
  {Garrison-Kimmel}, {Loebman}, {El-Badry}, {Hopkins}, {Boylan-Kolchin},
  {Faucher-Gigu{\`e}re}, {Bullock}, {Benincasa}, \& {Bailin}}]{Samuel2019a}
{Samuel}, J., {Wetzel}, A., {Tollerud}, E., {et~al.} 2020,
  \href{http://dx.doi.org/10.1093/mnras/stz3054}{\mnras, 491, 1471}

\bibitem[{{Schaye} {et~al.}(2015){Schaye}, {Crain}, {Bower}, {Furlong},
  {Schaller}, {Theuns}, {Dalla Vecchia}, {Frenk}, {McCarthy}, {Helly},
  {Jenkins}, {Rosas-Guevara}, {White}, {Baes}, {Booth}, {Camps}, {Navarro},
  {Qu}, {Rahmati}, {Sawala}, {Thomas}, \& {Trayford}}]{Schaye2015}
{Schaye}, J., {Crain}, R.~A., {Bower}, R.~G., {et~al.} 2015,
  \href{http://dx.doi.org/10.1093/mnras/stu2058}{\mnras, 446, 521}

\bibitem[{{Schlafly} \& {Finkbeiner}(2011)}]{Schlafly2011}
{Schlafly}, E.~F. \& {Finkbeiner}, D.~P. 2011,
  \href{http://dx.doi.org/10.1088/0004-637X/737/2/103}{\apj, 737, 103}

\bibitem[{{Schlegel} {et~al.}(1998){Schlegel}, {Finkbeiner}, \&
  {Davis}}]{schlegel98}
{Schlegel}, D.~J., {Finkbeiner}, D.~P., \& {Davis}, M. 1998,
  \href{http://dx.doi.org/10.1086/305772}{\apj, 500, 525}

\bibitem[{Seabold \& Perktold(2010)}]{statsmodels}
Seabold, S. \& Perktold, J. 2010, in 9th Python in Science Conference

\bibitem[{{Shectman} {et~al.}(1996){Shectman}, {Landy}, {Oemler}, {Tucker},
  {Lin}, {Kirshner}, \& {Schechter}}]{1996ApJ...470..172S}
{Shectman}, S.~A., {Landy}, S.~D., {Oemler}, A., {et~al.} 1996,
  \href{http://dx.doi.org/10.1086/177858}{\apj, 470, 172}

\bibitem[{{Simpson} {et~al.}(2018){Simpson}, {Grand}, {G{\'o}mez}, {Marinacci},
  {Pakmor}, {Springel}, {Campbell}, \& {Frenk}}]{Simpson17:1705.03018}
{Simpson}, C.~M., {Grand}, R. J.~J., {G{\'o}mez}, F.~A., {et~al.} 2018,
  \href{http://dx.doi.org/10.1093/mnras/sty774}{\mnras, 478, 548}

\bibitem[{{Sinha} \& {Holley-Bockelmann}(2012)}]{Sinha11031675}
{Sinha}, M. \& {Holley-Bockelmann}, K. 2012,
  \href{http://dx.doi.org/10.1088/0004-637X/751/1/17}{\apj, 751, 17}

\bibitem[{{Smercina} {et~al.}(2018){Smercina}, {Bell}, {Price}, {D'Souza},
  {Slater}, {Bailin}, {Monachesi}, \& {Nidever}}]{Smercina2018}
{Smercina}, A., {Bell}, E.~F., {Price}, P.~A., {et~al.} 2018,
  \href{http://dx.doi.org/10.3847/1538-4357/aad2d6}{\apj, 863, 152}

\bibitem[{{Somerville} \& {Dav{\'e}}(2015)}]{2015ARA&A..53...51S}
{Somerville}, R.~S. \& {Dav{\'e}}, R. 2015,
  \href{http://dx.doi.org/10.1146/annurev-astro-082812-140951}{\araa, 53, 51}

\bibitem[{{Spencer} {et~al.}(2014){Spencer}, {Loebman}, \&
  {Yoachim}}]{Spencer2014}
{Spencer}, M., {Loebman}, S., \& {Yoachim}, P. 2014,
  \href{http://dx.doi.org/10.1088/0004-637X/788/2/146}{\apj, 788, 146}

\bibitem[{{Strauss} {et~al.}(2002){Strauss}, {Weinberg}, {Lupton}, {Narayanan},
  {Annis}, {Bernardi}, {Blanton}, {Burles}, {Connolly}, {Dalcanton}, {Doi},
  {Eisenstein}, {Frieman}, {Fukugita}, {Gunn}, {Ivezi{\'c}}, {Kent}, {Kim},
  {Knapp}, {Kron}, {Munn}, {Newberg}, {Nichol}, {Okamura}, {Quinn}, {Richmond},
  {Schlegel}, {Shimasaku}, {SubbaRao}, {Szalay}, {Vanden Berk}, {Vogeley},
  {Yanny}, {Yasuda}, {York}, \& {Zehavi}}]{Strauss2002}
{Strauss}, M.~A., {Weinberg}, D.~H., {Lupton}, R.~H., {et~al.} 2002,
  \href{http://dx.doi.org/10.1086/342343}{\aj, 124, 1810}

\bibitem[{{Tanaka} {et~al.}(2018){Tanaka}, {Chiba}, {Hayashi}, {Komiyama},
  {Okamoto}, {Cooper}, {Okamoto}, \& {Spitler}}]{Tanaka2018}
{Tanaka}, M., {Chiba}, M., {Hayashi}, K., {et~al.} 2018,
  \href{http://dx.doi.org/10.3847/1538-4357/aad9fe}{\apj, 865, 125}

\bibitem[{{Tanoglidis} {et~al.}(2020){Tanoglidis}, {Drlica-Wagner}, {Wei},
  {Li}, {S{\'a}nchez}, {Zhang}, {Peter}, {Feldmeier-Krause}, {Prat}, {Casey},
  {Palmese}, {S{\'a}nchez}, {DeRose}, {Conselice}, {Abbott}, {Aguena}, {Allam},
  {Avila}, {Bechtol}, {Bertin}, {Bhargava}, {Brooks}, {Burke}, {Carnero
  Rosell}, {Carrasco Kind}, {Carretero}, {Chang}, {Costanzi}, {da Costa}, {De
  Vicente}, {Desai}, {Diehl}, {Doel}, {Eifler}, {Everett}, {Evrard},
  {Flaugher}, {Frieman}, {Garc{\'\i}a-Bellido}, {Gerdes}, {Gruendl},
  {Gschwend}, {Gutierrez}, {Hartley}, {Hollowood}, {Huterer}, {James},
  {Krause}, {Kuehn}, {Kuropatkin}, {Maia}, {March}, {Marshall}, {Menanteau},
  {Miquel}, {Ogand o}, {Paz-Chinch{\'o}n}, {Romer}, {Roodman}, {Sanchez},
  {Scarpine}, {Serrano}, {Sevilla-Noarbe}, {Smith}, {Suchyta}, {Tarle},
  {Thomas}, {Tucker}, \& {Walker}}]{Tanoglidis200604294}
{Tanoglidis}, D., {Drlica-Wagner}, A., {Wei}, K., {et~al.} 2020,
  \href{http://arxiv.org/abs/2006.04294}{arXiv:2006.04294}

\bibitem[{{Taylor} {et~al.}(2011){Taylor}, {Hopkins}, {Baldry}, {Brown},
  {Driver}, {Kelvin}, {Hill}, {Robotham}, {Bland -Hawthorn}, {Jones}, {Sharp},
  {Thomas}, {Liske}, {Loveday}, {Norberg}, {Peacock}, {Bamford}, {Brough},
  {Colless}, {Cameron}, {Conselice}, {Croom}, {Frenk}, {Gunawardhana},
  {Kuijken}, {Nichol}, {Parkinson}, {Phillipps}, {Pimbblet}, {Popescu},
  {Prescott}, {Sutherland}, {Tuffs}, {van Kampen}, \&
  {Wijesinghe}}]{Taylor2011}
{Taylor}, E.~N., {Hopkins}, A.~M., {Baldry}, I.~K., {et~al.} 2011,
  \href{http://dx.doi.org/10.1111/j.1365-2966.2011.19536.x}{\mnras, 418, 1587}

\bibitem[{{Tollerud} {et~al.}(2011){Tollerud}, {Boylan-Kolchin}, {Barton},
  {Bullock}, \& {Trinh}}]{Tollerud2011}
{Tollerud}, E.~J., {Boylan-Kolchin}, M., {Barton}, E.~J., {Bullock}, J.~S., \&
  {Trinh}, C.~Q. 2011,
  \href{http://dx.doi.org/10.1088/0004-637X/738/1/102}{\apj, 738, 102}

\bibitem[{{Tollerud} {et~al.}(2014){Tollerud}, {Boylan-Kolchin}, \&
  {Bullock}}]{Tollerud2014}
{Tollerud}, E.~J., {Boylan-Kolchin}, M., \& {Bullock}, J.~S. 2014,
  \href{http://dx.doi.org/10.1093/mnras/stu474}{\mnras, 440, 3511}

\bibitem[{{Tully} {et~al.}(2015){Tully}, {Libeskind}, {Karachentsev},
  {Karachentseva}, {Rizzi}, \& {Shaya}}]{1503.05599}
{Tully}, R.~B., {Libeskind}, N.~I., {Karachentsev}, I.~D., {et~al.} 2015,
  \href{http://dx.doi.org/10.1088/2041-8205/802/2/L25}{\apjl, 802, L25}

\bibitem[{{Tully} {et~al.}(2009){Tully}, {Rizzi}, {Shaya}, {Courtois},
  {Makarov}, \& {Jacobs}}]{2009AJ....138..323T}
{Tully}, R.~B., {Rizzi}, L., {Shaya}, E.~J., {et~al.} 2009,
  \href{http://dx.doi.org/10.1088/0004-6256/138/2/323}{\aj, 138, 323}

\bibitem[{{van der Walt} {et~al.}(2011){van der Walt}, {Colbert}, \&
  {Varoquaux}}]{numpy}
{van der Walt}, S., {Colbert}, S.~C., \& {Varoquaux}, G. 2011,
  \href{http://dx.doi.org/10.1109/MCSE.2011.37}{Computing in Science and
  Engineering, 13, 22}

\bibitem[{{van Dokkum} {et~al.}(2015){van Dokkum}, {Abraham}, {Merritt},
  {Zhang}, {Geha}, \& {Conroy}}]{vandokkum2015}
{van Dokkum}, P.~G., {Abraham}, R., {Merritt}, A., {et~al.} 2015,
  \href{http://dx.doi.org/10.1088/2041-8205/798/2/L45}{\apjl, 798, L45}

\bibitem[{{van Leeuwen}(2007)}]{Hipparcos}
{van Leeuwen}, F. 2007,
  \href{http://dx.doi.org/10.1051/0004-6361:20078357}{\aap, 474, 653}

\bibitem[{{Virtanen} {et~al.}(2020){Virtanen}, {Gommers}, {Oliphant},
  {Haberland}, {Reddy}, {Cournapeau}, {Burovski}, {Peterson}, {Weckesser},
  {Bright}, {van der Walt}, {Brett}, {Wilson}, {Millman}, {Mayorov}, {Nelson},
  {Jones}, {Kern}, {Larson}, {Carey}, {Polat}, {Feng}, {Moore}, {VanderPlas},
  {Laxalde}, {Perktold}, {Cimrman}, {Henriksen}, {Quintero}, {Harris},
  {Archibald}, {Ribeiro}, {Pedregosa}, {van Mulbregt}, \& {SciPy 1. 0
  Contributors}}]{2020SciPy-NMeth}
{Virtanen}, P., {Gommers}, R., {Oliphant}, T.~E., {et~al.} 2020,
  \href{http://dx.doi.org/10.1038/s41592-019-0686-2}{Nature Methods, 17, 261}

\bibitem[{{Wechsler} \& {Tinker}(2018)}]{WechslerTinker}
{Wechsler}, R.~H. \& {Tinker}, J.~L. 2018,
  \href{http://dx.doi.org/10.1146/annurev-astro-081817-051756}{\araa, 56, 435}

\bibitem[{{Weinberg} {et~al.}(2015){Weinberg}, {Bullock}, {Governato}, {Kuzio
  de Naray}, \& {Peter}}]{Weinberg2015}
{Weinberg}, D.~H., {Bullock}, J.~S., {Governato}, F., {Kuzio de Naray}, R., \&
  {Peter}, A. H.~G. 2015,
  \href{http://dx.doi.org/10.1073/pnas.1308716112}{Proceedings of the National
  Academy of Science, 112, 12249}

\bibitem[{{Weisz} {et~al.}(2012){Weisz}, {Johnson}, {Johnson}, {Skillman},
  {Lee}, {Kennicutt}, {Calzetti}, {van Zee}, {Bothwell}, {Dalcanton}, {Dale},
  \& {Williams}}]{Weisz2012}
{Weisz}, D.~R., {Johnson}, B.~D., {Johnson}, L.~C., {et~al.} 2012,
  \href{http://dx.doi.org/10.1088/0004-637X/744/1/44}{\apj, 744, 44}

\bibitem[{{Wetzel} {et~al.}(2016){Wetzel}, {Hopkins}, {Kim},
  {Faucher-Gigu{\`e}re}, {Kere{\v{s}}}, \& {Quataert}}]{Wetzel2016}
{Wetzel}, A.~R., {Hopkins}, P.~F., {Kim}, J.-h., {et~al.} 2016,
  \href{http://dx.doi.org/10.3847/2041-8205/827/2/L23}{\apjl, 827, L23}

\bibitem[{{Wetzel} {et~al.}(2015){Wetzel}, {Tollerud}, \&
  {Weisz}}]{Wetzel:1503.06799}
{Wetzel}, A.~R., {Tollerud}, E.~J., \& {Weisz}, D.~R. 2015,
  \href{http://dx.doi.org/10.1088/2041-8205/808/1/L27}{\apjl, 808, L27}

\bibitem[{{Wheeler} {et~al.}(2014){Wheeler}, {Phillips}, {Cooper},
  {Boylan-Kolchin}, \& {Bullock}}]{1402.1498}
{Wheeler}, C., {Phillips}, J.~I., {Cooper}, M.~C., {Boylan-Kolchin}, M., \&
  {Bullock}, J.~S. 2014, \href{http://dx.doi.org/10.1093/mnras/stu965}{\mnras,
  442, 1396}

\bibitem[{{Wheeler} {et~al.}(2019){Wheeler}, {Hopkins}, {Pace},
  {Garrison-Kimmel}, {Boylan-Kolchin}, {Wetzel}, {Bullock}, {Kere{\v{s}}},
  {Faucher-Gigu{\`e}re}, \& {Quataert}}]{Wheeler2018}
{Wheeler}, C., {Hopkins}, P.~F., {Pace}, A.~B., {et~al.} 2019,
  \href{http://dx.doi.org/10.1093/mnras/stz2887}{\mnras, 490, 4447}

\bibitem[{{Willick} {et~al.}(1997){Willick}, {Courteau}, {Faber}, {Burstein},
  {Dekel}, \& {Strauss}}]{1997ApJS..109..333W}
{Willick}, J.~A., {Courteau}, S., {Faber}, S.~M., {et~al.} 1997,
  \href{http://dx.doi.org/10.1086/312983}{\apjs, 109, 333}

\bibitem[{{Willmer}(2018)}]{Willmer2018}
{Willmer}, C. N.~A. 2018,
  \href{http://dx.doi.org/10.3847/1538-4365/aabfdf}{\apjs, 236, 47}

\bibitem[{{Wu}(2020)}]{Wu:2001.00018}
{Wu}, J.~F. 2020, \href{http://dx.doi.org/10.3847/1538-4357/abacbb}{\apj, 900,
  142}

\bibitem[{{Xi} {et~al.}(2018){Xi}, {Taylor}, {Massey}, {Rhodes}, {Koekemoer},
  \& {Salvato}}]{Xi:1805.07407}
{Xi}, C., {Taylor}, J.~E., {Massey}, R.~J., {et~al.} 2018,
  \href{http://dx.doi.org/10.1093/mnras/sty1333}{\mnras, 478, 5336}

\bibitem[{{Zaritsky} {et~al.}(1997){Zaritsky}, {Smith}, {Frenk}, \&
  {White}}]{Zaritsky:1997ApJ...478...39Z}
{Zaritsky}, D., {Smith}, R., {Frenk}, C., \& {White}, S. D.~M. 1997,
  \href{http://dx.doi.org/10.1086/303784}{\apj, 478, 39}

\bibitem[{{Zibetti} {et~al.}(2009){Zibetti}, {Charlot}, \& {Rix}}]{Zibetti2009}
{Zibetti}, S., {Charlot}, S., \& {Rix}, H.-W. 2009,
  \href{http://dx.doi.org/10.1111/j.1365-2966.2009.15528.x}{\mnras, 400, 1181}

\bibitem[{{Zonca} {et~al.}(2019){Zonca}, {Singer}, {Lenz}, {Reinecke},
  {Rosset}, {Hivon}, \& {Gorski}}]{Zonca2019}
{Zonca}, A., {Singer}, L., {Lenz}, D., {et~al.} 2019,
  \href{http://dx.doi.org/10.21105/joss.01298}{The Journal of Open Source
  Software, 4, 1298}

\end{thebibliography}

\end{document}